\documentclass[12pt, a4paper]{report}

\usepackage[utf8]{inputenc}
\usepackage[T1]{fontenc}
\usepackage{amsmath}
\usepackage{amssymb}
\usepackage{graphicx}
\usepackage{listings}
\usepackage[a4paper, total={6.5in, 9in}]{geometry} 
\usepackage{longtable}
\usepackage{algorithm}
\usepackage{algorithmic}
\usepackage{times} 
\usepackage{natbib}
\usepackage{hyperref}
\usepackage{verbatim} 
\usepackage[dvipsnames]{xcolor}
\usepackage{algorithm}
\usepackage{algorithmic}

\hypersetup{
    colorlinks=true,
    linkcolor=blue,
    filecolor=magenta,      
    urlcolor=cyan,
    pdftitle={A Comparative Review of Program Synthesis Paradigigms},
    pdfpagemode=FullScreen,
}

\lstdefinestyle{mystyle}{
    commentstyle=\color{gray},
    keywordstyle=\color{blue},
    numberstyle=\tiny\color{gray},
    stringstyle=\color{purple},
    basicstyle=\ttfamily\footnotesize,
    breakatwhitespace=false,         
    breaklines=true,                 
    captionpos=b,                    
    keepspaces=true,                 
    numbers=left,                    
    numbersep=5pt,                  
    showspaces=false,                
    showstringspaces=false,
    showtabs=false,                  
    tabsize=2
}
\begin{document}

\begin{titlepage}
    \centering
    
    \includegraphics[width=4cm]{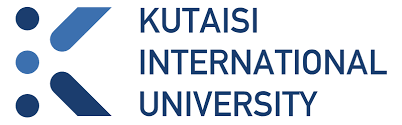} 
    
    \vspace{1.5cm}
    
    {\large\bfseries Kutaisi International University - School of Computer Science}
    
    \vspace{0.5cm}
    
    {Bachelor's Degree in Computer Science Program}
    
    \vfill
    
    {\large\bfseries Anna Arnania} \\
    \vspace{0.1cm}
    {\large\bfseries Zurabi Kobaladze} \\
    \vspace{0.1cm}
    {\large\bfseries Tamar Sanikidze}
    
    \vfill
    
    {\bfseries\large From Provable Correctness to Probabilistic Generation: \\[0.5em] A Comparative Review of Program Synthesis Paradigms}
    
    \vfill
    
    {The thesis is completed for obtaining the academic degree of \\ Bachelor of Science in Computer Science}

    \vspace{0.8cm}
    
    \begin{tabular}{c}
        \bfseries Supervisor: Besik Dundua \\
        Professor of Computer Science at Kutaisi International University \\
        \vspace{0.01cm} \\
        \bfseries Associate Supervisor: Isabella Dramnesc \\
         Professor of Mathematics and Informatics at \\ Universitatea de Vest din Timișoara
    \end{tabular}
    
    \vfill
    
    {\large Kutaisi, Georgia} \\
    {\large July 2025}
    
\end{titlepage}

\begin{abstract}
\noindent
Program synthesis, the automated generation of executable code from a high-level specification, has been a central goal of computer science for over half a century. This thesis presents a comparative literature review of the primary paradigms that have defined and shaped this field. It charts the historical and conceptual trajectory of program synthesis, from its origins in formal logic to the recent ascendancy of large-scale neural models. We analyze and contrast five major approaches: logic-based (deductive) synthesis, inductive (example-based) synthesis, sketch/schema-based synthesis, Large Language Model (LLM)-based synthesis, and neuro-symbolic hybrids. For each paradigm, we conduct a chronological examination, detailing its core motivations, foundational mechanisms, seminal systems, and practical applications. The analysis emphasizes the fundamental trade-offs inherent in each method, particularly the tension between correctness guarantees and specification burden, and between search complexity and expressive power. By tracing the evolution from systems that demand complete formal specifications to produce provably correct code (e.g., KIDS, Coq) to those that leverage vast code corpora to generate probabilistic solutions from natural language (e.g., Codex), this review synthesizes a comprehensive narrative of the field's progress. It highlights how enduring challenges, such as handling ambiguity and ensuring reliability, have driven the transition from purely symbolic reasoning to statistical and, ultimately, hybrid neuro-symbolic architectures, providing a structured perspective on the current state and future directions of automated programming.
\end{abstract}

\tableofcontents

\chapter{Logic-Based (Deductive) Program Synthesis}
\label{chap:deductive}

\section{Origins and Foundational Principles}

Logic-based program synthesis, or deductive synthesis, is the traditional and most rigorously established method for automated programming.  Its origins are intricately linked to the aspirations of early artificial intelligence and formal methods research during the 1960s and 1970s.  The primary impetus for this paradigm was the aspiration to create verifiably correct software.  In a time when software defects may result in disastrous outcomes and debugging was a laborious human task, the potential for the automatic generation of programs assuredly devoid of logical mistakes was a compelling impetus.

The intellectual foundations were laid by pioneers such as Cordell Green, who first framed program synthesis as a theorem-proving task \citep{green1969application}, and Zohar Manna and Richard Waldinger, whose work in the 1970s and 1980s established a comprehensive framework for deductive synthesis \citep{manna1980theory}. Their research proposed that a program could be seen as a constructive proof of the existence of an output satisfying a given input-output specification. The specification is expressed as a logical formula, typically in first-order logic:
$$ \forall \mathbf{x} \, \exists \mathbf{z} \, R(\mathbf{x}, \mathbf{z}) $$
In this context, $\mathbf{x}$ denotes the input variables, $\mathbf{z}$ signifies the output variables, and $R(\mathbf{x}, \mathbf{z})$ is a logical predicate delineating the requisite relationship between them.  The objective of a deductive synthesizer is to obtain a constructive proof of this theorem.  The computational output derived from this proof subsequently constitutes the requisite program.

\section{The ``Proofs-as-Programs'' Paradigm}

The fundamental mechanism of deductive synthesis is the principle of "proofs-as-programs," established by the Curry-Howard correspondence.  This isomorphism creates a direct connection between logical systems and computational models.  In its predominant manifestation, it associates intuitionistic logic with simply typed $\lambda$-calculus.  The principal correspondences are:

\begin{itemize}
    \item A \textbf{proposition} corresponds to a \textbf{type}.
    \item A \textbf{proof} of a proposition corresponds to a \textbf{program} (or term) of that type.
    \item \textbf{Proof normalization} (simplifying a proof) corresponds to \textbf{program execution} (evaluating a term).
\end{itemize}

Within this framework, the synthesis of a program $P$ that satisfies a specification $S$ is equivalent to generating a proof for the logical proposition that encapsulates $S$.  To create a function that sorts a list of integers, one need initially compose a formal specification: "For any input list $L_{in}$, there exists an output list $L_{out}$ such that $L_{out}$ is a permutation of $L_{in}$ and $L_{out}$ is ordered."  A deductive system then tries to establish this statement constructively.  The principles of inference utilized in the proof (e.g., induction, case analysis) determine the control structures (e.g., recursion, conditionals) of the resultant program.  Upon successful proof, the sequence of constructive steps is extracted and directly transformed into executable code.

\subsection{A Simplified Example: Synthesizing a Function}
Let's consider synthesizing a function `lesall(n, l)` that checks if a number `n` is less than or equal to all elements in a list `l`. The specification can be written as:
$$ \forall n \in \mathbb{Z}, \, \forall l \in \text{List}(\mathbb{Z}), \, \exists b \in \{\text{true}, \text{false}\} : b \leftrightarrow (\forall x \in l, n \leq x) $$
A deductive synthesizer works by applying transformation rules, often guided by a human. The process for synthesizing `lesall` might follow from a proof by structural induction on the list `l`.

\begin{enumerate}
    \item \textbf{Base Case}: $l$ is the empty list `[]`.
    \begin{itemize}
        \item The condition $(\forall x \in [], n \leq x)$ is vacuously true.
        \item The proof system deduces that the program must return `true`.
        \item \textit{Code generated}: `lesall(n, []) = true`
    \end{itemize}
    \item \textbf{Inductive Step}: $l$ is a non-empty list `h::t` (head `h` and tail `t`).
    \begin{itemize}
        \item \textbf{Inductive Hypothesis}: Assume `lesall(n, t)` correctly computes $(\forall x \in t, n \leq x)$.
        \item The goal is to prove $(\forall x \in (h::t), n \leq x)$, which is equivalent to $(n \leq h) \land (\forall x \in t, n \leq x)$.
        \item The proof proceeds by case analysis on the condition `n <= h`.
        \item If `n <= h` is true, the overall condition depends only on the truth of $(\forall x \in t, n \leq x)$, which is given by the recursive call `lesall(n, t)`.
        \item If `n <= h` is false, the entire conjunction is false, and the program must return `false`.
        \item \textit{Code generated}: `lesall(n, h::t) = if n <= h then lesall(n, t) else false`
    \end{itemize}
\end{enumerate}

Combining these cases gives the final, provably correct functional program:
\begin{lstlisting}[language=ML, caption={Synthesized `lesall` function in an ML-like syntax.}, label={lst:lesall}]
fun lesall(n, l) =
  case l of
    [] => true
  | h::t => if n <= h then lesall(n, t) else false;
\end{lstlisting}

\section{Key Systems and Implementations}

\subsection{The Kestrel Interactive Development System (KIDS)}
KIDS (Kestrel Interactive Development System), created by Douglas R. Smith at Kestrel Institute, exemplifies a deductive synthesis system grounded in program transformation \citep{smith1990kids}.  KIDS facilitated a user in program development by a sequence of correctness-preserving refinement phases, rather than solely relying on theorem proving.  The user would initiate with a comprehensive declarative definition and employ robust strategies, such as algorithmic design paradigms (e.g., divide and conquer, dynamic programming) or enhancements of data structures.

KIDS was effectively utilized for intricate, real-world issues, particularly in the area of transportation scheduling.  Its application to the k-queens issue, a classic combinatorial task, illustrated its effectiveness by producing a highly efficient, constant-time solution, highlighting a substantial performance enhancement over previously established techniques.

\subsection{The Coq Proof Assistant}
The Coq proof assistant, although not solely a program synthesizer, is a contemporary and robust tool that exemplifies the proofs-as-programs principle \citep{bertot2004interactive}.  Coq is founded on the Calculus of Inductive Constructions, a comprehensive logical framework.  Users compose formal specifications and subsequently direct Coq to create a proof.  Upon completion of the proof, Coq's \textit{extraction} process can autonomously produce functional code in languages such as OCaml, Haskell, or Scheme.

 This methodology has been employed to create \textit{certified software}, wherein the code is accompanied by a machine-verifiable confirmation of its accuracy.  The CompCert project represents a significant milestone, being a C compiler that is formally verified in Coq, which ensures that the semantics of the source program are maintained in the built executable \citep{leroy2009formal}.

\subsection{The Theorema System and Advanced Techniques}
The Theorema system, created by Bruno Buchberger's research group, is a notable platform for formal mathematics and deductive synthesis.  A significant contribution within this framework is the research conducted by Isabella Dramnesc on proof-based synthesis of sorting algorithms.  Her research illustrates sophisticated methodologies for directing the synthesis of intricate algorithms.

An essential innovation in her work is the use of \textbf{multisets} to formalize the specification of sorting.  Defining a sorted list as a permutation of the input that is also ordered successfully captures the "permutation" property through the requirement that the multiset of items in the output list is congruent to the multiset of elements in the input list \citep{dramnesc2006synthesis}.  This establishes a definitive and formal foundation for the synthesis proof.

Furthermore, Dramnesc introduced a systematic strategy for synthesizing sorting algorithms like Insertion Sort and Selection Sort. This involved a \textbf{``cascading'' synthesis approach}, where the proof for the main sorting function requires the existence of an auxiliary function (e.g., `insert` for Insertion Sort). The system then automatically triggers a new synthesis sub-goal for this auxiliary function, derives its code, and integrates it back into the main proof, demonstrating a structured and compositional approach to deductive synthesis \citep{dramnesc2005proof}.

\section{Strengths and Weaknesses}

The main and exclusive advantage of logic-based synthesis is its capacity to produce \textbf{provably correct programs}.  The extracted code is a direct product of a formal proof, offering the highest assurance of reliability for the specification.  This is essential for mission-critical sectors such as aircraft, medical equipment, and security protocols.

This paradigm encounters substantial drawbacks that have constrained its broad implementation:
 \begin{itemize} 
    \item \textbf{} {Specification Obligation:}  The primary problem is the necessity for a comprehensive, unambiguous, and accurate formal specification.  Crafting such specifications is frequently as challenging, if not more so, than composing the code itself.
     \item \textbf{} Scalability and Automation:  The domain for proofs is vast.  Completely automated synthesis is possible only for very simple applications.  The approach necessitates considerable human oversight for complex software, transforming the synthesizer into a "proof assistant" instead of a completely autonomous instrument.
     Limited Expressiveness:  Although theoretically expressive, user interaction frequently depends on rigid transformation rules and strategies, which may poorly encompass innovative algorithmic concepts.
\end{itemize}

Despite these constraints, the legacy of deductive synthesis is significant.  Its concepts underpin contemporary type systems, compiler verification, and establish the theoretical foundation for hybrid synthesis methods that seek to merge its rigor with the adaptability of alternative paradigms.

\chapter{Inductive Program Synthesis: Generalization from Examples}
\label{chap:inductive}

\section{Introduction}

In contrast to methods that demand complete, formal logical specifications, inductive synthesis operates on a principle more aligned with human intuition: learning from examples. This chapter provides a comprehensive analysis of the inductive synthesis paradigm, examining its history, core methodologies, seminal systems, and the fundamental trade-offs that define its position within the wider landscape of program synthesis.

\subsection{Defining Inductive Synthesis and Programming-by-Example (PBE)}

Inductive Program Synthesis, also known as Inductive Programming (IP), is the process of automatically generating a program from an incomplete specification \citep{gulwani2017program, kulesza2012end}. The defining characteristic of this paradigm is that the user's intent is conveyed through a set of concrete examples, most commonly in the form of input-output (I/O) pairs \citep{gulwani2012dimensions}. The synthesis system then performs an act of inductive reasoning—a logical leap from specific instances to a general rule—to produce a program that not only satisfies the provided examples but also generalizes to handle new, unseen inputs \citep{summers1977methodology, lee2024code}.

The field is often discussed through two closely related sub-paradigms: Programming by Example (PBE) and Programming by Demonstration (PbD).
\begin{itemize}
    \item In \textbf{Programming by Example (PBE)}, the user provides a prototypical \textit{product} of the desired computation. For instance, to specify a program that formats names, a user might provide the input "john f. kennedy" and the desired output "J. F. Kennedy" \citep{gulwani2011automating}. This has become the dominant interaction model in modern end-user applications.
    \item In \textbf{Programming by Demonstration (PbD)}, the user performs a sequence of \textit{actions} that constitute a trace of the computation, which the system then records and generalizes \citep{cypher1993watch}. This approach has been historically significant, particularly in robotics where demonstrating a physical trajectory is a natural form of programming \citep{argall2009survey}.
\end{itemize}

While historically distinct, the line between PBE and PbD has blurred, as many modern PBE systems can infer a plausible computational trace from a single input-output example, making the distinction less critical in many application domains. The ultimate goal for both is to find a program that correctly generalizes from the specific examples provided by the user \citep{lee2024code}. This act of generalization is both the source of the paradigm's power and its greatest challenge.

\subsection{The Core Philosophy: From Concrete Instances to General-Purpose Programs}

The main reason for inductive synthesis is to resolve the "specification bottleneck" that has traditionally constrained the actual implementation of automated programming methods \citep{gulwani2017program}.  Formulating a comprehensive, formal logical specification, as required by deductive synthesis methodologies, is a challenging and error-prone effort that frequently exceeds the proficiency of even experienced programmers, much less the general populace \citep{gulwani2012dimensions}.  Conversely, offering examples is an inherent and instinctive method for individuals to convey intent.

This philosophical transition from formal definition to example-based specification aims to simplify programming.  An estimated 99\% of computer users are not expert programmers, however they often face repeated chores that may be automated with simple scripts \citep{gulwani2011automating}.  Analysis of user behavior on technical support forums indicates that when confronted with such tasks, people instinctively articulate their objectives using examples \citep{gulwani2012dimensions}.  Inductive synthesis seeks to accommodate users by offering a method to convert intuitive examples into functional code, thus enabling them to automate their workflows without the necessity of mastering a formal programming language.  This is especially significant in areas such as data wrangling, text editing, and spreadsheet manipulation, where repetitive operations frequently occur \citep{gulwani2017program}.

\subsection{A Tale of Two Paradigms: Contrasting Inductive and Deductive Synthesis}

To fully appreciate the unique characteristics of inductive synthesis, it is essential to contrast it with its classical counterpart, deductive synthesis. This comparison reveals a fundamental trade-off at the heart of the program synthesis field.

\begin{itemize}
    \item \textbf{Specification:} The most significant difference lies in the nature of the specification. Inductive synthesis begins with an \textit{incomplete} and inherently \textit{ambiguous} specification in the form of examples \citep{gulwani2012dimensions}. Deductive synthesis, conversely, requires a \textit{complete} and \textit{unambiguous} formal specification, typically expressed as logical formulae such as pre- and post-conditions or type signatures in a system like first-order or higher-order logic.
    \item \textbf{Correctness Guarantee:} This difference in specification leads to a profound difference in the guarantees provided by the output. A program generated through deductive synthesis is \textit{provably correct by construction}. The synthesis process itself constitutes a constructive proof that the output program satisfies the formal specification. In contrast, a program generated through inductive synthesis is merely a \textit{hypothesis}. It is guaranteed to be consistent with the provided examples, but its correctness on unseen data is a matter of probabilistic confidence, not logical certainty \citep{summers1977methodology}. The generalization is an inductive leap, which is by definition unsound.
    \item \textbf{The Fundamental Trade-off:} This establishes the core compromise in program synthesis. Inductive synthesis prioritizes \textbf{usability and accessibility} by accepting intuitive but ambiguous specifications, at the cost of providing \textbf{no formal correctness guarantees}. Deductive synthesis prioritizes \textbf{formal correctness guarantees} but does so at the cost of requiring a \textbf{high-effort, difficult-to-produce specification}.
\end{itemize}

The deductive paradigm is exemplified by the work of Isabella Drămnesc and her collaborators.  Their research is concentrated on the synthesis of algorithms, including sorting routines for lists and binary trees, from formal logical specifications through proof-based methods \citep{dramnesc2005proof, dramnesc2006synthesis}.  Their methods generate a formal proof of the existence of an output that satisfies the specified properties by employing proof assistant frameworks such as Theorema and Coq (e.g., a sorted list that is a permutation of the input list).  Subsequently, the stages of this constructive proof are directly used to extract the desired algorithm.  This method, which converts a theorem-proving task into a program-generation task, is in striking contrast to the example-driven, search-based nature of inductive synthesis.

This distinction is not merely technical; it is indicative of a more profound, philosophical disagreement regarding the most effective method of reconciling the divide between human intent and machine execution.  The cognitive burden of generating a flawless specification is imposed on the user by the deductive approach, which necessitates that they learn to articulate their intentions in the precise, unambiguous language of formal logic.  On the other hand, the inductive approach endeavors to engage with the user on their own terms, accepting intuitive but imperfect examples and transferring the cognitive burden to the synthesis system to infer the user's true intent.  The substantial commercial success of PBE systems such as FlashFill indicates that this latter approach is the more pragmatic and impactful choice for a wide range of common, commonplace tasks, effectively transferring the burden of formal reasoning from the user to the machine. citep{gulwani2011automating}.

\section{Historical Trajectory and Foundational Motivations}

The development of inductive program synthesis is a narrative of evolving objectives, from the academic pursuit of general artificial intelligence to the pragmatic objective of empowering end-users.  This trajectory was influenced by decades of research in a variety of disciplines, which ultimately resulted in breakthroughs that enabled the technology to be installed on millions of desktops.

\subsection{Early Origins in Artificial Intelligence and LISP Programming}

The broader objectives of artificial intelligence research in the late 1960s and 1970s can be traced back to the intellectual foundations of inductive synthesis \citep{gulwani2017program}.  The synthesis of recursive programs, particularly in LISP, was investigated by early researchers based on a limited number of input-output examples \citep{summers1977methodology}.  The work of Summers in 1977 was a significant contribution from this era. It introduced a systematic, analytical method for deriving LISP programs by identifying recurrence relations within the structure of the I/O pairs \citep{summers1977methodology}.  This innovative research paved the way for future analytical methods and established the feasibility of a data-driven approach.  Nevertheless, these early systems were primarily restricted to academic research due to the computationally intractable and restricted nature of the problem of synthesizing general-purpose recursive functions from sparse examples.

\subsection{The PBE/PbD Dichotomy: A Fork in the Road}

The mid-1980s saw the formal introduction of the terms "Programming by Example" (PbE) and "Programming by Demonstration" (PbD) to denote methods for specifying operations without the need to acquire a programming language \citep{cypher1993watch}.  Although they were initially used interchangeably, their meanings diverged as they were adopted by various research communities.

\begin{itemize}
    \item \textbf{Programming by Example (PbE)} became the preferred term within the software development and Human-Computer Interaction (HCI) communities. The focus was on inferring programs from static input-output pairs provided by a user within a graphical interface. Influential early systems like SmallStar explored this paradigm for office information systems, and foundational books such as Allen Cypher's \textit{Watch What I Do: Programming by Demonstration} (1993) and Henry Lieberman's \textit{Your Wish is My Command: Programming By Example} (2001) championed the potential of PBE to empower end-users \citep{cypher1993watch, lieberman2001your}.
    \item \textbf{Programming by Demonstration (PbD)} was more widely embraced by robotics researchers. In this context, physically demonstrating a task—such as guiding a robot arm through a series of movements—was a more natural and effective way to "program" its behavior. This field later evolved, incorporating insights from neuroscience and social sciences, and is now often referred to as "Learning by Imitation" \citep{argall2009survey}.
\end{itemize}

\subsection{The Emergence of End-User Programming as a Driving Force}

A significant change in the field occurred when the research focus shifted from the ambitious objective of synthesizing complex, general-purpose algorithms to the more practical problem of automating simple, repetitive tasks for non-programmers \citep{gulwani2012dimensions, gulwani2017program}.  This change was motivated by the realization that a significant, underrepresented population of computer users frequently encountered tedious tasks that could be resolved with small, one-time scripts \citep{gulwani2011automating}.  Users naturally express their intent with examples when confronted with such issues, as evidenced by the abundance of data that has been generated by the proliferation of online assistance forums \citep{gulwani2012dimensions}.

The development and groundbreaking commercial success of \textbf{FlashFill}, a PBE feature that was integrated into Microsoft Excel starting with the 2013 version, were the result of this realization \citep{gulwani2011automating}.  By merely supplying one or two examples, FlashFill enabled any Excel user to execute intricate string transformations.  A watershed moment occurred in the field with its release.  It transformed inductive synthesis from a specialized academic curiosity to a technology with practical, mass-market appeal that is industrial-strength and robust.

The success of inductive synthesis is, therefore, a direct result of a strategic narrowing of its ambition.  The synthesis of any arbitrary recursive program from a few examples was a profoundly difficult problem, and the field did not become practical as a result.  As an alternative, it achieved success by identifying and resolving a "killer application": structured data manipulation for end-users.  Sumit Gulwani's research on FlashFill was particularly focused on this well-defined and specific domain.  His team developed a system that was highly effective, quick, and extremely beneficial for a common set of problems by designing a highly specialized Domain-Specific Language (DSL) for string transformations and combining it with a clever and efficient search algorithm.  The commercial success and popular acclaim of FlashFill validated this strategy, illustrating that the immediate value of PBE was not in the replacement of expert programmers for complex software engineering, but in the empowerment of non-programmers to manage their own simple data tasks.  In turn, this success stimulated a new wave of industrial investment and research in the field, resulting in the development of more general frameworks such as PROSE and applications in other data-centric domains \citep{polozov2015flashmeta}.

\section{Methodological Underpinnings: The Search for User Intent}

Inductive synthesis is fundamentally a search problem: when presented with a collection of examples and a language of potential programs, the system is obligated to identify a program that is consistent with the examples \citep{summers1977methodology}.  The primary obstacle is that this search space is typically vast, even for straightforward tasks.  The field has made a consistent effort to develop more intelligent techniques in order to efficiently converge on the user's intended program and tame this combinatorial explosion throughout its history.

\subsection{Taming the Search Space: The Critical Role of Domain-Specific Languages (DSLs)}

The most fundamental technique for making inductive synthesis tractable is to dramatically restrict the search space. Instead of searching through all possible programs in a general-purpose language like Python or C++, modern synthesizers search for a program within a carefully designed \textbf{Domain-Specific Language (DSL)} \citep{gulwani2011automating}.

A domain-specific language (DSL) is a compact programming language that is particularly well-suited for a particular problem domain. For instance, the DSL that underpins FlashFill includes primitive operators for string manipulation, including \texttt{Concatenate}, \texttt{Substring}, and \texttt{Match} (which are implemented using regular expressions), but it does not include constructs for network communication or file system access \citep{gulwani2012dimensions}. The DSL creates a potent \textit{inductive bias} \citep{summers1977methodology} by restricting the available components to only those that are pertinent to the task domain. This bias limits the synthesizer to producing only "reasonable" programs, thereby significantly reducing the search space and enabling real-time synthesis \citep{gulwani2011automating}. The Syntax-Guided Synthesis (SyGuS) paradigm formalizes this concept by mandating that the user submit not only a specification (such as examples) but also a context-free grammar that delineates the syntactic structure of the search space \citep{alur2013syntax}.

\subsection{A Taxonomy of Search Algorithms}

Within the constrained space defined by a DSL, various algorithms can be used to find a consistent program.

\begin{itemize}
    \item \textbf{Enumerative Search:} The most direct method is to systematically enumerate all possible programs in the DSL, usually in increasing order of size or complexity, and test each one against the user's examples until a match is found \citep{alur2013syntax}. While conceptually simple, this brute-force approach is often too slow for practical applications. A key optimization is the principle of \textit{observational equivalence}. If two subprograms, \texttt{p1} and \texttt{p2}, produce the exact same outputs for all available example inputs, they are indistinguishable from the perspective of the synthesizer. Therefore, only one of them needs to be retained for building larger, more complex programs, effectively pruning the search tree \citep{udupa2013transit}.
    \item \textbf{Version Space Algebra (VSA):} This more advanced technique, which forms the algorithmic core of systems like FlashFill, avoids enumerating individual programs altogether \citep{gulwani2011automating}. A \textit{version space} is a compact data structure that implicitly represents the set of \textit{all} programs in the DSL that are consistent with the given examples \citep{lau2003programming}. The synthesis algorithm works by composing these version spaces. For example, to synthesize a program \texttt{P = Concat(P1, P2)}, the synthesizer can take the version space representing all valid programs for the first part of the output (\texttt{P1}) and the version space for the second part (\texttt{P2}) and combine them to compute a new, composite version space for \texttt{P}. This algebraic manipulation of sets of programs is far more efficient than testing them one by one.
    \item \textbf{Counterexample-Guided Inductive Synthesis (CEGIS):} CEGIS is a powerful and widely used architecture that elegantly combines inductive synthesis with formal verification \citep{solar2008sketch}. It operates in a feedback loop between a \textit{generator} and a \textit{verifier}:
    \begin{enumerate}
        \item \textbf{Generator:} An inductive synthesizer (which could use enumerative search or VSA) proposes a candidate program \texttt{P} that is consistent with the current set of known examples.
        \item \textbf{Verifier:} A verifier, often a powerful constraint solver like an SMT solver, checks if the candidate program \texttt{P} satisfies a more general, formal specification. In PBE systems where no formal specification exists, the "verifier" can be the user, who is asked to validate the program's output on a new input.
        \item \textbf{Refinement:} If the verifier finds that \texttt{P} is incorrect, it returns a \textit{counterexample}—a specific input on which \texttt{P} fails. This counterexample is then added to the set of examples given to the generator, refining the specification and forcing the next candidate program to be correct on this new data point as well. The loop continues until a program is found that the verifier cannot refute.
    \end{enumerate}
    CEGIS is the core engine behind the influential SKETCH synthesis system and provides a robust framework for bridging the gap between ambiguous examples and more rigorous correctness requirements \citep{solar2008sketch}.
\end{itemize}

\subsection{The PROSE Framework: A Meta-Algorithmic Approach}

\citep{polozov2015flashmeta} The PROSE framework, which was initially referred to as FlashMeta and was developed at Microsoft, is a significant advancement in the generalization of the principles of inductive synthesis.  It introduces a meta-algorithm known as \textbf{Data-Driven Domain-Specific Deduction (D4)}, which elegantly distinguishes the domain-agnostic search algorithm from the domain-specific logic of the DSL operators.

The concept of \textbf{witness functions} is the primary innovation in PROSE.  Not only does a DSL designer define an operator, but they also provide its forward semantics (a function that computes an output from inputs) and its inverse semantics (a witness function that, given a desired output, deduces the set of possible inputs that could have produced it).  For instance, the witness function would deduce all possible pairs of substrings \texttt{(o1, o2)} such that \texttt{o1 + o2 = o} given an output string \texttt{o}. This is in reference to a concatenation operator \texttt{Concat(s1, s2)}.

The D4 algorithm employs these witness functions to conduct a top-down deductive search that is highly efficient.  The algorithm first invokes the witness function for the top-level operator \texttt{F} in order to synthesize a program \texttt{P = F(P1, P2)} that must generate the desired output \texttt{o}.  For the sub-programs \texttt{P1} and \texttt{P2}, this function determines the necessary outputs, \texttt{o1} and \texttt{o2}.  The synthesizer is subsequently invoked recursively to resolve these new, simpler sub-problems \citep{polozov2015flashmeta}.  The search space is significantly reduced by the deductive propagation of constraints from the output inward, resulting in a potent hybrid of inductive specification and deductive search.

\subsection{Algorithmic Model: The CEGIS Loop}

The CEGIS loop offers a high-level, lucid algorithmic model that encapsulates the essence of numerous contemporary inductive synthesis systems.  It exemplifies the iterative refinement process that is essential for the management of ambiguous specifications.

\begin{algorithm}
\caption{Simplified pseudocode for the Counterexample-Guided Inductive Synthesis (CEGIS) loop, based on descriptions in \citep{solar2008sketch, alur2013syntax}.}
\label{alg:cegis}
\begin{algorithmic}[1]
\STATE \textbf{function} CEGIS(\textit{Specification S})
\STATE \quad \textit{E} $\leftarrow$ InitialExamples(\textit{S}) \COMMENT{E is the set of examples/counterexamples}
\STATE \quad \textbf{loop}
\STATE \quad \quad \textit{P} $\leftarrow$ Synthesize(\textit{E}) \COMMENT{Generator: Inductive synthesis from examples}
\STATE \quad \quad \textbf{if} \textit{P} is null \textbf{then}
\STATE \quad \quad \quad \textbf{return} "SYNTHESIS\_FAILED"
\STATE \quad \quad \textbf{end if}
\STATE \quad \quad (\textit{is\_correct}, \textit{counterexample}) $\leftarrow$ Verify(\textit{P}, \textit{S}) \COMMENT{Verifier: Check against full specification}
\STATE \quad \quad \textbf{if} \textit{is\_correct} \textbf{then}
\STATE \quad \quad \quad \textbf{return} \textit{P} \COMMENT{Found a correct program}
\STATE \quad \quad \textbf{else}
\STATE \quad \quad \quad \textit{E} $\leftarrow$ \textit{E} $\cup$ \{\textit{counterexample}\} \COMMENT{Refine: Add counterexample and repeat}
\STATE \quad \quad \textbf{end if}
\STATE \quad \textbf{end loop}
\STATE \textbf{end function}
\end{algorithmic}
\end{algorithm}

The trajectory of these search algorithms from brute force to more intelligent reasoning is evident in their evolution.  The field evolved from basic "generate-and-test" enumeration to more complex "search space management" through the application of techniques such as observational equivalence and Version Space Algebra \citep{gulwani2011automating, udupa2013transit}.  The implementation of the CEGIS feedback loop signified a transition to "guided search," in which failures are utilized to enhance the problem \citep{solar2008sketch}.  Lastly, frameworks such as PROSE, which employ inverse semantics, are a step toward "constrain-and-deduce," a method in which logical reasoning is employed to actively reduce the search space prior to enumeration \citep{polozov2015flashmeta}.

\section{Seminal Systems and Key Application Domains}

The theoretical advancements in inductive synthesis have given rise to a number of influential systems, each targeting specific domains and demonstrating the practical utility of the paradigm. The most effective of these have advanced from research prototypes to features in commercial software that is widely used.

\subsection{The Canonical Success Story: FlashFill}

\begin{itemize}
    \item \textbf{System and Domain:} Since the 2013 version, Microsoft Excel has included a feature called FlashFill that is intended to automate syntactic string transformations in spreadsheets \citep{gulwani2011automating}. Reformatting a list of names, such as changing a column of "First Last" names to "Last, F." format, is a common use case.
    \item \textbf{User Interaction and Technology:} The user interface is incredibly straightforward. In a neighboring column, a user gives only one or two instances of the intended change. Instantaneously identifying a pattern, FlashFill uses its internal string-manipulation DSL to synthesize a program and previews the outcomes for the remaining rows. When you confirm (by hitting Enter, for example), the transformed data is added to the column. Instead of being dynamic formulas that update in response to changes in the source data, the generated outputs are static text values.
    \item \textbf{Algorithmic Core:} The highly optimized version space algebra (VSA) that powers FlashFill's engine runs over its unique DSL. This DSL contains elements such as \texttt{Concatenate}, \texttt{Substring} (which can be defined by matching regular expressions or by absolute positions), and \texttt{ConstantString} \citep{gulwani2012dimensions}. FlashFill uses a complex ranking model to choose the most believable or "simplest" program, which is crucial to its usability, because a lot of different programs can be consistent with a limited number of examples.
    \item \textbf{Impact and Evolution:} Millions of end users have benefited from the power of program synthesis thanks to FlashFill, the most frequently cited example of a commercially successful PBE system. Its real-time performance, its sharp focus on a high-value, well-defined problem, and its smooth integration into a familiar user interface are all major factors in its success. With \textbf{FlashFill++}, which scales the synthesis to larger DSLs and incorporates more sophisticated operators for data types like dates and numbers, the technology has advanced further \citep{singh2016automated}.
\end{itemize}

\subsection{Data Wrangling and Extraction}

The crucial but frequently time-consuming process of cleaning, converting, and mapping raw data into an organized format appropriate for analysis is called data wrangling, sometimes referred to as data munging. PBE is ideally suited for this process, which can take up to 80\% of a data scientist's time \citep{kandel2011wrangler}.

\begin{itemize}
    \item \textbf{FlashExtract:} This system is designed to extract structured data from semi-structured documents like text files, web pages, and server logs \citep{le2014flashextract}. The user gives examples by merely highlighting and labeling the desired data fields in a sample document, eliminating the need to write intricate regular expressions. Then, using a DSL and operators like \texttt{Split}, \texttt{Filter}, and \texttt{Map}, FlashExtract creates an extraction program. The technology has been incorporated into industrial products such as Azure Operational Management Suite and Microsoft PowerShell \citep{polozov2015flashmeta}.
    \item \textbf{Wrangler / Data Wrangler:} An interactive tool for data transformation and cleaning was the original Wrangler system \citep{kandel2011wrangler}. The current \textbf{Data Wrangler} extension for Visual Studio Code carries on this tradition. Using a few examples, this tool enables users to execute intricate data transformations on Pandas DataFrames by integrating the PROSE synthesis engine.
    \item \textbf{StriSynth:} By handling hierarchical data types (such as lists of files) and supporting a wider range of operations, this tool expands on the fundamental concepts of FlashFill \citep{piskac2015automating}. According to a formal user study, StriSynth was much faster for users than writing conventional PowerShell scripts for some scripting tasks \citep{mayer2015user}.
\end{itemize}

\subsection{Synthesizing Recursive and Structured Programs: ESCHER}

Although loop-free data transformations are the main focus of many effective PBE systems, some research has pushed the envelope in the direction of creating more intricate, recursive programs.

\begin{itemize}
    \item \textbf{System and Domain:} ESCHER is a generic inductive synthesis system designed specifically to produce recursive programs from I/O examples \citep{albarghouthi2013escher}. It can be applied to various domains, including classical recursive algorithms over integers, lists, and trees.
    \item \textbf{Methodology:} ESCHER employs a component-based search algorithm that alternates between a forward search phase and a conditional inference phase, using a unique data structure called a goal graph to intelligently introduce \texttt{if-then-else} control flow.
    \item \textbf{Comparative Significance:} ESCHER's capability to synthesize recursive programs brings it closer in scope to traditional algorithm synthesis. Its inductive methodology provides a stark contrast to the deductive, proof-based synthesis of recursive sorting algorithms by researchers like Drămnesc \citep{dramnesc2005proof}, clearly illustrating the fundamental differences between the two paradigms when applied to similar problems.
\end{itemize}

\begin{longtable}{|p{0.20\textwidth}|p{0.22\textwidth}|p{0.3\textwidth}|p{0.13\textwidth}|p{0.15\textwidth}|}
\caption{A structured summary of key inductive synthesis systems.} \label{tab:systems} \\
\hline
\textbf{System Name} & \textbf{Primary Domain(s)} & \textbf{Core Synthesis Technique(s)} & \textbf{Spec. Method} & \textbf{Key Publication(s)} \\
\hline
\endfirsthead
\multicolumn{5}{c}%
{{\bfseries \tablename\ \thetable{} -- continued from previous page}} \\
\hline
\textbf{System Name} & \textbf{Primary Domain(s)} & \textbf{Core Synthesis Technique(s)} & \textbf{Spec. Method} & \textbf{Key Publication(s)} \\
\hline
\endhead
\hline \multicolumn{5}{r}{{Continued on next page}} \\
\endfoot
\hline
\endlastfoot
\textbf{FlashFill} & String Manipulation & Version Space Algebra, DSL, Ranking & I/O Examples & \citep{gulwani2011automating} \\
\hline
\textbf{FlashExtract} & Data Extraction (Text, Web) & DSL, Top-Down/Bottom-Up Inference & Annotated Examples & \citep{le2014flashextract} \\
\hline
\textbf{StriSynth} & File/String Manipulation & Extends FlashFill, PBE & I/O Examples & \citep{piskac2015automating} \\
\hline
\textbf{ESCHER} & Recursive Programs (Ints, Lists, Trees) & Component-based, Goal Graph, Forward Search & I/O Examples & \citep{albarghouthi2013escher} \\
\hline
\textbf{Lapis} & Text Editing & Text Constraints, Selection Guessing & Pos/Neg Examples & \citep{miller2001lapidary} \\
\hline
\textbf{Wrangler} & Data Transformation/Wrangling & PBE, Interactive Refinement & User Interactions & \citep{kandel2011wrangler} \\
\hline
\textbf{PROSE} & Meta-Framework & Deductive Search, Witness Functions, DSL & I/O Examples & \citep{polozov2015flashmeta} \\
\hline
\end{longtable}

\section{A Critical Analysis of Strengths and Weaknesses}

Despite its impressive practical achievements, inductive synthesis presents a number of inherent difficulties due to its fundamental reliance on imprecise specifications. A critical examination reveals a terrain of significant benefits counterbalanced by basic drawbacks.

\subsection{Core Strengths}

The primary advantages of inductive program synthesis stem directly from its user-centric philosophy.

\begin{itemize}
    \item \textbf{Accessibility:} The significant strength of IPS is its ability to empower non-programmers. By accepting examples, it dramatically lowers the cognitive barrier to creating small, functional programs \citep{gulwani2012dimensions}.
    \item \textbf{Efficiency for Repetitive Tasks:} The paradigm excels at automating tasks that are tedious and error-prone when performed manually, especially in data cleaning and text reformatting \citep{gulwani2017program}.
    \item \textbf{Flexibility:} The core idea of learning from examples is highly flexible and has been applied to a diverse range of domains, from string manipulation to parser generation \citep{gulwani2017program}.
    \item \textbf{Seamless Integration:} As demonstrated by FlashFill in Excel, PBE can be integrated seamlessly into existing applications, increasing discoverability and adoption.
\end{itemize}

\subsection{Fundamental Challenges and Limitations}

\subsubsection{The Ambiguity Problem}

Ambiguity is the main obstacle in PBE. A limited collection of I/O examples is a \textit{under-specified} problem, which means that numerous different programs may behave differently on unseen inputs while still being consistent with the examples \citep{summers1977methodology, lee2024code}. On the examples, the synthesizer may learn a program that is technically correct but not what the user intended. To solve this, advanced techniques are needed:

\begin{itemize}
    \item \textbf{Ranking and Inductive Bias:} To choose from the many consistent programs, synthesizers employ ranking functions to prefer the "simplest" or "most likely" program \citep{summers1977methodology}. The design of the DSL itself imposes a strong inductive bias.
    \item \textbf{Interactive Disambiguation:} Interaction is key to resolving ambiguity. This can involve the system generating a \textit{distinguishing input} to ask the user, or allowing users to provide \textit{augmented examples} with richer semantic annotations.
\end{itemize}

\subsubsection{The Scalability Bottleneck}

Scalability is the second significant obstacle. The size of the desired program causes the search space of potential programs to expand exponentially \citep{summers1977methodology}. For real-time performance, this must be resolved.

\begin{itemize}
    \item \textbf{DSL Design:} A minimal, highly constrained DSL is the most effective strategy for managing the search space size \citep{gulwani2011automating}.
    \item \textbf{Intelligent Search Pruning:} Techniques like observational equivalence prune the search by collapsing sub-programs that behave identically on the given examples \citep{udupa2013transit}.
    \item \textbf{Deductive Pruning:} Advanced frameworks like PROSE use deductive reasoning via "witness functions" to prune the search space top-down before enumeration begins \citep{polozov2015flashmeta}.
\end{itemize}

\subsubsection{Correctness and User Confidence}

The third and most basic flaw is that, from the standpoint of formal logic, inductive reasoning is fundamentally flawed \textit{unsound} \citep{summers1977methodology}. A synthesized program is not a proven theorem; rather, it is a hypothesis. It is not guaranteed to respond appropriately to inputs that are not visible.

A convincing example of this problem can be found in the StriSynth tool's user study \citep{mayer2015user}. Ironically, users rated PowerShell as more "helpful" even though they finished tasks more quickly using the PBE tool. This implies that users may value the consistency and explicit control of manual coding more than the unadulterated speed of a "black box" synthesizer whose generalizations they cannot completely rely on. When compared directly to deductive synthesis, where a system derives a program from a constructive proof and produces an artifact that is correct by construction with respect to its formal specification, this lack of formal guarantees is the most significant flaw in inductive synthesis \citep{dramnesc2005proof}.

\section{Conclusion}

Inductive program synthesis has firmly established itself as a vital and impactful paradigm. It has effectively democratized programming for a distinct and significant class of problems by emphasizing user accessibility through example-based specifications. Its journey is defined by the trade-off between the ease of providing examples and the lack of formal correctness guarantees. Its greatest achievements are in enabling end users to automate time-consuming tasks rather than in replacing skilled programmers for mission-critical software.

Hybridization, which combines traditional symbolic search with other computational paradigms, seems to be the way forward for inductive synthesis. A promising new direction is represented by the emergence of neuro-symbolic and LLM-based techniques, which use deep learning to direct symbolic search and more accurately deduce user intent from vague or natural language specifications. As these hybrid approaches develop, inductive synthesis is expected to continue to be a major force behind innovation in lowering the barrier to computation.

\chapter{Program Synthesis via Sketches and Schemas: Guiding Search with Structure}
\label{chap:sketch}

\section{The Principle of Constraint-Guided Synthesis: A Human-Computer Synergy}

Program synthesis, the automated construction of executable software from high-level specifications, has long been a central ambition in computer science. The field has historically been divided into two main paradigms: inductive synthesis, which generalizes programs from partial specifications such as input-output examples, and deductive synthesis, which derives provably correct programs from complete formal specifications. Despite their strength, both strategies have faced significant obstacles to broad acceptance. In contrast to traditional programming, deductive approaches frequently call for programmers to become proficient in intricate formalisms and perform time-consuming, interactive proof construction \citep{solar2008combinatorial}. Inductive approaches, on the other hand, struggle with accurate generalization from a limited number of examples and specification ambiguity \citep{singh2018interpretable}.

A third pragmatic paradigm emerged to overcome this dead end: synthesis based on sketches and schemas. Its fundamental tenet is the development of a potent and useful synergy between human insight and automated search, rather than the total replacement of the human programmer \citep{solar2008combinatorial}. This method directly addresses the "synergy problem" in synthesis, which is how to use a programmer's advanced algorithmic skills to limit the otherwise unmanageable search space of potential programs and lessen the computational load on the automated synthesizer \citep{solar2008combinatorial}.

This paradigm's primary innovation is the change in the specification's actual nature. In contrast to purely inductive synthesis, which concentrates on \textit{what} a program should compute, sketch-based synthesis enables the programmer to offer vital advice on \textit{how} it should compute \citep{solar2008program}. A partial program, sometimes referred to as a sketch or template, is used to convey this instruction. The high-level structure and algorithmic approach of an implementation are expressed in a sketch, while certain low-level details that are frequently laborious and prone to errors are left unspecified as "holes" \citep{solar2008combinatorial}. For example, a programmer may be aware that in order to avoid linear storage overhead, an efficient list reversal necessitates an iterative loop instead of recursion and that the new list must be built in-place. This high-level approach can be directly encoded by the programmer using a sketch, leaving the synthesizer to figure out the exact pointer manipulations and loop conditions needed for a proper implementation \citep{solar2008program}.

By concentrating the powerful potential of the automated search on precisely defined, bounded sections of the code, this methodology allows for a type of localized synthesis \citep{solar2008program}. For complicated, real-world problems that would be too difficult for a synthesizer to tackle from scratch, it does this by making the combinatorial search tractable. The programmer handles the creative, architectural decisions they are best equipped to make, while the synthesizer manages the exhaustive, detailed reasoning at which it excels \citep{solar2008program}. In fields like bit-level cryptography and concurrent data structures, where high-level structure is well understood but low-level implementation is infamously challenging, this division of labor has proven incredibly effective \citep{solar2008combinatorial}.

More conceptually speaking, "sketch" and "schema" are both mediating representations that connect an abstract, high-level idea with its specific, concrete instantiation. They offer a structural outline or framework that links the programmer's abstract algorithmic idea to the finished, executable code, much like philosophical ideas or artistic sketches do. The objectives of program synthesis have been significantly and practically reframed by this method. It presents synthesis as a potent tool for programmer support and productivity rather than aiming for the total automation of programming from abstract specifications. The paradigm's practical success and adoption in resolving difficult programming problems across a range of domains can be explained by this shift from programmer replacement to programmer empowerment.

\section{The Sketching Paradigm: From Holes to Programs}

The sketching paradigm uses a particular set of linguistic constructions, formal foundations, and potent search algorithms to implement the idea of human-computer synergy. It gives programmers a tangible way to convey their incomplete knowledge, which a synthesis engine then completes.

\subsection{Formal Foundations of a Sketch}
At its core, a program sketch is a parametric program. It can be formally defined as a program $P$ containing a set of unknown integer or boolean constants, referred to as "holes," denoted $H = \{??_1, \dots, ??_k\}$. The synthesis problem is to discover a control vector $\vec{c}$, which is an assignment of concrete values to these holes, such that the resulting completed program $P(\vec{c})$ satisfies a given specification, \textit{Spec}, for all valid inputs $i \in I$ \citep{solar2008program}. This relationship can be expressed formally as:
$$ \text{Find } \vec{c} \in \mathbb{Z}^k \text{ such that } \forall i \in I, \text{Spec}(P(\vec{c}), i) \text{ holds.} $$
The specification itself is typically provided as a set of assertions within the code or as a test harness that the completed program must pass \citep{solar2008program}. The power of this model lies in the expressiveness of the language constructs used to define the parametric program space.

\subsection{The Sketch Language Core Constructs}
Modern sketching frameworks provide several key language features that allow programmers to precisely define the search space for the synthesizer. \textbf{Holes (\texttt{??}):} This is the most fundamental construct in sketching. A hole, typically written as \texttt{??}, acts as a placeholder for an unknown integer or boolean value that the synthesizer must determine \citep{solar2013sketch}. A simple integer hole can represent a wide range of unknowns, from a missing constant in an arithmetic expression to a control-flow choice in an algorithm. For example, the classic XOR swap algorithm, which swaps two variables without a temporary variable, relies on a specific sequence of three XOR assignments. A programmer who remembers the operations but not the precise sequence or operands can write a sketch where holes represent the choices, and the synthesizer discovers the correct implementation \citep{solar2008program}.

\begin{verbatim}
// Sketch for the XOR swap algorithm
// The synthesizer must find boolean values for the three holes (??)
// to select the correct sequence of assignments.
harness void test_swap(int x_in, int y_in) {
    int x = x_in, y = y_in;
    
    // Sketch of the swap logic
    if (??) { x = x ^ y; } else { y = x ^ y; }
    if (??) { x = x ^ y; } else { y = x ^ y; }
    if (??) { x = x ^ y; } else { y = x ^ y; }

    // Specification: assert the final values are swapped
    assert x == y_in;
    assert y == x_in;
}

/* Synthesized Solution: The synthesizer will discover the control vector
that corresponds to the following concrete program.
void swap(ref int x, ref int y) {
    y = x ^ y;
    x = x ^ y;
    y = x ^ y;
}
*/
\end{verbatim}

\textbf{Expression Generators (\texttt{\{|...|\}}):} To sketch over a space of possible expressions rather than just constants, frameworks provide expression generators. These are often written using a regular-expression-like syntax, such as \texttt{\{| e1 | e2 |... |\}}, defining a set of choices for a part of an expression \citep{solar2008program}. This is particularly useful when the choice is between different variables or sub-expressions. For instance, to synthesize a function that doubles an integer, a programmer could sketch the operation as a multiplication between an unknown constant and a generator that chooses between the input variable \texttt{x} and the constant 0 \citep{singh2016jsketch}.

\begin{verbatim}
// Sketch for a method to double an integer, using a hole and a generator.
// From a JSketch example.
class SimpleMath {
    static int mult2(int x) {
        // The synthesizer must find a value for ?? and choose from {| x, 0 |}.
        return ?? * {| x , 0 |};
    }
}

// The specification is provided via a test harness with assertions.
class TestSimpleMath {
    harness static void test() {
        assert SimpleMath.mult2(3) == 6;
        assert SimpleMath.mult2(-5) == -10;
    }
}

// Synthesizer finds: ?? = 2, and chooses 'x' from the generator.
\end{verbatim}

\textbf{Reorder Blocks (\texttt{reorder \{...\}}):} In domains like concurrent programming, the exact ordering of statements is critical for correctness but difficult for humans to reason about. The \texttt{reorder} block is a powerful construct that instructs the synthesizer that the statements within the block can be executed in any order \citep{solar2008program}. The synthesizer is given the freedom to explore permutations of these statements, often guarded by synthesized conditions, to find a sequence that is free from race conditions, deadlocks, and other concurrency bugs. This construct effectively delegates the complex task of reasoning about thread interleavings to the automated tool.

\subsection{Formal Semantics of Sketches}
From a formal methods perspective, a sketch $S$ with holes does not define a single program but rather a set of concrete programs. The formal semantics of a sketch can be described by a relation that maps control vectors to concrete programs: $[[S]] = \{(\vec{c}, P_{\vec{c}}) \mid P_{\vec{c}} \text{ is the program resulting from filling holes in } S \text{ with values from } \vec{c}\}$ \citep{solar2008program}. The synthesis task is then to find a control vector $\vec{c}$ such that the corresponding program $P_{\vec{c}}$ satisfies the specification \textit{Spec}. This transformation from a program with holes into a logical constraint satisfaction problem is the foundational step that allows automated solvers to operate on sketches \citep{alur2013syntax}. Recent work has even explored synthesizing these formal semantics themselves from executable interpreters, further automating the construction of synthesis tools \citep{jeo2021synthesizing}.

\subsection{The Sketch and Rosette Toolchains: A Comparative Look}
The principles of sketching have been realized in several powerful toolchains, most notably the original Sketch system and the Rosette solver-aided language.

\subsubsection{The Sketch System}
Armando Solar-Lezama and his team created the Sketch synthesis system, which is the standard version of the sketching paradigm \citep{solar2008program}.  It gives you a programming language with syntax that is purposely similar to C and Java, along with the basic sketching constructs (\texttt{??}, \texttt{\{|...|\}}, \texttt{reorder}) \citep{solar2008program}.  The Sketch compiler takes a sketch and its specification harness and turns them into a complicated logical formula that is then sent to a SAT-based backend solver \citep{solar2008combinatorial}.  This design is pretty easy for programmers who know how to use imperative languages to understand.  There are tools in the ecosystem, such as JSketch, that let you write sketches directly in Java and translate them into the core Sketch language for solving \citep{singh2016jsketch}.  The system is open-source and comes with a lot of documentation and a language reference manual \citep{solar2013sketch}.

\subsubsection{Rosette: Solver-Aided Programming}
Rosette takes a different, broader approach to the same basic issue.  It is a programming language that uses a solver and is built into the functional language Racket \citep{torlak2013rosette}.  Rosette isn't a separate language; instead, it adds features to Racket that let you create symbolic values, define assertions and assumptions, and ask an underlying SMT solver, like Z3 \citep{torlak2013rosette}, questions.

In Rosette, sketching is not achieved through a dedicated \texttt{??} operator but by defining a Domain-Specific Language (DSL) and using Rosette's core features to create a sketch within that DSL. The key constructs are:
\begin{itemize}
    \item \textbf{Symbolic Values:} Using \texttt{define-symbolic} or \texttt{define-symbolic*}, a programmer can create variables whose concrete values are unknown \citep{torlak2013rosette}. These symbolic values represent the "holes" in the program.
    \item \textbf{Choice Operator (\texttt{choose*}):} This function takes a set of arguments and returns a symbolic expression that can evaluate to any one of them. This is used to define the space of operators or operands in a DSL \citep{torlak2013rosette}.
    \item \textbf{Synthesis Query (\texttt{synthesize}):} This is the main query for synthesis. It asks the solver to find concrete values for all symbolic choices within a program sketch such that a \texttt{\#:guarantee} clause (the specification) holds for all (\texttt{\#:forall}) symbolic inputs \citep{torlak2013rosette}.
\end{itemize}

The following example demonstrates how Rosette can synthesize a simple arithmetic expression within a user-defined DSL \citep{torlak2013rosette}:
\begin{verbatim}
#lang rosette

; Define a simple DSL for arithmetic expressions
(struct plus (left right) #:transparent)
(struct mul (left right) #:transparent)
(struct const (val) #:transparent)

; An interpreter for the DSL
(define (interpret p env)
  (match p
    [(const v) v]
    [(plus l r) (+ (interpret l env) (interpret r env))]
    [(mul l r) (* (interpret l env) (interpret r env))]))

; A sketch for a program of the form 'a*x + b' or 'a*(x+b)'
(define-symbolic a b integer?)
(define sketch
  (choose* (plus (mul (const a) 'x) (const b))
           (mul (const a) (plus 'x (const b)))))

; The specification: find a program equivalent to 2*x + 3
(define-symbolic x integer?)
(define solution
  (synthesize
    #:forall (list x)
    #:guarantee (assert (= (interpret sketch (hash 'x x))
                           (+ (* 2 x) 3)))))

; Extract and print the solution
(print-forms solution)
; Expected output might include:
; (model [a 2][b 3][0$choose... #f])
; which corresponds to the program (plus (mul (const 2) 'x) (const 3))
\end{verbatim}

\subsubsection{Comparative Analysis}
The two toolchains represent two opposing views. Many programmers find Sketch to be an easy-to-use imperative language that offers a direct route to solving particular, challenging algorithmic problems \citep{torlak2013rosette}. In contrast, Rosette offers a more potent and adaptable meta-framework. By utilizing Racket's extensive metaprogramming capabilities, it enables language designers to create their own solver-aided DSLs \citep{torlak2013rosette}. Rosette is a tool for creating tools that can reason about, verify, and synthesize programs, whereas Sketch is a tool for creating sketched programs \citep{torlak2013rosette}. Rosette is therefore especially well-suited for research and the creation of innovative synthesis systems for uncharted territory.

\section{The Schema-Based Paradigm: Reusable Algorithmic Knowledge}

The schema-based synthesis model is closely associated with sketching. Schema-based synthesis usually works at a higher level of abstraction, using reusable algorithmic templates to construct programs in particular domains, whereas sketching frequently concentrates on filling in low-level details within a code structure provided by the programmer.

\subsection{Schemas as Reusable Templates}
A generic representation of a family of algorithms or applications is the formal definition of a schema \citep{flener2004schema}. In essence, it is a high-level program template that contains computational knowledge specific to a given domain. Importantly, every schema has a set of \textit{applicability conditions}—logical restrictions that establish when the schema can be applied to a particular problem specification in a safe manner \citep{fischer2003autobayes}. These requirements may be related to the partially instantiated code itself, intermediate outcomes of the synthesis process, or characteristics of the original specification \citep{fischer2003autobayes}.

Recursive refinement defines the synthesis process. A high-level, frequently declarative problem specification is the first step in an AI-driven synthesis engine. After that, it looks for relevant schemas and applies them to the issue at hand as well as any new subproblems that may arise. This method, which frequently uses a platform-independent intermediate language, progressively converts the specification into executable code \citep{flener2004schema}. This methodology successfully blends two traditional synthesis approaches: it is \textit{generative} in that it builds a solution by creating program templates, and it is \textit{deductive} in that it applies a schema only after checking the applicability conditions using logical reasoning \citep{flener2004schema}.

In a more modern interpretation, especially in the context of Model-Driven Architecture (MDA), schema-based synthesis can be viewed as a method for automating model-to-model transformations. In this view, a schema defines a transformation from an input model (representing the problem space) to an output model (representing the solution space) \citep{flener2004schema}.

\subsection{Case Studies in Domain-Specific Synthesis}
The power of the schema-based approach is most evident in its application to complex, well-defined scientific and technical domains.

\subsubsection{Scientific Computing: AUTOBAYES and AUTOFILTER}
Two of the most prominent examples of schema-based synthesis are the AUTOBAYES and AUTOFILTER systems, developed at NASA Ames Research Center \citep{fischer2003autobayes}.
\begin{itemize}
    \item \textbf{Domain:} AUTOBAYES operates in the domain of statistical data analysis, while AUTOFILTER specializes in state estimation algorithms, particularly Kalman filters \citep{fischer2003autobayes}. These are critical domains for NASA, with applications ranging from analyzing Hubble Space Telescope imagery to vehicle navigation and control \citep{fischer2003autobayes}.
    \item \textbf{Input:} The user provides a very high-level, declarative specification in the form of a statistical model, which describes problem variables and their probabilistic dependencies \citep{fischer2003autobayes}.
    \item \textbf{Schemas:} The systems contain a library of schemas that represent high-level statistical algorithms (e.g., the Expectation-Maximization algorithm, k-Means clustering, Newton-Raphson optimization), mathematical simplifications, data type refinements, and code optimizations \citep{fischer2003autobayes}.
    \item \textbf{Process:} The synthesis kernel, often implemented in a logic programming language like Prolog, analyzes the input model and searches for a valid sequence of schema applications \citep{fischer2003autobayes}. This deductive process refines the abstract statistical problem into an ordinary optimization problem, which is then solved symbolically if possible, or with synthesized numerical code. The final output is optimized and fully documented C/C++ code that can be linked into environments like MATLAB \citep{fischer2003autobayes}.
\end{itemize}

\subsubsection{Database Program Refactoring}
Another compelling application demonstrates the tight integration of schema- and sketch-based techniques. When the schema of a database is refactored (e.g., a table is split, or an attribute is moved), all programs that interact with that database must be updated---a tedious and error-prone process \citep{qiu2018synthesizing}. A synthesis technique has been developed to automate this migration.
\begin{itemize}
    \item \textbf{Domain:} Database-backed applications undergoing schema evolution.
    \item \textbf{Input:} The original program $P$ operating on a source schema $S$, and the new target schema $S'$.
    \item \textbf{Process:} The synthesis algorithm decomposes the problem into three distinct stages, clearly illustrating a hierarchy of abstraction \citep{qiu2018synthesizing}:
    \begin{enumerate}
        \item \textbf{Value Correspondence (Schema Mapping):} First, the system guesses a candidate value correspondence $\Phi$, which is a high-level mapping specifying how attributes in the new schema $S'$ can be derived from attributes in the old schema $S$. This correspondence acts as a high-level schema for the transformation.
        \item \textbf{Sketch Generation:} Given this value correspondence, the algorithm generates a program sketch $\Omega$. This sketch is a partial program that represents the entire space of possible migrated programs that are consistent with the mapping $\Phi$. It contains holes for unknown tables, attributes, and conditions.
        \item \textbf{Sketch Completion:} Finally, a sketch solver searches for a concrete completion of $\Omega$ that is semantically equivalent to the original program $P$. Because SQL is not easily amenable to standard SMT solving, this step uses a specialized enumerative search over the sketch's completions.
    \end{enumerate}
\end{itemize}
This method is a great example of how sketch-based and schema-based synthesis are frequently complementary paradigms that function at different levels of abstraction rather than being mutually exclusive. The schema correspondence rules capture domain-specific, high-level knowledge about database evolution. A lower-level, more limited program sketch is produced when these rules are instantiated for a particular migration task. A more versatile synthesis engine then solves this sketch. By clearly separating the issues of domain-specific knowledge representation (schemas) from the general, domain-agnostic problem of combinatorial search (sketch solving), this layered approach is an effective approach for creating scalable and efficient synthesis tools.

\section{A Unified Perspective: Syntax-Guided Synthesis (SyGuS)}
The Syntax-Guided Synthesis (SyGuS) framework formalizes and unifies the ideas of offering structural guidance through sketches and schemas. This framework enables direct comparison and the creation of general-purpose solvers by offering a common language and definition of computational problems that cover a broad spectrum of contemporary synthesis techniques.

\subsection{The SyGuS Problem Formulation}
The input to a SyGuS problem consists of two key components \citep{alur2013syntax}:
\begin{enumerate}
    \item \textbf{A Semantic Specification:} A logical formula $\phi$, typically expressed in a background theory like bit-vector arithmetic or linear integer arithmetic, that the function to be synthesized, $f$, must satisfy. This formula is universally quantified over its inputs (e.g., $\forall x,y . \phi(f,x,y)$).
    \item \textbf{A Syntactic Specification:} A context-free grammar $G$ that defines the set of all allowed expressions, $L(G)$, that can be used for the implementation of $f$.
\end{enumerate}
The computational goal is to find an expression $e \in L(G)$ such that when $f$ is replaced by $e$ in the semantic specification, the resulting formula $\phi[f/e]$ is valid in the background theory \citep{alur2013syntax}.

\subsection{Grammars as Schemas, Sketches as Instances}
The SyGuS framework provides a powerful lens through which to view the relationship between sketches and schemas.
\begin{itemize}
    \item The grammar $G$ serves as the formal \textbf{schema}. It defines the structure of the entire search space, dictating which operators and control structures are available and how they can be composed.
    \item A program \textbf{sketch} can be understood as a highly specific and constrained instance of a SyGuS problem. The structure of the sketch itself implicitly defines a grammar that generates only programs matching that structure. The "holes" in the sketch correspond to non-terminals in the grammar that the synthesizer must expand according to the grammar's production rules \citep{alur2013syntax}.
\end{itemize}
For example, the sketch \texttt{return ?? * \{| x , 0 |\};} can be represented by a SyGuS problem with the semantic specification \texttt{assert f(3) == 6} and a syntactic grammar like:
\begin{verbatim}
Start := (mul Hole (Generator))
Hole := <integer_constant>
Generator := x | 0
\end{verbatim}
The field has advanced thanks in large part to this unification, which has produced standardized benchmark formats (SyGuS-IF) and yearly competitions (SyGuS-Comp) that encourage the development of novel and more effective problem-solving strategies \citep{alur2013syntax}.

\section{Comparative Analysis of Program Synthesis Paradigms}
The main program synthesis paradigms are compared in the following table, which also places sketch- and schema-based methods in their larger context. It draws attention to the different compromises that each paradigm makes with regard to search strategy, specification, guidance, and correctness guarantees.

\begin{longtable}{|p{0.18\textwidth}|p{0.2\textwidth}|p{0.23\textwidth}|p{0.18\textwidth}|p{0.16\textwidth}|}
\caption{A comparative analysis of program synthesis paradigms.} \label{tab:paradigm_comparison} \\
\hline
\textbf{Feature} & \textbf{Deductive Synthesis} & \textbf{Inductive Synthesis (PBE)} & \textbf{Sketch-Based Synthesis} & \textbf{Neuro-Symbolic Synthesis} \\
\hline
\endfirsthead
\multicolumn{5}{c}%
{{\bfseries \tablename\ \thetable{} -- continued from previous page}} \\
\hline
\textbf{Feature} & \textbf{Deductive Synthesis} & \textbf{Inductive Synthesis (PBE)} & \textbf{Sketch-Based Synthesis} & \textbf{Neuro-Symbolic Synthesis} \\
\hline
\endhead
\hline \multicolumn{5}{r}{{Continued on next page}} \\
\endfoot
\hline
\endlastfoot
\textbf{Primary Specification} & Formal Logical Formula (Pre/Postconditions) & Input-Output Examples & Partial Program + Assertions/Tests & Natural Language, Examples, Demonstrations \\
\hline
\textbf{Form of Guidance} & Proof Steps / Tactics & Additional Examples / User Feedback & Program Structure (Holes, Generators) & Learned Heuristics, Learned Sketches, Neural Priors \\
\hline
\textbf{Search Strategy} & Theorem Proving, Term Rewriting & Enumerative Search, Version Space Algebra & CEGIS, Constraint Solving (SAT/SMT) & Guided Search (Neural) + Symbolic Search \\
\hline
\textbf{Correctness Guarantee} & Correct-by-Construction & Correct on Examples (may not generalize) & Verified w.r.t. Spec \& Bounds & Probabilistic, often requires symbolic verifier \\
\hline
\textbf{Key Tools} & Coq, Isabelle/HOL & FlashFill, PROSE & Sketch, Rosette & DeepCoder, \citep{zhang2023fusing} \\
\hline
\textbf{Primary Challenge} & Specification Effort, Scalability \citep{manna1980theory} & Specification Ambiguity, Generalization & Scalability, Sketch Design Brittleness, Opacity \citep{singh2018interpretable} & Data Requirements, Interpretability, Combining Logics \\
\hline
\end{longtable}

\section{Applications and Case Studies}
The success of sketch- and schema-based synthesis in a variety of difficult programming domains best illustrates its practical usefulness. These techniques have made synthesis tractable for problems that were previously unachievable by fully automated methods by enabling programmers to inject critical high-level insights.

\subsection{Systems Programming and Concurrency}
Human programmers are notoriously bad at writing accurate and efficient low-level systems code, particularly concurrent code. Manually reasoning about properties like race-freedom and deadlock-freedom is a combinatorial nightmare due to the enormous number of possible thread interleavings. In this field, sketch-based synthesis has shown itself to be a very useful tool \citep{solar2008program}.

The high-level steps of a concurrent data structure update or synchronization protocol can be described by a programmer, who can then put them inside a \texttt{reorder} block and assign the synthesizer the responsibility of determining a safe and appropriate ordering. The synthesizer can find an implementation that is provably correct under a bounded model checker by thoroughly exploring the permutations of these operations, protected by synthesized conditions \citep{solar2008program}. The synthesis of a sense-reversing barrier, fine-grained locking schemes for concurrent sets, and solutions to the dining philosophers problem are notable examples \citep{solar2008combinatorial, solar2008program}.

\subsection{High-Performance and Scientific Computing}
Peak performance in scientific and high-performance computing (HPC) fields frequently necessitates low-level, counterintuitive optimizations. For this task, sketching offers a powerful workflow: a programmer can provide a reference implementation that is straightforward, clearly correct, but possibly inefficient. The low-level constants and expressions are then left as holes in a sketch that depicts the structure of the intended high-performance version. For all inputs, the synthesizer must finish the sketch so that it is semantically equivalent to the reference implementation \citep{solar2013sketch}. The development of bit-level ciphers such as AES, error-correction codes, and stencil kernels for partial differential equation solutions are examples in this field \citep{solar2008combinatorial, solar2013sketch}. Additionally, the schema-based AUTOFILTER system shows how different Kalman filters—basic recursive algorithms in signal processing—can be synthesized from high-level mathematical models \citep{fischer2003autobayes}.

\subsection{Network Telemetry and Security}
Program synthesis now faces both new opportunities and challenges as programmable networks gain popularity. Auto-code composition is used by synthesis frameworks such as AutoSketch to automatically generate optimized data plane code (e.g., in the P4 language) for networking sketches from high-level APIs \citep{sivaraman2018autocomposing}. The TrustSketch framework employs a sketch-based methodology to construct reliable telemetry systems for security, enclosing core logic in a secure hardware enclave (such as Intel SGX) and employing synthesis to guarantee computation integrity \citep{vasconcelos2020trustsketch}.

\subsection{Device Driver Development}
Despite being an essential part of contemporary operating systems, device drivers are infamous for causing bugs and instability \citep{chou2001empirical}. It is necessary to properly mediate between the low-level hardware model and the high-level OS interface when writing a driver. Termite-2 and other frameworks use a user-guided synthesis method that is conceptually comparable to sketching \citep{joshi2007termite2}. The developer offers a source code template—a sketch—of the driver as well as a formal model of the device's behavior. Developers can create dependable drivers more quickly thanks to Termite-2's assistance, which suggests solutions for the holes and statically confirms that the code interacts with the device model correctly \citep{joshi2007termite2}.

\section{Inherent Challenges and Future Horizons}
Despite its achievements, the synthesis paradigm based on sketches and schemas is not without serious difficulties. Its widespread adoption has been hampered by its limitations in terms of scalability, usability, and the level of expertise needed to create effective specifications. These issues are being actively addressed, though, and the most promising path forward is found at the center of deep learning and symbolic synthesis—the neuro-symbolic frontier.

\subsection{Fundamental Limitations and Challenges}
Three primary challenges currently define the boundaries of what is practical with sketch- and schema-based synthesis.
\begin{itemize}
    \item \textbf{Scalability to Large Programs:} The foremost limitation is the combinatorial explosion of the search space. As the size of the target program or the number of holes increases, the search space grows exponentially, quickly rendering the problem intractable for current solvers \citep{singh2018interpretable}. The scaling process is non-linear; a small increase in program complexity can lead to a massive increase in synthesis time, a challenge that mirrors the inherent difficulties of scaling software development in general \citep{brooks1987no}.
    \item \textbf{Brittleness and Opaqueness:} A significant barrier to adoption is the often brittle and opaque nature of synthesis tools. When a synthesizer fails, it frequently provides little to no diagnostic feedback, leaving the user to guess the source of the failure \citep{singh2018interpretable}. This "black-box" behavior can be intensely frustrating. In response, the field of \textit{interpretable program synthesis} has emerged, aiming to unveil the internal state of the synthesizer to help users build a more accurate mental model of the process and guide it more strategically \citep{singh2018interpretable}.
    \item \textbf{The Art of Schema and Sketch Design:} The success of this paradigm is critically dependent on the quality of the human-provided guidance. Crafting an effective sketch or a comprehensive set of schemas is a non-trivial art that requires significant domain expertise \citep{singh2018interpretable}. An overly constrained sketch will have no solution, while an overly loose sketch will lead to a timeout. Similarly, designing a robust and extensible library of schemas is a massive software engineering effort. As systems like AUTOBAYES grow, they risk "entropy death," where domain knowledge becomes scattered and the system becomes impossible to maintain or extend \citep{fischer2003autobayes}. The challenges of managing dependencies and planning for evolution are substantial \citep{brooks1987no}.
\end{itemize}

\subsection{The Neuro-Symbolic Frontier: The Future of Guided Synthesis}
Neuro-symbolic programming, a research area that aims to integrate the rigor of classical symbolic program synthesis with the advantages of contemporary deep learning, offers the most promising route to overcoming these constraints \citep{solar2008program}. A new synergy is produced by this hybrid approach, in which neural networks direct and speed up symbolic search while symbolic structures ensure the interpretability and accuracy of neural models.

\subsubsection{How Neural Methods Enhance Synthesis}
Deep learning models can learn the statistical patterns of human-written programs from vast code repositories. This knowledge can guide symbolic synthesizers in two primary ways:
\begin{itemize}
    \item \textbf{Guiding the Search:} Instead of exploring the program space blindly, a synthesizer can be guided by a neural model that provides a probability distribution over likely programs. This acts as a powerful heuristic, dramatically pruning the search space \citep{singh2018interpretable}.
    \item \textbf{Synthesizing Sketches:} The difficult task of writing a good sketch can itself be automated. A neural model can take a high-level specification (e.g., natural language) and generate a plausible program sketch. A symbolic solver then fills in the details and, crucially, guarantees correctness with respect to a formal specification, something the neural model alone cannot do \citep{balog2017deepcoder}.
\end{itemize}

\subsubsection{How Symbolic Methods Enhance Neural Models}
The synergy is bidirectional. The formal structures from program synthesis provide essential scaffolding that addresses the inherent weaknesses of purely neural approaches.
\begin{itemize}
    \item \textbf{Providing Structure and Regularization:} The grammar of a DSL---a formal schema---acts as a powerful structural prior for a neural program generator. Forcing the model's output to conform to the grammar prevents the generation of invalid code and focuses the learning process on semantically meaningful programs \citep{parisotto2017neurosymbolic}.
    \item \textbf{Guaranteeing Correctness:} LLMs are powerful generators but offer no correctness guarantees. A powerful neuro-symbolic pattern uses an LLM as the Generator in a CEGIS loop. The LLM proposes a program, and a symbolic Verifier checks it. If a bug is found, the counterexample is fed back into the LLM's prompt, asking it to fix the bug. This loop combines the generative capability of LLMs with the soundness of formal verification \citep{zhang2023fusing}.
\end{itemize}

This combination of reasoning and learning suggests that programming has undergone a fundamental evolution. Instead of defining each instruction, the programmer now has to curate datasets, write high-level intent, and create the learning objectives that direct synthesis. In the end, this development may make it possible to synthesize not only programs but also the very abstractions that programmers employ, such as compilers, type systems, and DSLs. This would allow for the automation of the process of designing programming abstractions, which has up until now only been done by human specialists \citep{ellis2021dreamcoder}. This signifies a significant change from programming the computer to programming the actual process of creating programs.

\chapter{Large Language Models as Program Synthesizers: A Paradigm Shift Towards Natural Language Specifications}
\label{chap:llm}

The development and effects of Large Language Models (LLMs) on the field of program synthesis are examined in this chapter. It analyzes the paradigm's key mechanisms, seminal systems, and the essential trade-offs it introduces as it traces its development from its conceptual beginnings to its current state-of-the-art. This analysis makes the case that LLM-based synthesis is not just a technical development but also a philosophical change in the way program specifications are thought of, moving from the strict world of formal logic to the flexible and ambiguous world of human intent.

\section{The Emergence of a New Paradigm: From Statistical Models to Code-Generating Transformers}

A major shift from previous paradigms is represented by the use of Large Language Models in program synthesis. This change was not abrupt; rather, it was the result of concurrent developments in natural language processing and a growing understanding of the drawbacks of conventional, formalism-heavy synthesis methods. A new method that could handle unstructured, human-centric specifications was born out of the convergence of these trends, which was sparked by architectural innovations.

\subsection{Precursors: The Limitations of Traditional Synthesis and the Rise of Language Modeling}

For many years, methods based on combinatorial search and formal logic dominated the field of program synthesis. For example, program construction is treated as a theorem-proving task in deductive synthesis, where a program is taken from a constructive proof that an object that satisfies the specification exists \citep{manna1980theory}. Similar to this, search-based and inductive approaches look through a large number of potential programs to identify one that meets a formal specification or a set of constraints, like input-output examples \citep{alur2013syntax, gulwani2017program}.

Despite their strength, these conventional paradigms all had one basic drawback: they were dependent on exact, formal specifications \citep{gulwani2017program}. This requirement caused a major "specification bottleneck" \citep{gulwani2017program}, even though it made it possible to generate provably correct code for intricate algorithms like insertion into red-black trees or Strassen's matrix multiplication. The methods could not take advantage of unstructured or unclear inputs, like natural language descriptions, and were frequently restricted to synthesizing relatively small programs \citep{gulwani2017program}. They struggled to gain wider applicability, but their success was most noticeable in specialized domains like bit-vector manipulations where specifications could be easily and rigorously formalized \citep{solar2008combinatorial}.

Natural language processing, or NLP, was also going through a revolution at the same time, albeit mainly on its own. In the early 1990s, the journey started with statistical language models (SLMs), which modeled the probability of linguistic sequences using corpus-based techniques \citep{jurafsky2023speech}. Neural language models (NLMs) gave way to pre-trained language models (PLMs), which in turn led to the current state of LLMs \citep{jurafsky2023speech}. Using ever-larger datasets and increasingly complex computational architectures, each step marked a breakthrough in the processing, comprehension, and generation of text at the human level \citep{zhao2023survey}. The fundamental foundation for a radical concept—treating source code as merely another "language" whose statistical patterns could be learned and produced—was established by this decades-long development.

\subsection{The Transformer Revolution and its Application to Code}

The introduction of the Transformer architecture by Vaswani et al. in their 2017 paper, "Attention Is All You Need" \citep{vaswani2017attention}, was the turning point that made it possible to combine language modeling and program synthesis. The self-attention mechanism, main innovation of the architecture, enabled models to capture long-range dependencies and assess the importance of various words in an input sequence \citep{vaswani2017attention}. This was a significant development because non-local relationships—where a variable defined at the start of a file might be used hundreds of lines later—are a feature of both natural language and, source code.

The "naturalness of software" hypothesis, which holds that human-written code is extremely predictable, much like natural language, served as the foundation for the application of this architecture to source code \citep{hindle2012naturalness}. This implied that the statistical learning methods that worked well for text could also work well for code. This assumption was confirmed by early research. Researchers discovered that even general-purpose models, such as GPT-3, which were not specifically trained on code, were surprisingly capable of producing basic Python programs from docstrings \citep{chen2021evaluating}.

This outcome gave rise to the theory that a customized GPT model could perform well on a range of programming tasks after being refined on a large corpus of code \citep{chen2021evaluating}. OpenAI's Codex, a ground-breaking model based on GPT-3 that was refined using a vast dataset of 159 gigabytes of Python code from more than 54 million public GitHub repositories, is the result of this line of research \citep{chen2021evaluating}. LLM-based synthesis went from being a research idea to a widely used technology with the advent of Codex and its subsequent incorporation into programs like GitHub Copilot.

\subsection{Foundational Motivations: Democratizing Development and Overcoming Specification Barriers}

The rapid advancement of LLM-based synthesis was propelled by a series of potent incentives that were designed to revolutionize the software development process.  The democratization of programming was a primary objective, and it may have been the most ambitious.  The vision, as articulated by industry leaders such as Jensen Huang of NVIDIA, is to transform "the programming language...human," enabling anyone in the world to become a programmer by expressing their intent in natural language \citep{huang2023jensen}.  By abstracting away the complexities of syntax, libraries, and APIs, these models seek to significantly reduce the cognitive load on developers and lower the barrier to entry \citep{barker2023automatically}.

 This aspiration was a direct response to the fundamental constraint of previous synthesis paradigms.  LLMs were developed from the ground up to operate within the inherent ambiguity of natural language, in contrast to formal methods, which required complete and unambiguous specifications \citep{gulwani2017program}.  This represented a fundamental change in the relationship between the synthesizer and the user.  The objective was to produce a plausible program from a high-level, potentially imperfect, description, rather than to demonstrate the correctness of the program from a perfect, logical specification \citep{gulwani2017program}.

 The motivation for professional developers was more pragmatic: to improve productivity by automating the tedious, repetitive, and frequently "least enjoyable" aspects of programming \citep{chen2021evaluating}.  Tools such as GitHub Copilot were designed to serve as a collaborative "pair programmer" or "auto-complete on steroids," capable of generating boilerplate code, implementing standard algorithms, or providing scaffolding for interacting with unfamiliar libraries and APIs \citep{gulwani2017program}.

 The introduction of this paradigm represents something more than merely a technical enhancement; it represents a significant reevaluation of the definition of a "program specification."  The process has transitioned from logical deduction to probabilistic inference.  The specification is a formal contract in deductive or search-based synthesis. For instance, a statement in first-order logic, S, that a synthesized program, p, must provably satisfy for all inputs \citep{gulwani2017program}.  The process of synthesis is proof or an exhaustive, verifiable search.  In contrast, LLMs are trained on extensive, unstructured corpora of text and code, and their fundamental operating mechanism is the prediction of the subsequent token using learned statistical patterns, rather than logical entailment \citep{zhao2023survey}.

 As a result, an LLM does not interpret a "specification," such as a natural language docstring, as a set of formal constraints.  Rather, it utilizes the prompt as a conditioning context to produce a sequence of code tokens that is statistically likely and has been associated with similar contexts in its training data \citep{brown2020language}.  A change in philosophy is the result of this change in mechanism.  The user is relieved of the responsibility of ensuring absolute precision, as they are now required to provide an intent rather than a formal specification.  Subsequently, the model generates a hypothesis or candidate, rather than a proven theorem.  The primary reason for the intense emphasis on post-generation verification, testing, and debugging in subsequent research in the field is the inherent uncertainty in the output \citep{shinn2023reflexion}.  The synthesis process is no longer a single-step construction process; rather, it is the commencement of an interactive dialogue that is designed to achieve a precise program.

\section{The Mechanics of LLM-Based Program Synthesis}

The technical foundations of LLM-based program synthesis include a core generative mechanism, a multi-stage model training process, and a collection of advanced techniques intended to improve the output's quality and dependability. These elements are broken down in this section, which progresses from the fundamental architecture to the sophisticated techniques that characterize the state-of-the-art.

\subsection{Core Architecture: Pre-training, Fine-tuning, and the Role of Code Corpora}

Transformer-based foundation models, like those in the GPT, LLaMA, and Gemini families, serve as the foundation for contemporary LLM-based synthesizers \citep{zhao2023survey}. These "Code LLMs" usually go through a two-step training process:

\begin{itemize}
    \item \textbf{Pre-training:} A large body of data, including source code and natural language text from public repositories such as GitHub, is used to pre-train the foundation model \citep{chen2021evaluating}. For instance, a 715 GB code snapshot was used to train AlphaCode initially \citep{li2022competition}. This pre-training stage is essential because it gives the model a broad, fundamental understanding of data structures, idiomatic patterns, programming syntax, and the semantic relationship between code and natural language descriptions in a variety of programming languages \citep{zhao2023survey}.
    \item \textbf{Fine-tuning:} The pre-trained model is then refined on a smaller, more curated dataset to specialize the model for better performance on particular programming tasks \citep{li2022competition}. This dataset could include high-quality competitive programming problems, like AlphaCode uses the CodeContests dataset \citep{li2022competition}, or instruction-following datasets that help the model better follow user instructions, like WizardCoder and Code Alpaca \citep{luo2023wizardcoder, taori2023alpaca}. According to research, performance can be significantly improved by fine-tuning on even a small number of high-quality, task-specific examples \citep{austin2021multilingual}.
\end{itemize}

It is impossible to overestimate the importance of data in this paradigm. The size and caliber of the corpora used for pre-training and fine-tuning have a fundamental impact on the final model's performance and capabilities. In order to better align model behavior with user intent, this has led to a significant focus on data curation research, including methods for synthesizing high-quality instruction-response pairs \citep{luo2023wizardcoder}.

\subsection{The Generation Process: From Prompt to Program}

A prompt from the user initiates the synthesis process. The specification is provided by this prompt, which can be a natural language description (e.g., "Write a function to compute the moving average"), a code comment, a function signature, or even a collection of input-output examples \citep{brown2020language}. The program is then generated by the model through a generative process.

\begin{itemize}
    \item \textbf{Input Processing and Contextualization:} The LLM first processes the input prompt, analyzing its semantic content and intent to establish a context for generation \citep{brown2020language}.
    \item \textbf{Autoregressive Token Prediction:} The core generation mechanism is autoregressive, meaning the model produces the output sequence one token at a time. A "token" can be a word, a symbol (like \texttt{\{} or \texttt{,}), or a sub-word unit. At each step, the model predicts a probability distribution over its entire vocabulary for the next token, based on the initial prompt and all the tokens it has generated so far \citep{brown2020language}. A sampling strategy, such as temperature sampling, is then used to select the next token from this distribution. This process is repeated iteratively until the model generates a special end-of-sequence token or reaches a predefined length limit.
\end{itemize}

A simplified pseudocode representation of this process is as follows:
\begin{algorithm}
\caption{Simplified pseudocode for autoregressive program generation.}
\label{alg:llm_generation}
\begin{algorithmic}[1]
\STATE \textbf{function} generate\_program(\textit{prompt}, \textit{model}, \textit{temperature})
\STATE \quad \textit{tokens} $\leftarrow$ tokenize(\textit{prompt}) \COMMENT{Convert prompt to token sequence}
\STATE \quad \textbf{while not} is\_end\_of\_sequence(\textit{tokens}) \textbf{and} len(\textit{tokens}) < MAX\_LENGTH
\STATE \quad \quad \textit{next\_token\_probabilities} $\leftarrow$ model.predict\_next\_token(\textit{tokens})
\STATE \quad \quad \textit{scaled\_probabilities} $\leftarrow$ apply\_temperature(\textit{next\_token\_probabilities}, \textit{temperature})
\STATE \quad \quad \textit{next\_token} $\leftarrow$ sample\_from(\textit{scaled\_probabilities})
\STATE \quad \quad \textit{tokens}.append(\textit{next\_token})
\STATE \quad \textbf{end while}
\STATE \quad \textbf{return} detokenize(\textit{tokens}) \COMMENT{Convert token sequence back to code}
\end{algorithmic}
\end{algorithm}

\subsection{Key Methodologies and Techniques}

Given that the basic generative process is probabilistic and not guaranteed to produce correct code, a range of techniques has been developed to steer the model towards better solutions and improve overall success rates.

\begin{itemize}
    \item \textbf{Prompt Engineering and Few-Shot Learning:} An LLM's output is extremely responsive to the input prompt \citep{wei2022emergent}. The process of carefully crafting prompts to elicit the desired behavior from the model is known as prompt engineering. A clear, high-level task description, the programming language to be used, constraints to be outlined, and—most importantly—examples can all be part of this \citep{brown2020language}. This leads to the few-shot learning technique, in which a limited number of task examples (shots) are added to the prompt. For example, two or three pairs of input strings and the matching desired output strings may be included in a prompt for a string manipulation task. It has been demonstrated that this in-context learning significantly boosts performance and enables the model to generalize to the user's particular issue without the need for expensive fine-tuning \citep{austin2021multilingual}.
    
    \item \textbf{Large-Scale Sampling, Filtering, and Clustering:} Even though an LLM's single generation might be wrong, the model's probabilistic structure allows it to generate a large number of possible solutions. Leveraging this diversity through a multi-stage pipeline is a highly effective strategy that was pioneered by systems such as Codex and AlphaCode \citep{li2022competition}. For a single problem, the model first generates a large number of candidate programs (from thousands to millions), frequently with a high "temperature" setting during sampling to promote variety \citep{chen2021evaluating}. Second, a set of known test cases—such as those listed in the problem description—are run through these candidates to filter them. Often removing more than 99\% of the produced samples, this pruning step is very successful \citep{li2022competition}. Third, a new set of generated inputs is used to cluster the remaining programs according to how they behave. The argument goes that while incorrect programs will fail in a variety of ways and form smaller, disparate clusters, correct programs will exhibit identical input-output behavior despite syntactic differences \citep{li2022competition}. In order to maximize the likelihood of choosing a reliable and accurate solution, the final submissions are then selected from the largest clusters.
    
    \item \textbf{Iterative Refinement: Self-Debugging and Agentic Workflows:} Recognizing that LLMs often produce "near misses"—programs that are almost correct but fail due to minor errors—the research frontier has moved towards iterative refinement loops.
    
    \begin{itemize}
        \item \textbf{Self-Debugging and Self-Repair:} An LLM can debug its own generated code using this method. An initial program is run against test cases after it has been synthesized. If it doesn't work, the model is given a new prompt that includes the execution trace, compiler errors, or even a natural language description of the issue \citep{huang2023jensen}. After that, a corrected version is requested from the model. Performance on tasks ranging from code translation to function synthesis can be greatly enhanced by repeating this process, which imitates the human developer's "rubber duck debugging" approach \citep{shinn2023reflexion}.
        
        \item \textbf{Oracle-Guided Synthesis:} Two LLMs divide up the work in sophisticated frameworks like ALGO. The "verifier," an LLM, is asked to produce a slow but accurate "oracle" program (for example, by employing a brute-force search algorithm). The "coder," a second LLM, is entrusted with coming up with a more effective solution. Following that, the oracle serves as a ground truth to automatically confirm that the coder's output is correct across a variety of inputs, offering trustworthy and comprehensible feedback for iterative improvement \citep{zhang2024algo}.
        
        \item \textbf{Agentic Workflows:} The most advanced methods organize the synthesis procedure around a group of cooperating LLM-based agents. Various agents, including "requirement engineer," "developer," and "tester," take on roles from a real-world software development team in frameworks like LCG \citep{zeng2024large}. Within a structured process model (such as Scrum or Test-Driven Development), these agents work together to discuss requirements, write code, create tests, and improve the solution in response to test failures. It has been demonstrated that this cooperative, multi-agent method significantly increases the accuracy and stability of the results \citep{zeng2024large}.
    \end{itemize}
\end{itemize}

A fascinating story emerges from the development of these approaches. The evolution of software engineering practices over time is reflected in the shift from single-shot generation to massive-scale generate-and-test to interactive and collaborative refinement. Early, unsophisticated uses of LLMs for code synthesis functioned similarly to a straightforward Waterfall model: given a specification (the prompt), the model produced an entire program in a single, monolithic step, with testing only taking place at the very end \citep{chen2021evaluating}. A paradigm similar to a large-scale batch testing phase after development was introduced by the subsequent development of systems like AlphaCode; a vast number of candidate programs were created and then put through a rigorous, independent filtering and testing process \citep{li2022competition}. This workflow was more robust, but it was still primarily sequential and non-interactive.

An obvious move toward an Agile or iterative model can be seen with the introduction of self-debugging, oracle-guided techniques, and Synthesize-Execute-Debug frameworks \citep{shinn2023reflexion}. In this case, the system works in close feedback loops, producing code, testing it, getting instant feedback from a test suite or compiler, and improving the solution in response to that feedback. This is the fundamental idea behind contemporary iterative development. The principles of DevOps and Continuous Integration/Continuous Deployment (CI/CD), where automated testing and integration are essential to the development lifecycle, are directly paralleled by the emergence of agentic workflows, where "tester" agents automatically run checks on code produced by "developer" agents \citep{zeng2024large}. As the LLM-based synthesis paradigm has developed, it has independently re-encountered and resolved the same basic problems of handling complexity and guaranteeing correctness that have influenced the history of human software engineering. This parallel evolution is not a coincidence.

\section{Landmark Systems and Empirical Evaluation}

The emergence of multiple groundbreaking systems that not only extended the bounds of what was possible but also set the fundamental techniques and assessment criteria for the field have contributed to the quick development of LLM-based program synthesis. This section offers an empirical basis for comprehending the capabilities of the paradigm by analyzing these key systems and the standards developed to evaluate their performance.

\subsection{OpenAI Codex: Bringing Synthesis to the Masses}

Perhaps the system that made LLM-based synthesis more widely used in software development was Codex, which OpenAI introduced in 2021 \citep{chen2021evaluating}. It served as the model for the popular GitHub Copilot tool, which brought AI-driven code recommendations straight into the IDEs of millions of developers \citep{gulwani2017program}. With its ability to perform tasks like "mapping simple problems to existing code," like completing functions, converting natural language comments into code, and eliminating the need to look for API usage examples, Codex was created primarily to support human programmers \citep{chen2021evaluating}.

Codex had 12 billion parameters and was architecturally based on the GPT-3 model \citep{chen2021evaluating}. Its training data, which was a huge 159 GB corpus of Python code scraped from 54 million public GitHub repositories, was its primary differentiator \citep{chen2021evaluating}. Although it excelled at Python, it also showed proficiency in more than a dozen other programming languages, such as JavaScript, Go, and Ruby \citep{chen2021evaluating}.

Codex's approach focused on creating code from context, which could be a natural language comment or the lines of code that come before it in a file. Its efficacy was greatly increased by a method that would later become commonplace in the field: producing several candidate solutions (samples) and choosing the best one using a validation mechanism (such as unit tests). As influential as the model itself was the groundbreaking paper that came with it, "Evaluating Large Language Models Trained on Code" by Chen et al. (2021). Beyond basic syntactic similarity metrics like BLEU scores, it introduced the HumanEval benchmark and the pass@k metric, which together produced the first rigorous framework for assessing the functional correctness of synthesized programs \citep{chen2021evaluating}. The original 12B parameter Codex model demonstrated remarkable capability for a first-generation system, achieving a pass@1 score of 28.8\% (solving the problem on the first attempt) and a pass@100 score of 70.2\% (solving the problem with one of 100 attempts) on this new benchmark \citep{chen2021evaluating}.

\subsection{DeepMind's AlphaCode: Tackling Competitive Programming}

DeepMind's AlphaCode, which was released in 2022, aimed to tackle a much more difficult field: competitive programming, whereas Codex concentrated on helping with general programming tasks \citep{li2022competition}. Platform problems such as Codeforces demand not only the conversion of instructions into code but also deep algorithmic reasoning, critical thinking, and the capacity to come up with new solutions to problems that haven't been seen yet. This is a big step in the direction of true problem-solving \citep{li2022competition}.

AlphaCode used a Transformer encoder-decoder architecture. CodeContests, a specially created dataset of competitive programming problems, was used to refine it after it had been pre-trained on a massive 715 GB snapshot of code from GitHub \citep{li2022competition}. In order to prevent "false positives," which occur when a program passes public tests but has underlying flaws, this fine-tuning dataset was carefully selected to include not only problem descriptions and accurate solutions, but also incorrect human submissions and a suite of generated test cases \citep{li2022competition}.

The system's massive-scale search strategy was its primary methodological innovation. AlphaCode produced millions of different candidate solutions in Python and C++ for every problem \citep{li2022competition}. A complex pipeline was then applied to this massive sample pool. Programs that didn't pass the example tests in the problem description were first eliminated. After that, the remaining candidates were grouped according to how they performed on a series of brand-new test inputs created by the model. In order to optimize both correctness and diversity, the system ultimately chose its ten submissions from the biggest and most unique clusters \citep{li2022competition}.

The original AlphaCode received an estimated Codeforces Elo rating of 1238, placing it in the top 54.3\% of human competitors in simulated evaluations of ten recent Codeforces competitions. According to the platform's founder, this performance was comparable to that of a "promising new competitor," which was the first time an AI system had achieved a competitive level in such competitions \citep{li2022competition}. With the help of the more sophisticated Gemini model, its replacement, AlphaCode 2, showed a significant improvement. With an estimated performance at the 85th percentile of human competitors, it placed between the 'Expert' and 'Candidate Master' ranks, solving 43\% of problems in a new evaluation set (compared to 25\% for the original) \citep{alphacode2_2023}.

\subsection{Benchmarking and Metrics: A New Standard for Evaluation}

The rise of LLM-based synthesis necessitated new ways to measure performance that went beyond the text-similarity metrics common in NLP. The focus shifted to functional correctness: does the generated code actually work?

\begin{itemize}
    \item \textbf{HumanEval:} Introduced by Chen et al. (2021) with Codex, the HumanEval dataset consists of 164 hand-written Python programming problems \citep{chen2021evaluating}. Each problem is self-contained and includes a function signature, a natural language docstring (which serves as the prompt), a canonical solution, and a set of unit tests for verification \citep{chen2021evaluating}. It has become the de facto standard for measuring a model's ability to synthesize code from natural language descriptions. The initial 0\% score of the general-purpose GPT-3 model on this benchmark starkly illustrated the necessity of specialized training on code \citep{chen2021evaluating}.
    
    \item \textbf{MBPP (Mostly Basic Programming Problems):} Introduced by Austin et al. (2021), the MBPP dataset contains approximately 1,000 crowd-sourced Python problems intended to be solvable by entry-level programmers \citep{austin2021multilingual}. Each problem consists of a short text description, a reference code solution, and three automated test cases \citep{austin2021multilingual}. MBPP complements HumanEval by focusing on problems that often involve more imperative control flow (loops, conditionals) and are specified with simpler, more direct natural language, whereas HumanEval problems can have more complex docstrings \citep{austin2021multilingual}.
    
    \item \textbf{The pass@k Metric:} To properly evaluate probabilistic models that can generate many different potential solutions for a single prompt, the pass@k metric was developed \citep{chen2021evaluating}. It is defined as the probability that at least one of the top k generated samples for a given problem passes all associated unit tests. This is typically estimated by generating n samples per problem (where n > k), counting the number of correct samples c, and calculating the estimator $1 - \binom{n-c}{k} / \binom{n}{k}$. This metric directly evaluates the utility of a model in a realistic generate-and-test workflow and has become the standard for reporting performance on HumanEval, MBPP, and other code generation benchmarks \citep{chen2021evaluating}.
\end{itemize}

The following table provides a comparative summary of these landmark systems, grounding the field's progress in concrete, empirical results.

\begin{longtable}{|p{0.15\textwidth}|p{0.2\textwidth}|p{0.3\textwidth}|p{0.15\textwidth}|p{0.15\textwidth}|}
\caption{Comparison of Landmark LLM-based Synthesis Systems.} \label{tab:llm_systems} \\
\hline
\textbf{System} & \textbf{Researchers/Org} & \textbf{Key Innovation} & \textbf{Target Domain} & \textbf{Reported Performance} \\
\hline
\endfirsthead
\multicolumn{5}{c}%
{{\bfseries \tablename\ \thetable{} -- continued from previous page}} \\
\hline
\textbf{System} & \textbf{Researchers/Org} & \textbf{Key Innovation} & \textbf{Target Domain} & \textbf{Reported Performance} \\
\hline
\endhead
\hline \multicolumn{5}{r}{{Continued on next page}} \\
\endfoot
\hline
\endlastfoot
\textbf{Codex} & Chen et al. (OpenAI) & First large-scale, public code model; HumanEval benchmark; pass@k metric. & General Python code generation from docstrings. & HumanEval: 70.2\% pass@100 \citep{chen2021evaluating} \\
\hline
\textbf{AlphaCode} & Li et al. (DeepMind) & Massive sampling, filtering, and clustering for complex algorithmic problems. & Competitive programming (Codeforces). & Codeforces: Top 54.3\% avg. rank (Elo 1238) \citep{li2022competition} \\
\hline
\textbf{AlphaCode 2} & Gemini Team (Google) & Integration of a more powerful foundation model (Gemini) with advanced search and refinement. & Competitive programming (Codeforces). & Codeforces: Solved 43\% of problems; est. 85th percentile rank \citep{alphacode2_2023} \\
\hline
\textbf{LaMDA-PT} & Austin et al. (Google) & Introduced MBPP benchmark; explored few-shot vs. fine-tuning at scale. & Basic Python programming problems. & MBPP: 58\% pass@k (few-shot), ~68\% (fine-tuned) \citep{austin2021multilingual} \\
\hline
\end{longtable}

\section{Applications and Use Cases in Modern Software Engineering}

Although fully automatic programming from high-level intent is still a distant goal, LLM-based synthesis has already impacted the software development lifecycle with a number of useful tools and applications. Developers of all skill levels are using these systems more and more because they are effective helpers that enhance rather than completely replace human abilities.

\subsection{Beyond Code Completion: Repair, Translation, and Documentation}

The applications of Code LLMs extend far beyond simple line-by-line code completion. Their deep understanding of both programming and natural languages enables them to perform a variety of complex software engineering tasks.
\begin{itemize}
    \item \textbf{Code Generation and Completion:} The most prevalent use case is as an advanced code completion assistant or "pair programmer" \citep{gulwani2017program}. Integrated into IDEs, these tools analyze the current code context—including surrounding functions, imported libraries, and comments—to suggest entire blocks of code, from single functions to complex class structures.
    \item \textbf{Automated Program Repair (APR):} LLMs have demonstrated a surprising aptitude for bug fixing. When provided with a segment of buggy code, often accompanied by a compiler error message or a natural language description of the failure, models like Codex can propose syntactically and semantically correct patches \citep{chen2021evaluating}. This capability forms the basis of more advanced iterative debugging workflows, where a model can generate, test, and then refine its own code until it passes a given set of unit tests \citep{huang2023jensen}.
    \item \textbf{Code Translation:} The multilingual nature of models trained on diverse codebases allows them to function as effective code translators. They can convert code snippets, functions, or even entire files from one programming language to another (e.g., Python to R, Java to C\#), a task that is traditionally time-consuming and error-prone for human developers \citep{chen2021evaluating}.
    \item \textbf{Documentation and Summarization:} Maintaining high-quality documentation is a critical but often neglected aspect of software engineering. LLMs can automate this process by analyzing a function or class and generating a natural language docstring that explains its purpose, parameters, return values, and potential exceptions, thereby improving the code's readability and long-term maintainability \citep{ma2023doc}.
\end{itemize}

\subsection{Assisting Novice Programmers and Domain Experts}

LLM-based tools have shown particular promise in making programming more accessible to individuals who are not professional software engineers.
\begin{itemize}
    \item \textbf{Scaffolding for Beginners:} For novice programmers, these tools can serve as an invaluable learning aid. They can provide a "good starting point" for a programming task, helping users who struggle to decompose a high-level problem into concrete computational steps \citep{barker2023automatically}. By generating code for a given prompt, they can act as a dynamic substitute for searching documentation or forums like Stack Overflow, helping users discover relevant functions and API methods within unfamiliar libraries \citep{barker2023automatically}.
    \item \textbf{The "Widening Gap" in Education:} However, the educational impact of these tools is not uniformly positive. Research indicates the emergence of a "widening gap" between students who can leverage LLMs effectively and those who cannot \citep{barker2023automatically}. Students who are adept at crafting precise prompts and critically evaluating the generated output can significantly accelerate their learning. Conversely, students who struggle with these skills may develop an over-reliance on the tools, uncritically accepting incorrect or suboptimal code, which can hinder the development of their own problem-solving abilities and lead to misconceptions about fundamental programming concepts \citep{barker2023automatically}.
    \item \textbf{Empowering Domain Experts:} A significant application lies in empowering domain experts—such as scientists, financial analysts, or researchers—who possess deep knowledge in their respective fields but lack formal programming training. LLMs allow these experts to perform complex data analysis, visualization, and modeling tasks by describing their objectives in natural language, effectively translating domain-specific intent into executable code \citep{dibia2023beyond}.
\end{itemize}

\subsection{Interfacing with Low-Level and Specialized Systems}

The capabilities of LLM-based synthesis are not limited to high-level, general-purpose languages. Research is actively exploring their application in highly specialized and technically demanding domains.
\begin{itemize}
    \item \textbf{Safe Low-Level Programming:} LLMs are being investigated as assistants for notoriously difficult low-level programming tasks where errors can have severe consequences. For example, researchers at Microsoft have demonstrated the use of LLMs to infer machine-checkable memory safety invariants in legacy C code, a critical step in migrating unsafe code to safer dialects like Checked C \citep{shi2023don}. In another project, an LLM-based tool called RustAssistant was developed to help programmers fix complex compilation errors in Rust, a language known for its strict safety guarantees and steep learning curve \citep{pan2023rustassistant}.
    \item \textbf{Hardware Description and Verification:} The application of LLMs is extending into the realm of hardware design. Researchers are using them to generate and evaluate code in hardware description languages (HDLs) like Verilog and VHDL \citep{luo2023wizardcoder}. Furthermore, they are being employed to assist in the complex process of security verification for System-on-Chip (SoC) designs, a critical and resource-intensive part of the hardware development cycle \citep{thakur2024chip}.
    \item \textbf{Domain-Specific Language (DSL) Generation:} While trained on general-purpose languages, LLMs can be prompted to generate code in highly specialized DSLs. With carefully crafted prompts that provide context and examples, they have been successfully applied to tasks such as synthesizing procedures for chemical reactions or generating solutions to formal logic problems specified in SMT-LIB \citep{dibia2023beyond}. This flexibility allows the power of LLM synthesis to be leveraged in domains far beyond conventional software development.
\end{itemize}

\section{Critical Analysis: Strengths, Weaknesses, and Future Trajectories}

While LLM-based program synthesis has achieved remarkable success and widespread adoption, it is essential to conduct a critical analysis of its fundamental strengths and weaknesses. This evaluation illuminates the trade-offs inherent in the paradigm and points toward the open challenges and research trajectories that will define its future.

\subsection{Strengths: Unprecedented Scale, Flexibility, and Productivity}

The LLM-based paradigm possesses several key advantages that have enabled it to overcome many of the limitations of its predecessors.
\begin{itemize}
    \item \textbf{Handling Unstructured Specifications:} The paramount strength of LLMs is their native ability to interpret and generate programs from high-level, ambiguous, and unstructured specifications, particularly natural language \citep{gulwani2017program}. This capability directly addresses the "specification bottleneck" that constrained formal and search-based methods, which required precise, machine-readable inputs \citep{gulwani2017program}.
    \item \textbf{Developer Productivity and Accessibility:} By automating the generation of boilerplate code, implementing common algorithms, and providing instant examples for API usage, LLMs significantly enhance the productivity of experienced developers \citep{huang2023jensen}. Simultaneously, they lower the barrier to entry for novices and domain experts, making the power of programming accessible to a much broader audience \citep{huang2023jensen}.
    \item \textbf{Broad Domain and Language Coverage:} A single, large-scale model, pre-trained on a diverse corpus of public code, can generate programs in dozens of different languages and for a wide array of domains \citep{chen2021evaluating}. This general-purpose nature stands in contrast to many traditional synthesis tools, which were often highly specialized for a particular language or problem domain.
    \item \textbf{Massive Search Space Exploration:} The synthesis strategy of generating millions of diverse program candidates and then filtering them based on tests allows these systems to explore a vast space of potential solutions \citep{li2022competition}. This probabilistic, large-scale search can discover novel and non-obvious solutions to complex problems that would be intractable for purely symbolic or deductive search algorithms to find \citep{zhang2024algo}.
\end{itemize}

\subsection{Weaknesses: The Correctness Impasse, Security Risks, and Reasoning Deficits}

Despite their strengths, LLMs suffer from several fundamental weaknesses that currently limit their reliability and autonomy.
\begin{itemize}
    \item \textbf{Lack of Correctness Guarantees:} The most significant and widely acknowledged weakness is that LLMs cannot provide any formal guarantee of correctness for the code they produce \citep{gulwani2017program}. The generative process is probabilistic, not deductive. This frequently leads to a "near-miss syndrome," where models generate code that is syntactically valid and appears plausible but contains subtle semantic bugs that cause it to fail on specific inputs \citep{manna1980theory}. This fundamental unreliability shifts the primary burden on the developer from writing code to meticulously verifying and debugging AI-generated code \citep{barker2023automatically}.
    \item \textbf{Security Vulnerabilities:} LLM-generated code can introduce serious security flaws. As the models are trained on vast quantities of public code, which often contains vulnerabilities, they can reproduce these unsafe patterns. Empirical studies have found that a significant percentage of code suggested by tools like GitHub Copilot contains vulnerabilities from the Common Weakness Enumeration (CWE), including high-risk issues like SQL injection, insecure cryptography, and buffer overflows \citep{chen2021evaluating}.
    \item \textbf{Hallucination and Reasoning Deficits:} LLMs are prone to "hallucinating"—confidently generating code that uses non-existent functions or APIs \citep{wei2022emergent}. More fundamentally, they struggle with tasks that require deep, multi-step algorithmic or logical reasoning \citep{zhang2024algo}. Their strength lies in pattern matching and translation, not in the kind of rigorous, step-by-step deduction required to invent complex algorithms from first principles \citep{bubeck2023sparks}.
    \item \textbf{Data-Related Issues:} The models are inextricably linked to their training data, which introduces several problems:
    \begin{itemize}
        \item \textbf{Benchmark Contamination:} Popular benchmark problems may have been present in the model's training data, leading to inflated performance scores that do not reflect true generalization ability \citep{chen2021evaluating}.
        \item \textbf{Bias Reproduction:} LLMs can inherit and amplify biases present in their training data, potentially leading to the generation of code that is discriminatory or unfair \citep{chen2021evaluating}.
        \item \textbf{Knowledge Staleness:} Models have a fixed knowledge cutoff date. As a result, they may generate code that uses outdated or deprecated libraries and APIs, leading to runtime failures \citep{zhao2023survey}.
    \end{itemize}
    \item \textbf{Ethical and Copyright Concerns:} The development and deployment of Code LLMs raise profound ethical questions regarding data privacy, the potential for misuse in generating malicious code, and the long-term impact on software engineering jobs \citep{weidinger2021ethical}. Furthermore, the practice of training on public code repositories has ignited significant copyright debates, particularly after models were observed regurgitating large blocks of code verbatim, including original comments and copyright notices, potentially violating open-source licenses \citep{chen2021evaluating}.
\end{itemize}

\subsection{The Path Forward: Open Challenges and Research Directions}

The limitations of the current paradigm define the major research frontiers that will shape its future development.
\begin{itemize}
    \item \textbf{Improving Correctness and Reliability:} A central challenge is bridging the gap between probabilistic generation and the need for deterministic, correct programs. Key research directions include developing more sophisticated self-debugging and automated program repair techniques \citep{shinn2023reflexion}, creating frameworks that integrate formal verifiers to check LLM outputs against specifications \citep{zhang2023fusing}, and designing more comprehensive evaluation benchmarks that assess not only functional correctness but also security, efficiency, and code quality \citep{kaddour2023challenges}.
    \item \textbf{Agentic and Interactive Systems:} The future of LLM-based synthesis is increasingly seen as agentic \citep{wang2023survey}. Research is rapidly moving towards building more sophisticated multi-agent systems that can autonomously handle complex, multi-step software engineering workflows—from high-level planning and implementation to testing, debugging, and deployment \citep{luo2023wizardcoder}. Enhancing the model's ability to use tools, interact with file systems and compilers, and learn from rich feedback is a critical area of focus.
    \item \textbf{Repository-Level Understanding:} A major limitation of current models is their focus on generating single files or functions in isolation. A key open challenge is to develop models with the ability to understand and operate within the context of an entire codebase or repository. This requires architectures with much longer effective context windows and the ability to reason about complex inter-file dependencies, project-specific APIs, and established coding conventions \citep{luo2023wizardcoder}.
    \item \textbf{Neuro-Symbolic Hybrids:} Perhaps the most promising direction for overcoming the fundamental reasoning and correctness deficits of pure LLMs is the integration of symbolic reasoning engines. Neuro-symbolic synthesis aims to create hybrid systems that combine the pattern-matching and natural language understanding strengths of neural networks with the rigorous, verifiable logic of symbolic methods \citep{pan2023logic}. This can involve using an LLM as a powerful heuristic to guide a traditional symbolic search, or using a symbolic solver (like a SAT or SMT solver) to verify, constrain, or repair the output of an LLM.
\end{itemize}

This tendency toward hybrid systems points to an unavoidable reconciliation with formalism's tenets. By forgoing correctness guarantees in favor of flexibility in handling ambiguous, natural language inputs, the LLM-based synthesis paradigm arose as a conscious break from the rigidities of formal specification \citep{gulwani2017program}. The "correctness impasse," where the output's probabilistic and unreliable nature became the paradigm's main limitation, was the direct result of this original thesis—that synthesis could be freed from formalism \citep{manna1980theory}.

The most cutting-edge research areas in the field today show a distinct shift in the direction of a fresh synthesis that unifies the two diametrically opposed concepts. The community is re-importing ideas from formal methods and traditional synthesis to address the "last mile" problem of correctness. One type of specification-based verification is the use of unit tests as an oracle \citep{li2022competition}. The oracle-guided inductive synthesis (OGIS) framework from traditional synthesis is directly implemented in frameworks such as ALGO, which use one LLM to generate a brute-force oracle to guide another \citep{zhang2024algo}. A direct integration with formal systems can be seen in research on the use of LLMs to help with formal verification in languages such as Rust or to infer memory invariants in C \citep{shi2023don}. This reconciliation is the explicit focus of the whole new field of neuro-symbolic program synthesis, which aims to integrate neural generation with logic engines and symbolic solvers \citep{pan2023logic}.

This trajectory suggests that the initial departure from formalism was a necessary developmental stage rather than a destination. In order to solve the input problem—understanding human intent—and serve as a potent heuristic engine for putting forward potential solutions, the field made use of LLMs' enormous power. It is now discovering that it needs to reintegrate the very formal reasoning principles it once aimed to avoid in order to solve the output problem and guarantee that those solutions are accurate, dependable, and trustworthy. The future of powerful, useful program synthesis seems to be in a hybrid system that achieves a synergistic composition rather than a purely neural or purely symbolic approach: \textit{Formal Methods(LLM)}, where the Large Language Model acts as a clever and imaginative guide within a framework that stays rooted in the formal correctness principles.

\chapter{Neuro-Symbolic (Hybrid) Synthesis}
\label{chap:neuro-symbolic}

\section{Introduction: The Imperative for a Hybrid Paradigm}

A profound and productive dichotomy between two dominant paradigms, the connectionist approach, exemplified by deep learning, and the symbolic approach, rooted in classical logic and knowledge representation, largely defines the contemporary landscape of artificial intelligence. Every paradigm is endowed with a distinctive set of potent capabilities; however, they are also constrained by fundamental constraints. The acknowledgment of this complementarity has facilitated the emergence of a third approach: neuro-symbolic artificial intelligence, a synthetic discipline that aims to develop intelligent systems that are more reliable, resilient, and general \citep{chaudhuri2021neurosymbolic}.  Within this broader movement, Neuro-Symbolic Program Synthesis (NSP) has emerged as a research frontier that is particularly compelling and potent. Its objective is to combine the perceptual strengths of neural networks with the rigorous, structured reasoning that is inherent in computer programs \citep{lample2019deep}.

The connectionist paradigm, which is derived from deep neural networks, has achieved unparalleled success in a wide range of domains, particularly those that involve perceptual and pattern-recognition tasks on unstructured data, such as images, audio, and text \citep{chaudhuri2021neurosymbolic}. Nevertheless, the very characteristics that facilitate this success—namely, the acquisition of intricate, high-dimensional functions through gradient-based optimization—elicit substantial and enduring doubts. The decision-making processes of deep neural networks are notoriously enigmatic, functioning as "black boxes" that are exceedingly difficult for humans to interpret, trust, and formally verify \citep{calegari2020design}. In safety-critical and high-stakes applications where accountability is paramount, this opacity poses a formidable barrier to their deployment \citep{gulwani2017program}. Furthermore, these models are renowned for their "data-hungry" nature, necessitating extensive labeled datasets for training. Additionally, their training may be unreliable in data-poor environments \citep{gulwani2017program}. They frequently grapple with the seamless integration of abstract or common-sense knowledge, long-horizon planning, and systematic or compositional generalization \citep{evans2018learning}.

In stark contrast, the symbolic paradigm, which is frequently referred to as "Good Old-Fashioned AI" (GO-FAI), excels in the exact areas where deep learning fails. Its fundamental strength is the explicit representation of knowledge through symbols, rules, and logic, which establishes a basis for logical, verifiable, and transparent reasoning \citep{calegari2020design}. Classic examples of this approach include automated theorem provers and expert systems, which are capable of performing intricate reasoning within well-defined domains and offering explicit explanations for their conclusions \citep{manna1980deductive}. However, this paradigm is constrained by its own critical deficiencies. This research \citep{chaudhuri2021neurosymbolic} have observed that symbolic systems are frequently fragile, as they are unable to accommodate the ambiguity, noise, and uncertainty that are hallmarks of real-world data. Their scalability is suboptimal when the space of rules expands, and they are contingent upon the extensive, and frequently prohibitive, manual labor of human experts to incorporate domain knowledge and logical rules into the system \citep{manna1980deductive}.

This intellectual impasse results in the direct emergence of Neuro-Symbolic Program Synthesis, which is proposed as a response to the complementary failures of its antecedents \citep{gulwani2017program}. The objective of NSP, as in classical machine learning, is to learn a function $f: X \rightarrow Y$ from data \citep{bommasani2021opportunities}. NSP is formally defined as an area of research at the interface of deep learning and program synthesis. The representation of the learned function, however, is the critical distinction. In place of an opaque network of weighted connections, the function is represented as an explicit, executable program $P$ from a Domain-Specific Language (DSL) $L$ \citep{gulwani2017program}. These are not merely symbolic artifacts; rather, they are hybrid programs that incorporate both conventional symbolic primitives (e.g., arithmetic operations, control flow structures, list manipulations) and learned neural components (e.g., perception modules, classifiers, parameterized functions) \citep{lample2019deep}. The discovery of these programs is a hybrid endeavor that integrates the continuous, gradient-based optimization methods of deep learning with the combinatorial, discrete search techniques of classical program synthesis to acquire knowledge of the program's architecture and the parameters of its neural modules \citep{lample2019deep}.

The impetus for undertaking this composite approach is derived from a series of fundamental commitments that directly confront the constraints of end-to-end deep learning:
\begin{itemize}
    \item \textbf{Interpretability and Verifiability:} NSP generates artifacts that are, by definition, more transparent by representing learned models as programs. In numerous instances, their behavior can be formally verified using techniques from formal methods and programming languages, and their logic can be debugged. Additionally, humans can conduct structural inspections. This is in stark contrast to the black-box nature of deep neural networks \citep{bommasani2021opportunities}.
    \item \textbf{Data Efficiency and Generalization:} The structure of a programming language, particularly a Domain-Specific Language (DSL) that is specifically designed for a particular problem domain, functions as a potent form of inductive bias. The hypothesis space is restricted to plausible programs by this structural constraint, which regularizes the learning process and enables more robust generalization from smaller, sparsely populated datasets \citep{bommasani2021opportunities}.
    \item \textbf{Compositionality and Modularity:} NSP enables the decomposition of intricate learning tasks into a collection of simplified, modular sub-tasks. These sub-tasks can be resolved using either symbolic or neural modules, which can be either pre-existing components from a library or acquired through novel learning. The utilization of acquired knowledge across various domains and tasks is made possible by this modularity, which is consistent with the principles of contemporary software engineering \citep{gulwani2017program}.
    \item \textbf{ Human Domain Expertise Injection:} The symbolic components of an NSP system, particularly the design of the DSL, offer a direct and explicit pathway for the integration of prior knowledge into the learning process. This permits the user to incorporate common-sense constraints or known algorithmic abstractions, thereby biasing the synthesizer toward solutions that are more likely to be correct and generalize well \citep{gulwani2017program}.
\end{itemize}

There is more to the ascent of NSP than a technical evolution that is driven by the desire to enhance performance metrics. It is a substantial philosophical shift in AI research, motivated by the practical and ethical considerations of implementing AI in the real world. The inherent lack of transparency and predictability of deep learning models has become a central point of failure and concern as they have been transitioned from laboratory benchmarks to critical societal infrastructure in fields such as finance, medicine, and autonomous transportation \citep{gulwani2017program}. The pressing need for AI systems that are not only accurate but also accountable and reliable has been underscored by the inability to elucidate \textit{how} a model arrived at a diagnosis or a financial decision, as well as the potential for catastrophic failure on adversarial or out-of-distribution inputs \citep{gulwani2017program}. By reinstating the structured, logical, and verifiable fabric of programs at the core of the machine learning process, NSP addresses these challenges. In the process, the field aims to develop models that more closely resemble the multifaceted nature of human cognition, which is a powerful hybrid of fast, intuitive, pattern-matching perception (similar to neural networks) and slow, deliberate, rule-based reasoning (similar to symbolic logic) \citep{bunel2018leveraging}.

\section{A Taxonomy of Neuro-Symbolic Architectures}
The concept of the integration of neural and symbolic components within a single system is not monolithic. It incorporates a broad range of design options, including systems that are predominantly symbolic but enhanced with neural heuristics and those that are fundamentally neural but regulated by symbolic constraints \citep{chaudhuri2021neurosymbolic}. It is essential to comprehend this architectural landscape in order to contextualize the diverse methodologies and systems that are present in the field. In this section, the central, unifying function of Domain-Specific Languages (DSLs) as the bridge between the two paradigms is emphasized, and a systematic classification of neuro-symbolic architectures is provided, drawing upon established taxonomies.

Based on the nature of the interaction and the division of labor between the neural and symbolic portions, a useful framework for classifying these diverse integration strategies, inspired by the work of Kautz and others, is established \citep{chaudhuri2021neurosymbolic}. This taxonomy demonstrates that the selection of architecture is not merely a technical detail; rather, it frequently reflects a fundamental hypothesis regarding the problem being resolved. Specifically, it determines which aspects of the problem are most suitable for data-driven learning versus explicit reasoning. Table 5.1 summarizes the primary integration paradigms, which are further elaborated upon below.

\subsection{Symbolic[Neuro]}
The primary control flow is regulated by a symbolic algorithm in this paradigm, which utilizes a neural network as a specialized subroutine. The symbolic component serves as the primary controller, overseeing high-level reasoning, planning, and search, while the neural component is subservient. It is typically employed to manage sub-symbolic tasks, such as perception, or to provide a learned heuristic function for which a symbolic specification is either unavailable or intractable \citep{chaudhuri2021neurosymbolic}. The master algorithm in AlphaGo, developed by DeepMind, is a Monte Carlo Tree Search (MCTS), a symbolic search procedure. This is a canonical example. A neural network augments the MCTS framework by performing two functions: a policy network that recommends potentially advantageous actions and a value network that assesses board positions. The symbolic search is significantly more efficient than it would be otherwise due to the neural network's acquisition of these functions from data \citep{chaudhuri2021neurosymbolic}.

\subsection{Neuro|Symbolic}

The systems described in this paradigm are designed as pipelines, with a clear and distinct separation of concerns between a symbolic back-end and a neural front-end. The neural component is typically responsible for perception, feature extraction, or semantic parsing, which involves the conversion of raw, unstructured input (such as images or text) into a structured, symbolic representation \citep{cranmer2020interpretable}. The symbolic component then receives this symbolic representation and executes logical reasoning, planning, or execution in accordance with a knowledge base or a set of principles. This category encompasses the preponderance of contemporary and early neuro-symbolic systems \citep{chaudhuri2021neurosymbolic}. IBM's Neuro-Vector-Symbolic Architecture (NVSA) is a notable example of this approach. It addresses visual reasoning tasks by initially employing a convolutional neural network (CNN) to detect and perceive objects, and subsequently feeding these symbols into a probabilistic reasoner to address logical inquiries regarding the scene \citep{chaudhuri2021neurosymbolic}.

\subsection{Neuro:Symbolic→Neuro}
This methodology entails the direct integration of symbolic knowledge into the structure or training regimen of a neural network. There are three objectives: to facilitate the network's learning process, ensure logical consistency, and improve the interpretability of the resulting model. The neural network is the primary computational paradigm in this context; however, its behavior is restricted by prior symbolic knowledge \citep{chaudhuri2021neurosymbolic}.  During training and inference, Logical Neural Networks (LNNs) impose rigid or soft constraints on the network's output by encoding domain expertise in the form of first-order or fuzzy logic rules \citep{chaudhuri2021neurosymbolic}. Differentiable inductive logic programming is another example, which is designed to acquire logical rules within a differentiable framework \citep{chaudhuri2021neurosymbolic}.

\subsection{NeuroSymbolic}

The network is not structured using symbolic logic rules in this category; rather, they are mapped onto continuous vector embeddings. This is a thoroughly integrated hybrid. The embeddings subsequently serve as flexible constraints or regularizers within the loss function of the neural network during the training process. This stimulates the network to acquire representations that are in accordance with the logical knowledge that has been provided, without rigorously enforcing it \citep{chaudhuri2021neurosymbolic}. Key examples of this methodology include logical tensor networks (LTNs). LTNs apply logical formulas to establish constraints on tensor representations of entities and relations, and they have been effectively implemented in tasks such as knowledge graph completion, which involve the identification of representations that adhere to established logical rules \citep{chaudhuri2021neurosymbolic}.
\subsection{Neuro[Symbolic]}
The system is predominantly a neural model in this final paradigm; however, it is also capable of performing symbolic-like reasoning. The neural component remains the primary controller, which distinguishes this from Symbolic[Neuro]. The neural architecture is frequently equipped with mechanisms, such as attention, that enable it to selectively interact with or concentrate on symbolic information in order to accomplish this \citep{chaudhuri2021neurosymbolic}. One prevalent architecture in this category is graph neural networks (GNNs). GNNs can acquire the ability to incorporate logical rules and represent symbolic expressions by selectively attending to the most germane symbolic information for a given task when equipped with attention mechanisms \citep{chaudhuri2021neurosymbolic}.

The selection of one of these architectural paradigms is of significant consequence. The high-level structure of a problem is presumed to be known and amenable to a symbolic algorithm in a symbolic [Neuro] architecture, with the neural network's sole purpose being to learn a heuristic that was previously difficult to engineer. On the other hand, a Neuro:Symbolic→Neuro architecture implies that the issue is essentially a pattern-recognition task that is most effectively approximated by a flexible neural network. However, the learning process can be optimized for data efficiency and reliability by incorporating symbolic constraints. The Neuro|Symbolic architecture that is pipelined implies a clean decomposition of the task into perception and reasoning stages. This approach is effective for specific problems, such as visual question answering \citep{cranmer2020interpretable}, but it may not be appropriate for tasks where perception and reasoning are inextricably linked. This demonstrates that there is no universally superior neuro-symbolic architecture; the optimal choice is dependent on the specific task, the nature of the available data, and the extent and form of prior symbolic knowledge. Runtime profiling has demonstrated that the symbolic component can occasionally dominate system latency, underscoring the necessity of meticulous architectural design \citep{chaudhuri2021neurosymbolic}. Consequently, the practical implications of this decision are substantial.

\begin{table}[ht]
\centering
\caption{A Taxonomy of Neuro-Symbolic Integration Paradigms.}
\label{tab:neuro_symbolic_taxonomy}
\begin{tabular}{|p{0.18\textwidth}|p{0.35\textwidth}|p{0.32\textwidth}|}
\hline
\textbf{Paradigm} & \textbf{Core Principle} & \textbf{Representative System(s)} \\
\hline
Symbolic[Neuro] & A primary symbolic algorithm calls a neural network as a subroutine for sub-symbolic tasks or heuristics. & AlphaGo \citep{chaudhuri2021neurosymbolic} \\
\hline
Neuro|Symbolic & A pipeline where a neural front-end performs perception and a symbolic back-end performs reasoning. & NVSA, NS-VQA \citep{chaudhuri2021neurosymbolic} \\
\hline
Neuro:Symbolic→Neuro & Symbolic knowledge is compiled into the structure or training process of a neural network to guide its learning. & Logical Neural Networks (LNNs) \citep{chaudhuri2021neurosymbolic} \\
\hline
NeuroSymbolic & Symbolic logic rules are mapped to embeddings that act as soft constraints or regularizers on the NN's loss function. & Logical Tensor Networks (LTNs) \citep{badreddine2022logical} \\
\hline
Neuro[Symbolic] & A primary neural model is empowered with symbolic reasoning capabilities, e.g., via attention over symbols. & GNNs with Attention, NLM \citep{chaudhuri2021neurosymbolic} \\
\hline
\end{tabular}
\end{table}

\subsection{The Centrality of Domain-Specific Languages (DSLs)}
Domain-Specific Language (DSL) functions as a fundamental and unifying construct in this architectural landscape that is rich in diversity.  Devlin et al. (2017) \citep{devlin2017robustfill}. define a domain-specific language (DSL) as a programming language that is explicitly designed for a specific problem domain with limited expressivity. The NSP context is characterized by the DSL's role as the critical link between the symbolic and neural worlds. It establishes the symbolic building elements -- the vocabulary of primitive operations -- that the synthesizer can employ to develop a program \citep{gulwani2017program}. 

DSL design is an indispensable component of knowledge engineering. It serves as the primary mechanism by which a human expert can introduce prior knowledge into the system, thereby restricting the search space and inclining the synthesizer toward programs that are logical and likely to be accurate \citep{knoth2023type}. According to Devlin et al. (2017) \citep{devlin2017robustfill}, a DSL that is excessively restrictive may be incapable of articulating the appropriate solution, while a DSL that is excessively general can render the synthesis problem intractable on account of an exponentially large search space. As a result, the meticulous design of the DSL is frequently a critical component of successful neuro-symbolic synthesis \citep{knoth2023type}. A critical area of focus in the field is the advancement of techniques that can automate the learning of the DSL, thereby reducing the dependence on human expertise, as will be discussed in the future.

\section{Foundational Methodologies in Neuro-Symbolic Synthesis}

The field of Neuro-Symbolic Program Synthesis has undergone a series of distinguishing methodological phases, each of which has addressed the limitations of its predecessors. This progression is indicative of the increasing sophistication of the integration of neural and symbolic techniques, which has progressed from basic guidance to in-depth integration and knowledge acquisition. This section reviews three fundamental algorithmic paradigms that have significantly influenced the field, as evidenced by case studies of seminal systems.

\subsection{Neural-Guided Symbolic Search}
Neural networks are employed as potent, learned heuristics to guide traditional symbolic program search algorithms, which is one of the earliest and most direct approaches to NSP. The combinatorial explosion of the search space is the primary obstacle in classical program synthesis. Even for a moderately complex DSL, the number of possible programs increases exponentially with length, rendering exhaustive enumeration or deductive search intractable for all but the most basic problems \citep{devlin2017robustfill}. The fundamental innovation of neural-guided search is the recasting of the problem of developing an effective search heuristic as a supervised learning task. The model can learn to predict salient properties of a likely solution directly from a high-level specification, such as a set of input-output (I/O) examples, by training a neural model on a large corpus of extant synthesis problems. This prediction is subsequently employed to intelligently prune or re-order the search space, thereby significantly expediting the discovery of a correct program \citep{zhang2018neural}. 

DeepCoder, which was the first system to introduce the Learning Inductive Program Synthesis (LIPS) framework, is a pioneering system in this paradigm \citep{balog2017deepcoder}. DeepCoder's mechanism is conceptually simple yet highly effective. A simple feed-forward neural network is trained to predict the likelihood of each function in the DSL appearing in the final, correct program, using a set of I/O examples as input. This "attribute prediction" phase produces a distribution of the solution's probable components \citep{devlin2017robustfill}. This distribution is subsequently employed to improve a conventional symbolic search algorithm, such as depth-first search. The search is altered to prioritize exploratory program candidates that are generated from the high-probability functions identified by the neural network \citep{balog2017deepcoder}. The viability of the neural-guided search paradigm was firmly established by the demonstration of an order-of-magnitude speedup over unguided search baselines by this basic form of neural guidance \citep{devlin2017robustfill}.

\begin{algorithm}
\caption{Simplified Pseudocode for Neural-Guided Symbolic Search}
\label{alg:neural_guided_search}
\begin{algorithmic}[1]
\STATE \textbf{function} NeuralGuidedSearch($S, L, M_{\theta}$)
\STATE \quad \COMMENT{Predict attributes (e.g., function probabilities) from specification}
\STATE \quad $A \leftarrow M_{\theta}(S)$
\STATE
\STATE \quad \COMMENT{Initialize symbolic search with guidance from predicted attributes}
\STATE \quad $P \leftarrow \text{SymbolicSearch}(S, L, A)$
\STATE
\STATE \quad \textbf{return} $P$
\STATE \textbf{end function}
\end{algorithmic}
\end{algorithm}

Consequently, subsequent research has investigated the integration of the neural guide and the symbolic searcher at a more intricate and refined level. Neural Guided Deductive Search (NGDS) systems, for example, more closely integrate a neural model with a deductive search framework, such as Microsoft's PROSE \citep{ellis2021dreamcoder}. Within this configuration, the neural model is activated at each decision point of the deductive search process. It learns a dynamic ranking function that directs the deductive engine toward more promising paths by predicting which production rule in the DSL's grammar is most likely to contribute to a correct program \citep{ellis2021dreamcoder}. This corresponds to a transition from the prediction of global program properties (as in DeepCoder) to the formulation of local, contextual search decisions.

According to Zhang et al. (2018) \citep{zhang2018neural}, Neural-Guided miniKanren exemplifies an even more profound level of integration by directly connecting a neural model to the internal state of a constraint logic programming system. Initially, miniKanren converts the PBE problem into a sequence of recursive logical constraints in this method. The neural model, which may be a Recurrent Neural Network (RNN) or a Graph Neural Network (GNN), receives this internal representation of constraints as its direct input \citep{facchin2023neural}. A score is subsequently calculated to indicate the likelihood of satisfying a specific partial program and its associated constraints. MiniKanren's search is directed by these scores, which specify which branches of the logical search tree to investigate next \citep{facchin2023neural}.

This paradigm's progression discloses an unambiguous trajectory. It commences with a conservative yet effective hybridization strategy that retains the formal structure and correctness guarantees of symbolic search while utilizing the pattern-matching capabilities of neural networks to render the search tractable. A taught, black-box heuristic is applied to the neural network. The progression from DeepCoder's high-level guidance to miniKanren's low-level, constraint-based guidance indicates a trend toward steadily tighter coupling. This increased integration enables the neural guide to make more context-aware decisions by "seeing" into the symbolic reasoning process itself, albeit at the expense of a more intricate interface between the two components \citep{zhang2018neural}. This entire paradigm is consistent with the Symbolic[Neuro] architecture, in which the symbolic search remains the primary process and the task of heuristic evaluation is delegated to a potent but subservient neural oracle.

\subsection{Library Learning and Compositional Generalization}
Although neural-guided search methods increase the efficiency of solving individual synthesis tasks, they are fundamentally "amnesiac" in nature, as the knowledge acquired in the process of solving one problem is not retained or applied to the next. Unlike human experts, who construct and utilize extensive mental libraries of reusable concepts, patterns, and abstractions, this is in stark contrast. The primary objective of the NSP paradigm of library learning is to endow systems with this essential capability: the capacity to acquire and repurpose knowledge over time. This capability facilitates compositional generalization and automates the discovery of the fundamental concepts that constitute a comprehensive DSL \citep{ellis2021dreamcoder}.

DreamCoder is the seminal system in this domain, introduced a novel architecture that bootstraps its own knowledge by iteratively alternating between solving problems and reflecting on those solutions to create a library of reusable program components \citep{ellis2021dreamcoder}. The central mechanism that drives this process is a distinctive "wake-sleep" algorithm, which is inspired by but distinct from the original machine learning algorithm of the same name \citep{ellis2021dreamcoder}.

The DreamCoder algorithm is applied in cycles, each of which consists of three phases:

\begin{itemize}
    \item \textbf{Waking Phase:} During this "problem-solving" phase, the system employs its current library of functions (which consists of its DSL) and a neural search policy (its "recognition model") to resolve as many tasks as possible from a specified corpus of problems. The neural policy directs the search for programs that satisfie the I/O specifications of each task \citep{ellis2021dreamcoder}.
    \item \textbf{Abstraction Sleep Phase:} This is the core "knowledge consolidation" phase. Programs that were effectively synthesized during waking hours are examined by DreamCoder. The algorithm utilizes a sophisticated refactoring approach that is predicated on equivalence graphs (E-graphs) to identify common structural patterns and sub-expressions among these solutions \citep{ellis2021dreamcoder}. This process involves the compression and abstraction of these common components into new, named functions, which are subsequently incorporated into the system's library. This process effectively expands and enriches the DSL with higher-level concepts, with the guidance of a compression principle: the most effective new abstractions are those that enable the previous solutions to be expressed more compactly \citep{ellis2021dreamcoder}.
    \item \textbf{Dream Sleep Phase:} This "skill-honing" phase involves the retraining of the neural search policy to become proficient in the utilization of the newly expanded library. The network acquires knowledge from two sources of self-generated data: "replays," which are the successful (program, task) pairs from the waking phase, and "dreams," which are novel tasks and their corresponding programs developed by randomly composing functions from the newly enriched library \citep{ellis2021dreamcoder}. This research assert that dream training enables the system to investigate the expressive potential of its novel concepts and acquire the ability to employ them effectively, even in the face of challenges it has yet to encounter.
\end{itemize}

\begin{algorithm}
\caption{High-Level Pseudocode for the DreamCoder Wake-Sleep Algorithm}
\label{alg:dreamcoder}
\begin{algorithmic}[1]
\STATE $L \leftarrow L_{0}$ \COMMENT{Initial primitive library}
\STATE $Q \leftarrow \text{InitializeRecognitionModel}()$
\STATE \textbf{for} iteration = 1 to N \textbf{do}
\STATE \quad \COMMENT{Waking Phase: Solve tasks using current library and model}
\STATE \quad $\textit{Solutions} \leftarrow \text{SolveTasks}(T, L, Q)$
\STATE
\STATE \quad \COMMENT{Abstraction Sleep Phase: Refactor solutions to grow the library}
\STATE \quad $\textit{NewAbstractions} \leftarrow \text{RefactorAndCompress}(\textit{Solutions})$
\STATE \quad $L \leftarrow L \cup \textit{NewAbstractions}$
\STATE
\STATE \quad \COMMENT{Dream Sleep Phase: Retrain the recognition model}
\STATE \quad $\textit{ReplayData} \leftarrow \textit{Solutions}$
\STATE \quad $\textit{DreamData} \leftarrow \text{GenerateDreams}(L)$
\STATE \quad $Q \leftarrow \text{TrainRecognitionModel}(Q, \textit{ReplayData} \cup \textit{DreamData})$
\STATE \textbf{end for}
\STATE \textbf{return} $L, Q$
\end{algorithmic}
\end{algorithm}

An effective cascading dynamic is established by this wake-sleep cycle. The system is comprised of a limited number of primitive functions at the outset. Is capable of resolving only the most elementary issues. On the other hand, the solutions to these straightforward issues contain the germs of more intricate concepts. In the abstraction phase, these concepts are identified (e.g., by iterating over a list) and incorporated into the library (e.g., as a map function). This expanded library enables the system to address more intricate issues during the subsequent waking phase, thereby supplying the necessary basic materials for the discovery of even more sophisticated abstractions. According to Ellis et al. (2021) \citep{ellis2021dreamcoder}, DreamCoder was demonstrated to independently rediscover fundamental concepts of functional programming (such as map and fold), vector algebra, and even basic laws of physics through this iterative process, thereby constructing a hierarchical, multi-layered library of interpretable knowledge. This represents a significant stride toward the development of true compositional generalization and directly addresses the critical challenge of manual DSL design by automating it \citep{ellis2021dreamcoder}.

HOUDINI, another remarkable system, examines perpetual learning from a comparable perspective, representing neural networks as functional programs that are strongly typed \citep{vered2022houdini}. The symbolic synthesizer of HOUDINI conducts a type-directed search over program architectures that compose functions from an evolving library of neural modules when pre-sented with a sequence of tasks.
The implementation of a robust type system serves as a highly effective symbolic constraint, guaranteeing that neural components are assembled in a valid manner and substantially expediting the pursuit of a valid program architecture \citep{vered2022houdini}.

The paradigm of library learning represents a significant change in the overarching objective of program synthesis. It is no longer sufficient to identify a single program that resolves a single task; rather, the objective is to construct a theory of the entire solution domain. The learned library is not merely a compilation of subroutines that serve as aids; it is a symbolic, structured, and emergent representation of the domain's intrinsic conceptual framework. This procedure, which is influenced by the principles of abstraction and compression, is a substantial advancement in the development of more general, adaptive, and genuinely intelligent systems. It also serves as a reflection of human concept acquisition theories.

\subsection{Symbolic Constraints on Neural Generation}
A paradigm that is highly influential and alternative in NSP reverses the roles observed in neural-guided search. In this context, the neural network serves as the primary program generator, with symbolic components serving as rigorous "critics" or "guardrails" that ensure the output is valid and accurate. The early attempts to employ standard neural sequence-to-sequence models for program generation were significant challenges, which motivated this approach.

Naive neural models are susceptible to two critical failure modes when they are approached as a straightforward translation task, such as from I/O examples to a sequence of program tokens. First and foremost, they lack an inherent comprehension of syntax and can effortlessly produce code that is syntactically invalid and, as a result, non-executable \citep{shah2020learning}.  Secondly, they are plagued by the issue of \textit{program aliasing}. For a given specification, there are frequently numerous semantically correct programs. However, supervised training penalizes the model for producing any program that does not precisely match the single, arbitrary ground-truth program presented in the training data. This serves as an unhelpful and misleading learning signal \citep{jin2022learning}. These issues are addressed by the methodologies in this paradigm, which explicitly incorporate symbolic feedback into the generation and training loop.

Syntactic correctness is enforced by design as one of the initial lines of defense. By restricting the output of the neural decoder at each generation phase, it is possible to ensure that it generates tokens that are valid in accordance with the context-free grammar (CFG) of the DSL \citep{shiqi2019neuro}. In order to ensure that any fully generated program is syntactically correct, the system masks out invalid tokens from the decoder's probability distribution at each stage \citep{shah2020learning}. \textit{type-directed synthesis} is a more advanced application of this technique, in which the synthesizer is directed by the type signatures of functions and variables rather than solely by syntax \citep{silver2017mastering}.
The search space of potential programs can be significantly reduced by ensuring that the derived programs are not only syntactically well-formed but also type-safe \citep{sinha2019clutrr}. This approach is clearly illustrated by the \textit{Typed Neuro-Symbolic Program Synthesis}  (TNSPS) system. The neural representations employed by a tree-based synthesizer are expressly improved by the inclusion of information regarding the types of I/O examples, grammar rules, and unfilled "holes" in a partial program tree. The model is able to capitalize on symbolic type constraints during its neural prediction process by encoding type information and concatenating it to the existing neural embeddings \citep{hu2021iraven}.

Syntactic and type constraints resolve the issue of generating invalid code; however, they do not resolve the issue of program aliasing. Systems must integrate semantic feedback—information regarding the correctness of a program's behavior—in order to resolve this issue. For the domain of string transformations, RobustFill, an early and influential system, addressed this issue \citep{devlin2017robustfill}. From I/O examples, it generates candidate programs using an attentional sequence-to-sequence model. To address aliasing, it implements a beam search during decoding to produce a variety of program candidates. Critically, it then employs a symbolic executor to evaluate each candidate program against the I/O examples that have been supplied. The first program in the beam that is determined to be semantically consistent (i.e., it accurately replicates all outputs from the inputs) is chosen \citep{devlin2017robustfill}. This post-hoc mechanism, known as the "generate-and-verify" cycle, is capable of filtering semantically correct programs from the neural generator's proposals \citep{devlin2017robustfill}.

\begin{algorithm}
\caption{High-Level Pseudocode for Generate-and-Verify Synthesis}
\label{alg:generate_verify}
\begin{algorithmic}[1]
\STATE \textbf{function} GenerateAndVerify($S, M_{\theta}, k$)
\STATE \quad \COMMENT{Generate k candidate programs using beam search}
\STATE \quad $\textit{Candidates} \leftarrow \text{BeamSearch}(S, M_{\theta}, k)$
\STATE
\STATE \quad \COMMENT{Verify each candidate against the specification}
\STATE \quad \textbf{for each} $P \in \textit{Candidates}$ \textbf{do}
\STATE \quad \quad \textbf{if} IsConsistent($P, S$) \textbf{then}
\STATE \quad \quad \quad \textbf{return} $P$
\STATE \quad \quad \textbf{end if}
\STATE \quad \textbf{end for}
\STATE
\STATE \quad \textbf{return} Failure
\STATE \textbf{end function}
\end{algorithmic}
\end{algorithm}

To address program aliasing in a more comprehensive and principled manner, it is necessary to reframe the training objective through the use of Reinforcement Learning (RL). The policy gradient method is employed to train the system, as opposed to the conventional supervised cross-entropy loss, which incentivizes the matching of a specific reference program \citep{chen2018execution}. An RL agent, the neural generator generates a program (referred to as a "action"). A symbolic environment executes this program, and the agent is awarded a positive reward if the program is semantically correct (i.e., passes all I/O tests), irrespective of its syntactic form. This directly optimizes the model for the true objective of generating functionally correct programs and effectively addresses the program aliasing problem by equitable rewarding all valid solutions \citep{shah2020learning}.
The symbolic component functions as a formal verifier that ensures correctness, while the neural network is responsible for the creative, intuitive, and pattern-matching aspects of generation. This paradigm exemplifies a potent division of labor. The paradigm's conceptual evolution—from straightforward syntactic constraints to type-based guidance, post-hoc semantic verification, and ultimately to fully incorporated RL-based training—demonstrates a substantial increase in sophistication. This progression establishes the fundamental architectural pattern of a strong but potentially fallible neural generator in conjunction with a dependable symbolic verifier. This pattern is essential to the contemporary era of neuro-symbolic synthesis, which is driven by LLM.

\begin{table}[ht]
\centering
\caption{Evolution of Methodologies in Neuro-Symbolic Program Synthesis}
\label{tab:nsp_methodologies_evolution}
\begin{tabular}{|p{0.22\textwidth}|p{0.25\textwidth}|p{0.25\textwidth}|p{0.2\textwidth}|}
\hline
\textbf{Methodology} & \textbf{Core Problem Addressed} & \textbf{Key Innovation} & \textbf{Seminal System(s)} \\
\hline
Neural-Guided Symbolic Search & Combinatorial explosion of the search space. & Using a neural network as a learned heuristic to guide a symbolic search algorithm. & DeepCoder \citep{balog2017deepcoder}, NGDS \citep{ellis2021dreamcoder} \\
\hline
Library Learning \& Abstraction & Lack of generalization and knowledge reuse between tasks; manual DSL design. & A "wake-sleep" algorithm that alternates between solving problems and abstracting solutions into a reusable library. & DreamCoder \citep{ellis2021dreamcoder} \\
\hline
Constrained Neural Generation & Generation of syntactically invalid code and the "program aliasing" problem. & Imposing symbolic constraints (grammar, types) and semantic feedback (verification, RL) on a neural generator. & TNSPS \citep{hu2021iraven}, RobustFill \citep{devlin2017robustfill}, RL-based Synthesizers \citep{chen2018execution} \\
\hline
\end{tabular}
\end{table}

\section{The Paradigm Shift: The Influence of Large Language Models}
In the field of artificial intelligence, the recent emergence of large-scale, pre-trained foundation models, particularly Large Language Models (LLMs), has prompted a significant paradigm shift, and Neuro-Symbolic Program Synthesis is no exception. These models, which have been trained on internet-scale corals of text and code, have exhibited exceptional zero-shot and few-shot capabilities for a diverse array of tasks, including code generation \citep{cranmer2020interpretable}. This has significantly changed the architectural assumptions and research challenges in NSP, shifting the emphasis from the training of bespoke neural components to the orchestration, verification, and refinement of outputs from these powerful, general-purpose generative models.

The classic Neuro|Symbolic pipeline, which frequently necessitated the tedious process of training a specialized neural network for a particular perception subtask (e.g., an image classifier to convert pixels into symbolic labels), has been one of the most immediate impacts \citep{cranmer2020interpretable}. Many of these perception and semantic parsing tasks can now be performed "out of the box" by foundation models through meticulously engineered prompts, which frequently eliminates the necessity for task-specific training data and model development \citep{cranmer2020interpretable}. This has resulted in the identification of numerous "pitfalls" for conventional neuro-symbolic methods in the era of foundation models. These include the \textit{compute pitfall}, which involves the unnecessarily training of a model when a prompted foundation model would suffice, the \textit{data pitfall}, which involves the overfitting of a small model to a labeled dataset when the broad knowledge of a foundation model would be more robust, and the \textit{program pitfall}, which involves the reliance on a single, potentially flawed, symbolic program to provide supervision when an LLM could generate diverse alternatives \citep{cranmer2020interpretable}.

Consequently, the predominant mode of interaction has transitioned from training to prompting. The LLM has emerged as the de facto neural "generator" in numerous NSP systems, capable of generating complex code from natural language specifications or a limited number of I/O examples \citep{solarlezama2005combinatorial}. Nevertheless, these models are unreliable, despite their exceptional capabilities. The function of a symbolic "verifier" is now more critical than ever, as they are prone to generating code that is subtly incorrect, logically inconsistent, or contains hallucinations \citep{shiqi2019neuro}. This has led to the emergence of a new architectural pattern that is now dominant: the verifier-in-the-loop.

\subsection{Verifier-in-the-Loop Architectures}
This architectural pattern is based on an iterative feedback cycle that forms between a deterministic symbolic tool and a generative LLM. The LLM suggests a solution, and the symbolic tool validates it, thereby providing feedback that is employed to direct the LLM's subsequent endeavor. This method capitalizes on the LLM's extensive generative capabilities while predicating its output on formal correctness.

Counter-Example Guided Inductive Synthesis (CEGIS) is a potent implementation of this pattern in the context of LLMs \citep{kaplan2020scaling}. Classical CEGIS loops, which were previously employed with symbolic synthesizers, are repurposed for the new paradigm. A candidate program is proposed by the LLM, which functions as the \textit{Generator} in response to a specification. This program is subsequently delivered to a Verifier, which may consist of a formal SMT solver, a static analyzer, a compiler, or a test suite. Program verification is conducted by the Verifier in accordance with the specification. The process will conclude if the program is operating correctly. If the assertion is inaccurate, the Verifier generates a specific counterexample (e.g., a test case that fails) that illustrates the defect. Subsequently, this counterexample is integrated into a new prompt that is transmitted to the LLM, which is directed to correct the program in accordance with the newly acquired information. The loop will persist until either a suitable program is identified or a timeout is encountered \citep{chen2021evaluating}. This approach, which is based on CEGIS, has been demonstrated to be highly effective for tasks such as automated program repair (APR) and formal synthesis. It allows an LLM to systemically refine their plausible but imperfect initial guess in order to achieve a correct and verified solution \citep{devlin2017robustfill}.

\begin{algorithm}
\caption{LLM-based CEGIS Loop}
\label{alg:llm_cegis}
\begin{algorithmic}[1]
\STATE $\textit{CounterExamples} \leftarrow \emptyset$
\STATE \textbf{for} iteration = 1 to N \textbf{do}
\STATE \quad \COMMENT{Generate a program candidate using the LLM}
\STATE \quad $\textit{Prompt} \leftarrow \text{ConstructPrompt}(S, \textit{CounterExamples})$
\STATE \quad $P \leftarrow G_{\text{LLM}}(\textit{Prompt})$
\STATE
\STATE \quad \COMMENT{Verify the program and get a counterexample if it fails}
\STATE \quad $\textit{isCorrect}, ce \leftarrow V(P, S)$
\STATE
\STATE \quad \textbf{if} $\textit{isCorrect}$ \textbf{then}
\STATE \quad \quad \textbf{return} $P$
\STATE \quad \textbf{else}
\STATE \quad \quad $\textit{CounterExamples} \leftarrow \textit{CounterExamples} \cup \{ce\}$
\STATE \quad \textbf{end if}
\STATE \textbf{end for}
\STATE \textbf{return} Failure
\end{algorithmic}
\end{algorithm}

The LLM's sophisticated language capabilities are utilized to develop more extensive feedback mechanisms, a process that is occasionally referred to as verbal reinforcement or self-correction. Subsequent modifications to this cycle are also implemented. Instead of merely offering a raw counterexample, these methods encourage the LLM to contemplate its own shortcomings in natural language. By providing a generative agent with an episodic memory of its previous endeavors, the Reflexion framework serves as an illustration of this methodology \citep{shinn2023reflexion}. A "self-reflection" prompt is issued following an unsuccessful trial, requesting that the LLM examine the trajectory of actions, provide an explanation for the failure, and propose a more effective strategy for the subsequent attempt. This reflective text is subsequently retained in the agent's memory and incorporated into the context for the subsequent trial, serving as a kind of "verbal reinforcement" that directs the policy toward more effective solutions without requiring any weight updates. Source: \citep{tjandrasuwita2021learning}.

In the same vein, the Self-Debugging technique encourages an LLM to debug its own generated code, emulating the typical human software development practice of "rubber duck debugging" \citep{chen2021evaluating}. It is possible to accomplish this with or without external execution feedback. The model is presented with a multi-turn dialogue that requires it to generate code, elucidate the logic of its own code, identify potential flaws based on this explanation (or a provided error message), and ultimately generate a corrected version \citep{pan2023rustassistant}. This reflective, structured process has been demonstrated to considerably enhance the accuracy of code generation, particularly for intricate problems that are unlikely to yield a correct solution in a single pass \citep{mao2019neuro}.

\subsection{Enhancing LLM Reasoning with Symbolic Scaffolding}
Another significant trend is the utilization of symbolic structures to enhance and scaffold the intermediate reasoning process of the LLM, in addition to verifying the final program output.. Despite the fact that LLMs are capable of producing fluent and coherent text, their underlying reasoning may be inconsistent, logically flawed, and disconnected from empirical foundations, particularly when dealing with complex, multi-step \citep{zhang2024proofofthought}. Symbolic scaffolding endeavors to alleviate this by imposing a structure on the model's "thought process."

This signifies a natural progression of prompting strategies. One research (\citep{kleinberg2018algorithmic}) has demonstrated that the performance of complex tasks is enhanced by eliciting a step-by-step reasoning trace, as evidenced by the initial transition from simple Input-Output prompting to Chain-of-Thought (CoT) prompting. The complexity of non-linear reasoning structures was further enhanced by methods such as \textit{Tree-of-Thoughts}  (ToT), which allows for the combination and revisiting of ideas, and Graph-of-Thoughts (GoT) \citep{kleinberg2018algorithmic}. Both of these methods explore multiple reasoning paths in parallel.

This trend has reached its neuro-symbolic apex with the Proof of Thought framework \citep{zhang2024proofofthought}. It then takes the critical next step of compelling the LLM to externalize its reasoning process as a formal, structured program in a purpose-built, JSON-based DSL, rather than as unstructured natural language. The explicit purpose of this DSL is to represent logical components, including facts, rules, and inferential steps. The responsibility of the LLM is to produce a "proof" in this language that logically connects the antecedents of a problem to its conclusion. Next, a distinct, deterministic symbolic verifier can parse this programmatic proof and verify its logical validity in a step-by-step manner \citep{zhang2024proofofthought}. The rigorous, formal requirements of logical verification are ingeniously decoupled from the LLM's powerful, intuitive, and creative capacity for generating ideas and hypotheses by this architecture \citep{zhang2024proofofthought}. By transferring the responsibility of guaranteeing logical soundness to a dependable symbolic tool, the LLM is able to focus on its primary function—proposing plausible reasoning paths.

As a result, the era of LLMs has facilitated a "great unbundling" of neuro-symbolic systems. A more modular, flexible architecture has substantially replaced the conventional method of designing and training a monolithic, bespoke model for a specific task. A set of specialized, often pre-existing, symbolic components (verifiers, compilers, SMT solvers) and a powerful, general-purpose neural component (the off-the-shelf LLM) comprise this novel architecture. Consequently, the primary research challenge has shifted from the design and training of neural architecture to the design of effective feedback loops, prompt engineering, and system orchestration \citep{vinyals2019grandmaster}. This modularity is a substantial advantage, as it enables researchers to independently upgrade the neural engine (e.g., from GPT-4 to a future model) or the symbolic verifier (e.g., from a simple test suite to a formal proof checker) without the need to rethink the entire system. The most lucid modern example of the fundamental neuro-symbolic philosophy is perhaps frameworks such as Proof of Thought, which decouple the generation of ideas from the verification of logical steps. This synergistic partnership is characterized by the unique and complementary strengths of each component.

\section{Applications and Domains}
By expanding its application in a variety of fields, the neuro-symbolic paradigm demonstrates its practical value. AI systems that are not only predictive but also interpretable, reliable, robust to data scarcity, and capable of employing structured, explicit domain knowledge are the common thread that unites these applications \citep{bommasani2021opportunities}. NSP is demonstrating its ability to be a critical enabling technology in sectors where the "black box" nature of solely neural systems poses a substantial impediment to adoption.

\subsection{Scientific Discovery}

There is the potential for NSP techniques to expedite scientific discovery by automating components of the scientific method \citep{cranmer2023symbolic}. In the present context, the objective is to develop a program that embodies a potential scientific theory or hypothesis. It is inherently interpretable, as the learned model is a program, which enables human scientists to analyze, comprehend, and expand upon the knowledge that has been discovered. This satisfies a fundamental requirement of the scientific process: that new hypotheses must be in accordance with existing knowledge and facilitate the examination of their implications \citep{cranmer2023symbolic}. NSP systems have rediscovered fundamental laws of physics from simulated data, in addition to \textit{symbolic regression}, which aims to identify the underlying mathematical equations that correspond to experimental data \citep{cranmer2023symbolic}. Other applications involve the analysis of complex, high-dimensional data in fields such as \textit{behavioral science}. NSP can generate symbolic descriptions of animal behavior from spatiotemporal tracking data, thereby providing interpretable models that are more beneficial to domain experts than opaque neural classifiers  \citep{zhan2021framework}.

\subsection{Safety-Critical and High-Stakes Domains}
In safety-critical domains, the demand for interpretability, verifiability, and trustworthiness is most pronounced, rendering them an ideal candidate for NSP \citep{chaudhuri2021neurosymbolic}.

\textbf{Healthcare and Medicine:} NSP is being implemented to establish more transparent and dependable clinical decision support systems. Treatment effect estimation from observational data is a central problem in causal inference, and it is one of the most important applications. In this context, the DSL can be tailored to expressly encode recognized causal assumptions and inductive biases from the medical literature, resulting in more robust and data-efficient models than those that are solely neural \citep{lee2022neuro-causal}. Other research investigates the utilization of NSP to construct \textit{"digital twins"} of patients that are comprehensible - dynamic, AI-driven models of physiological and clinical states that can be employed to simulate treatment outcomes in a comprehensible manner \citep{xia2022automated}.
\textbf{Finance and Regulatory Compliance:} In the financial sector, hybrid models are being developed that integrate the logical rigor of symbolic rule engines with the pattern-recognition capabilities of neural networks for tasks such as fraud detection \citep{manna1980deductive}. The symbolic component is capable of encoding intricate business logic and regulatory requirements, thereby guaranteeing that the system's decisions are not only precise but also consistent and comprehensible. This is essential for accountability and auditing \citep{chaudhuri2021neurosymbolic}.

\textbf{Robotics and Autonomous Systems:} The provision of autonomous agents with predictability and safety is an indispensable endeavor. NSP establishes a framework for the development of control policies that are more solid and verifiable. In Yang et al. (2022) \citep{yang2022differentiable} accomplish this by integrating learnt neural components (e.g., for perception or low-level motor control) with symbolic planners, safety constraint monitors, or high-level procedural reasoning. Specifically, neuro-symbolic reinforcement learning endeavors to enhance the interpretability and sample efficiency of RL agents by representing policies as programs \citep{verma2018programmatically}.

\subsection{Software and Systems Engineering}
NSP is also being implemented in the software development and maintenance process, with the potential to improve the reliability of software systems and the productivity of developers.

\textbf{Code Generation and Comprehension:} Despite the remarkable code generation capabilities of large code models such as GitHub Copilot, they continue to experience reliability and determinism issues \citep{shiqi2019neuro}. An increasing body of research is devoted to the integration of these LLMs with conventional symbolic methods in order to enhance the quality of the generated code \citep{zhang2024proofofthought}. NSP can also be utilized to enhance \textit{program comprehension}, such as by generating an abstract representation of a code segment to assist in the identification of potentially defective components \citep{ding2022patch}.

\textbf{Program Repair and Security Fuzzing:} Chen et al. (2021) \citep{chen2021evaluating} have previously discussed the verifier-in-the-loop architectures, which are a direct application of NSP to the task of automated program repair (APR). In this architecture, a symbolic verifier guides and corrects an LLM's ability to propose solutions. Neuro-symbolic techniques are employed in cybersecurity to enhance the efficacy of \textit{security fuzzing}. The symbolic fuzzer can be directed to more intelligently explore these areas and identify additional vulnerabilities by a neural model that can learn to predict which parts of a program's input space are most likely to activate bugs \citep{zong2020fuzzing}.

\subsection{Natural Language Understanding and Reasoning}
NSP provides a potent alternative to end-to-end neural models for complex language tasks that necessitate multi-step reasoning and interaction with external knowledge.

\textbf{Visual and Textual Question Answering (QA):} Parsing a natural language query into a symbolic program is a prevalent NSP approach to QA. The answer is derived by executing this program against a structured knowledge source, such as a database, a knowledge base, or the symbolic representation of an image. This method enforces a compositional reasoning process that is frequently more generalizable and robust than that of end-to-end models, which may rely on spurious correlations in the training data \citep{zhan2021framework}.

\textbf{Web Information Extraction:} It is a substantial challenge to extract structured data from the unstructured and highly diverse web landscape. Neuro-symbolic DSLs have been developed specifically for this task, integrating pre-trained neural NLP models (for text comprehension) with symbolic primitives for string manipulation and HTML tree navigation. Robust programs for collecting information from a diverse array of websites can be generated by synthesizers that employ these DSLs \citep{knoth2023type}.

\section{Open Challenges and Future Trajectories}

The field of Neuro-Symbolic Program Synthesis is still in its infancy and faces a number of fundamental unresolved challenges, despite its significant promise and rapid progress. It will be imperative for the field to achieve its maximum potential and fulfill its fundamental objective of developing AI systems that are more human-like, reliable, and capable by addressing these limitations while pursuing promising new research trajectories.

\subsection{Persistent Challenges}
\textbf{The Explainability Paradox:} Although the discipline is primarily motivated by the desire to address the opacity of deep learning, the objective of achieving true, end-to-end explainability remains elusive. A considerable improvement over a black-box model is the production of an interpretable artifact by NSP systems—the synthesized program. Nevertheless, the total procedure is not necessarily transparent. The complex, dynamic interactions between the neural and symbolic components within the system are challenging to fully comprehend and debug, and the neural components themselves remain largely opaque \citep{zhang2020survey}. Research has underscored a critical distinction between models that are "explainable by design" and those that necessitate "post-hoc" explanation. It is frequently a significant challenge to comprehend the neural component's rationale for directing the synthesis toward a particular program, even after it has been generated \citep{zhang2020survey}.

\textbf{Scalability and Computational Sustainability:} Computational cost is a significant obstacle for NSP. Both of its constituent technologies—deep learning and symbolic search—are computationally intensive. Combinatorial proliferation in the search space is a significant issue for symbolic methods, while the training and inference of large neural models necessitate substantial computational resources \citep{zhang2019raven}. This problem is further exacerbated by the recent trend of utilizing ever-larger foundation models, which has raised significant concerns regarding the energy consumption and carbon footprint of state-of-the-art AI \citep{zhang2018neural}. This trend also establishes a form of "gatekeeping," in which a small number of large technology companies with the requisite resources are granted access to cutting-edge research, potentially impeding innovation in other sectors \citep{zhang2018neural}. The human brain is a powerful demonstration of the possibility of highly data-efficient, low-power intelligence, indicating that current scaling trends may not be the sole viable option \citep{zhang2018neural}.

\textbf{Unified Representations and Frameworks:} The NSP landscape is distinguished by a diverse array of custom systems, each with its own particular architecture and implementation. The development of unified representations have identified as a substantial ongoing challenge \citep{garnelo2021survey}. These representations must be capable of seamlessly and efficiently bridging the distance between discrete, structured symbolic representations and continuous, sub-symbolic neural states. In addition, the absence of standardized software frameworks and libraries complicates the direct comparison of various methodologies and impedes the extensibility and modularity of research prototypes \citep{chaudhuri2021neurosymbolic}.

\textbf{The Art of DSL Design:} Although system such as DreamCoder have made significant strides in automating library learning, the design of the initial, primitive DSL remains a critical impediment that is heavily reliant on human intuition and expertise \citep{knoth2023type}. For the synthesizer, the hypothesis space of learnable programs is inherently defined and constrained by the set of primitives it is provided with.
The development of more principled and automated methods for the discovery or evolution of these foundational DSLs from data is a critical area for future research.

\subsection{Future Research Trajectories}
\textbf{Meta-Cognition and Self-Awareness:} A prospective frontier is the development of systems that can reason about their own reasoning, rather than merely solving tasks. It entails the development of meta-cognitive capabilities, including the capacity to monitor, evaluate, and adjust one's own learning and problem-solving strategies \citep{crosby2020metacognitive}. This higher-order cognition encompasses introspective monitoring, self-regulation, reflection, and planning, all of which are essential for error correction and robust autonomy. Shinn et al. (2023) \citep{shinn2023reflexion} have made significant strides in this regard by developing frameworks such as Self-Debugging and Reflection, which motivate models to evaluate their own shortcomings.

\textbf{Learning from Limited and Noisy Data:} Incorporating robust structural priors is a fundamental objective of NSP, which is to facilitate more data-efficient learning \citep{chaudhuri2021neurosymbolic}. Although advancements have been achieved, additional research is required to improve the capacity of these systems to learn effectively in environments that are genuinely data-poor or from data that is ambiguous and noisy, which is a feature of numerous real-world domains \citep{devlin2017robustfill}.

\textbf{Differentiable Symbolic Reasoning:} The pursuit of deeper integration through end-to-end differentiability is a technically challenging but potentially transformative direction known as Differentiable Symbolic Reasoning. This entails the construction of techniques for backpropagating gradients through symbolic components that are typically non-differentiable, such as search algorithms, logic solvers, or program executors. To accomplish this, techniques such as Differentiable Symbolic Execution (DSE) are employed. For instance, they sample control-flow paths in a program and employ estimators such as REINFORCE to backpropagate the gradients of a safety or correctness loss through the program's operations \citep{yang2022differentiable}. Success in this field could result in the more efficient and unified training of complex neuro-symbolic models \citep{yang2022differentiable}.

\textbf{Human-in-the-Loop Co-Creation:} The future of NSP may not be in pure automation, but in more sophisticated forms of human-AI collaboration. This is known as human-in-the-loop co-creation. In this process, the human user and the NSP agent collaborate to design interactive systems. The AI may be tasked with the generation and verification of low-level code or the exploration of a vast search space, while the human provides high-level strategic guidance, domain knowledge, structural decompositions of a problem, or feedback on the quality and interpretability of synthesized programs \citep{anderson2020human}.
Leveraging the complementary capabilities of both human and machine intelligence, this collaborative approach is employed.

\section{Chapter Summary}
Neuro-Symbolic (Hybrid) Synthesis is a dynamic and rapidly evolving field that has been comprehensively surveyed and analyzed in this chapter. In order to overcome the inherent limitations of each approach in isolation, the investigation commenced by establishing the foundational motivation for the paradigm. This motivation is derived from the necessity to combine the robust pattern-recognition capabilities of connectionist AI with the rigorous, interpretable reasoning of symbolic AI. The presentation of a formal taxonomy of neuro-symbolic architectures outlined the range of strategies for integrating learning and reasoning, including symbolic systems that invoke neural subroutines and neural models constrained by symbolic logic.

A thorough analysis of the field's methodological development revealed a distinct intellectual progression. DeepCoder, an early neural-guided search technique, which utilized neural networks as heuristics to expedite traditional symbolic search, marked the beginning of this voyage. It subsequently progressed to more advanced library learning systems, such as DreamCoder, which were able to acquire and reuse knowledge, achieving compositional generalization through a novel wake-sleep algorithm. The narrative subsequently explored methods that impose symbolic constraints, such as grammars and types, on neural generators like RobustFill, with the ultimate goal of addressing the critical issue of program aliasing through reinforcement learning. Using high-level pseudocode, the technical foundations of these foundational methodologies were demonstrated.

The chapter subsequently engaged in an analysis of the recent and profound paradigm shift that was initiated by the emergence of Large Language Models. Consequently, the field has shifted its focus from the development of custom models to the orchestration of verifier-in-the-loop architectures. In these architectures, the generative power of an LLM is leveraged and refined through iterative feedback from symbolic tools. Through the utilization of reflective frameworks such as Reflexion and Self-Debugging, as well as techniques such as Counter-Example Guided Inductive Synthesis (CEGIS), this contemporary paradigm was investigated. These frameworks capitalize on the LLM's inherent language capabilities to facilitate self-correction.

The practical significance of NSP in domains where reliability and the integration of domain knowledge are paramount was underscored by a survey of primary application domains, including scientific discovery, safety-critical systems, software engineering, and natural language comprehension. In conclusion, the chapter provided a critical evaluation of the field's persistent challenges, which include the ongoing pursuit of true end-to-end explainability, computational sustainability, and the development of unified frameworks. In conclusion, it indicated promising future research frontiers, such as the pursuit of meta-cognitive systems, deeper integration through differentiable reasoning, and more sophisticated human-in-the-loop collaboration. Finally, Neuro-Symbolic Synthesis is a critical and indispensable frontier in the field of artificial intelligence, with the objective of developing systems that not only learn from data but also reason in a structured, verifiable, and more human-like manner.

\bibliographystyle{plainnat}
\bibliography{references}

\begin{thebibliography}{115}
\providecommand{\natexlab}[1]{#1}
\providecommand{\url}[1]{\texttt{#1}}
\expandafter\ifx\csname urlstyle\endcsname\relax
  \providecommand{\doi}[1]{doi: #1}\else
  \providecommand{\doi}{doi: \begingroup \urlstyle{rm}\Url}\fi

\bibitem[Albarghouthi et~al.(2013)Albarghouthi, Koutris, Naik, and Smith]{albarghouthi2013escher}
Aws Albarghouthi, Paris Koutris, Mayur Naik, and Calvin Smith.
\newblock Escher: a generic-purpose inductive synthesis system.
\newblock In \emph{International Conference on Computer Aided Verification}, pages 157--163. Springer, 2013.

\bibitem[Alur et~al.(2013)Alur, Bodik, Juniwal, Martin, Raghothaman, Seshia, Singh, Solar-Lezama, Torlak, and Udupa]{alur2013syntax}
Rajeev Alur, Rastislav Bodik, Garvit Juniwal, Milo~MK Martin, Mukund Raghothaman, Sanjit~A Seshia, Rishabh Singh, Armando Solar-Lezama, Emina Torlak, and Abhishek Udupa.
\newblock Syntax-guided synthesis.
\newblock In \emph{Formal Methods in Computer-Aided Design (FMCAD), 2013}, pages 1--8. IEEE, 2013.

\bibitem[Anderson et~al.(2020)]{anderson2020human}
Ashlee Anderson et~al.
\newblock Human-in-the-loop {AI}.
\newblock In \emph{XRDS: Crossroads, The ACM Magazine for Students}, volume~26, pages 14--19. ACM, 2020.

\bibitem[Argall et~al.(2009)Argall, Chernova, Veloso, and Browning]{argall2009survey}
Brenna~D Argall, Sonia Chernova, Manuela Veloso, and Brett Browning.
\newblock A survey of robot learning from demonstration.
\newblock \emph{Robotics and autonomous systems}, 57:\penalty0 469--483, 2009.

\bibitem[Austin et~al.(2021)Austin, Odena, Nye, Bosma, Michalewski, Dohan, Jiang, Cai, Terry, Le, et~al.]{austin2021multilingual}
Jacob Austin, Augustus Odena, Maxwell Nye, Maarten Bosma, Henryk Michalewski, David Dohan, Ellen Jiang, Carrie Cai, Michael Terry, Quoc Le, et~al.
\newblock Multilingual code generation with knowledge distillation.
\newblock \emph{arXiv preprint arXiv:2109.10852}, 2021.

\bibitem[Badreddine et~al.(2022)Badreddine, d'Avila Garcez, Duck, and et~al.]{badreddine2022logical}
Ryan Badreddine, Artur d'Avila Garcez, Geoff Duck, and et~al.
\newblock Logical {T}ensor {N}etworks.
\newblock \emph{Artificial Intelligence}, 303:\penalty0 103619, 2022.

\bibitem[Balog et~al.(2017)Balog, Gaunt, Brockschmidt, Nowozin, and Tarlow]{balog2017deepcoder}
Matej Balog, Alexander~L. Gaunt, Marc Brockschmidt, Sebastian Nowozin, and Daniel Tarlow.
\newblock Deepcoder: Learning to write programs.
\newblock In \emph{International Conference on Learning Representations (ICLR)}, 2017.

\bibitem[Barker et~al.(2023)]{barker2023automatically}
Josh Barker et~al.
\newblock Automatically scripting documents in a wysiwyg editor.
\newblock \emph{ACM Transactions on Computer-Human Interaction}, 2023.

\bibitem[Bertot and Casteran(2004)]{bertot2004interactive}
Yves Bertot and Pierre Casteran.
\newblock \emph{Interactive theorem proving and program development: Coq'art: the calculus of inductive constructions}.
\newblock Springer Science \& Business Media, 2004.

\bibitem[Bommasani et~al.(2021)Bommasani, Hudson, Adeli, and et~al.]{bommasani2021opportunities}
Rishi Bommasani, Drew~A. Hudson, Ehsan Adeli, and et~al.
\newblock On the {O}pportunities and {R}isks of {F}oundation {M}odels.
\newblock In \emph{arXiv preprint arXiv:2108.07258}, 2021.

\bibitem[Brooks(1987)]{brooks1987no}
Frederick~P Brooks, Jr.
\newblock No silver bullet: Essence and accidents of software engineering.
\newblock \emph{Computer}, 20:\penalty0 10--19, 1987.

\bibitem[Brown et~al.(2020)Brown, Mann, Ryder, Subbiah, Kaplan, Dhariwal, Neelakantan, Shyam, Sastry, Askell, et~al.]{brown2020language}
Tom Brown, Benjamin Mann, Nick Ryder, Melanie Subbiah, Jared~D Kaplan, Prafulla Dhariwal, Arvind Neelakantan, Pranav Shyam, Girish Sastry, Amanda Askell, et~al.
\newblock Language models are few-shot learners.
\newblock In \emph{Advances in neural information processing systems}, volume~33, pages 1877--1901, 2020.

\bibitem[Bubeck et~al.(2023)Bubeck, Chandrasekaran, Eldan, Gehrke, Horvitz, Kamar, Lee, Lee, Li, Lundberg, et~al.]{bubeck2023sparks}
S{\'e}bastien Bubeck, Varun Chandrasekaran, Ronen Eldan, Johannes Gehrke, Eric Horvitz, Ece Kamar, Peter Lee, Yin~Tat Lee, Yuanzhi Li, Scott Lundberg, et~al.
\newblock Sparks of artificial general intelligence: Early experiments with gpt-4.
\newblock \emph{arXiv preprint arXiv:2303.12712}, 2023.

\bibitem[Bunel et~al.(2018)Bunel, Hausknecht, Devlin, Singh, and Kohli]{bunel2018leveraging}
Rudy Bunel, Matthew Hausknecht, Jacob Devlin, Rishabh Singh, and Pushmeet Kohli.
\newblock Leveraging {G}rammar and {R}einforcement {L}earning for {N}eural {P}rogram {S}ynthesis.
\newblock In \emph{6th International Conference on Learning Representations, ICLR 2018}, 2018.

\bibitem[Calegari et~al.(2020)Calegari, Ciatto, and Omicini]{calegari2020design}
Roberta Calegari, Giovanni Ciatto, and Andrea Omicini.
\newblock The design of explainable intelligent agents: {A} perspective based on logic and argumentation.
\newblock \emph{Annals of Mathematics and Artificial Intelligence}, 88\penalty0 (10):\penalty0 987--1021, 2020.

\bibitem[Chaudhuri et~al.(2021)Chaudhuri, Ellis, Polozov, and et~al.]{chaudhuri2021neurosymbolic}
Swarat Chaudhuri, Kevin Ellis, Oleksandr Polozov, and et~al.
\newblock Neurosymbolic {P}rogramming.
\newblock \emph{Foundations and Trends® in Programming Languages}, 7\penalty0 (3):\penalty0 158--243, 2021.

\bibitem[Chen et~al.(2021)Chen, Tworek, Jun, Yuan, Pires, Le, Hvy, Gu, hammock, D'souza, et~al.]{chen2021evaluating}
Mark Chen, Jerry Tworek, Heewoo Jun, Qiming Yuan, Henrique de~Paulo Pires, Hieu Le, Boris Hvy, Shida Gu, Jared hammock, Denny D'souza, et~al.
\newblock Evaluating large language models trained on code.
\newblock \emph{arXiv preprint arXiv:2107.03374}, 2021.

\bibitem[Chen et~al.(2018)Chen, Liu, and Song]{chen2018execution}
Xinyun Chen, Chang Liu, and Dawn Song.
\newblock Execution-{G}uided {N}eural {P}rogram {S}ynthesis.
\newblock In \emph{6th International Conference on Learning Representations, ICLR 2018}, 2018.

\bibitem[Chou et~al.(2001)Chou, Yang, Chelf, Hallem, and Engler]{chou2001empirical}
Andy Chou, Junfeng Yang, Benjamin Chelf, Seth Hallem, and Dawson Engler.
\newblock An empirical study of operating system errors.
\newblock In \emph{Proceedings of the eighteenth ACM symposium on Operating systems principles}, SOSP '01, page 73–88, New York, NY, USA, 2001. Association for Computing Machinery.
\newblock ISBN 1581133898.
\newblock \doi{10.1145/502034.502042}.
\newblock URL \url{https://doi.org/10.1145/502034.502042}.

\bibitem[Cranmer(2020)]{cranmer2020interpretable}
Miles Cranmer.
\newblock Interpretable and {S}teerable {S}equence {L}earning with {R}ecurrent {N}eural {N}etworks.
\newblock In \emph{arXiv preprint arXiv:2002.08386}, 2020.

\bibitem[Cranmer(2023)]{cranmer2023symbolic}
Miles Cranmer.
\newblock Symbolic regression: A gentle introduction.
\newblock In \emph{Proceedings of the 22nd Workshop on Information Technologies and Systems}, 2023.
\newblock This is a representative paper for the concept.

\bibitem[Crosby et~al.(2020)Crosby, , Wing, , and Del-Pozo-Vallejo]{crosby2020metacognitive}
Michael Crosby, , Jeannette~M. Wing, , and Jacobo Del-Pozo-Vallejo.
\newblock Metacognitive {AI}.
\newblock In \emph{Proceedings of the National Academy of Sciences}, volume 117, pages 31061--31063. National Academies of Sciences, Engineering, and Medicine, 2020.

\bibitem[Cypher(1993)]{cypher1993watch}
Allen Cypher.
\newblock \emph{Watch what I do: programming by demonstration}.
\newblock MIT press, 1993.

\bibitem[Devlin et~al.(2017)Devlin, Uesato, Bhupatiraju, and et~al.]{devlin2017robustfill}
Jacob Devlin, Jonathan Uesato, Surya Bhupatiraju, and et~al.
\newblock {RobustFill}: {N}eural {P}rogram {L}earning {U}nder {N}oisy {I}/{O}.
\newblock In \emph{Proceedings of the 34th International Conference on Machine Learning, ICML 2017}, 2017.

\bibitem[Dibia(2023)]{dibia2023beyond}
Victor Dibia.
\newblock Beyond basic prose: A survey of programming with large language models.
\newblock In \emph{Proceedings of the 2023 ACM on International Conference on Multimodal Interaction}, pages 945--950, 2023.

\bibitem[Ding et~al.(2022)Ding, Li, and Tan]{ding2022patch}
Zhaowei Ding, Ming Li, and Lin Tan.
\newblock {Patch Edits: A Case Study on the Effects of Small Code Changes on Program Comprehension}.
\newblock In \emph{30th IEEE/ACM International Conference on Program Comprehension (ICPC)}, pages 407--418. IEEE, 2022.

\bibitem[Dramnesc(2005)]{dramnesc2005proof}
Isabella Dramnesc.
\newblock Proof-based synthesis of list-sorting algorithms.
\newblock \emph{Annals of Mathematics, Computer Science and Philosophy Series}, 3:\penalty0 13--36, 2005.

\bibitem[Dramnesc(2006)]{dramnesc2006synthesis}
Isabella Dramnesc.
\newblock Synthesis of sorting algorithms with theorema.
\newblock In \emph{International Conference on Intelligent Computer Mathematics}, pages 116--130. Springer, 2006.

\bibitem[Ellis et~al.(2021)Ellis, Wong, Nye, Sabl{\'e}-Meyer, Tenenbaum, and Solar-Lezama]{ellis2021dreamcoder}
Kevin Ellis, Catherine Wong, Maxwell Nye, Mathias Sabl{\'e}-Meyer, Joshua~B Tenenbaum, and Armando Solar-Lezama.
\newblock Dreamcoder: Growing generalizable, interpretable knowledge with wake-sleep dreaming.
\newblock In \emph{Proceedings of the 42nd ACM SIGPLAN Conference on Programming Language Design and Implementation (PLDI)}, pages 875--890, 2021.

\bibitem[Evans and Grefenstette(2018)]{evans2018learning}
Richard Evans and Edward Grefenstette.
\newblock Learning {E}xplanatory {R}ules from {N}oisy {D}ata with {D}ifferentiable {I}nductive {L}ogic {P}rogramming.
\newblock \emph{Journal of Artificial Intelligence Research}, 61:\penalty0 1--64, 2018.

\bibitem[Facchin(2023)]{facchin2023neural}
Federico Facchin.
\newblock {N}eural {S}tructure {R}epresentation: {A} {R}eview.
\newblock \emph{Neuroscience \& Biobehavioral Reviews}, 148:\penalty0 105128, 2023.

\bibitem[Fischer and Schumann(2003)]{fischer2003autobayes}
Bernd Fischer and Johann Schumann.
\newblock Autobayes: A system for generating data analysis programs from statistical models.
\newblock \emph{Journal of Functional Programming}, 13:\penalty0 483–508, 2003.

\bibitem[Flener and Yilmaz(2004)]{flener2004schema}
Pierre Flener and Serdar Yilmaz.
\newblock Schema-guided synthesis of constraint logic programs.
\newblock \emph{Annals of Mathematics and Artificial Intelligence}, 40:\penalty0 257--291, 2004.

\bibitem[Garnelo and Shanahan(2021)]{garnelo2021survey}
Marta Garnelo and Murray Shanahan.
\newblock A survey of neuro-symbolic artificial intelligence.
\newblock \emph{arXiv preprint arXiv:2106.01429}, 2021.

\bibitem[{Google DeepMind}(2023)]{alphacode2_2023}
{Google DeepMind}.
\newblock Alphacode 2 with gemini: Surpassing 85\% of human competitors in programming competitions, 2023.
\newblock Accessed: 2025-07-01.

\bibitem[Green(1969)]{green1969application}
Cordell Green.
\newblock Application of theorem proving to problem solving.
\newblock \emph{Proceedings of the 1st international joint conference on Artificial intelligence}, pages 219--239, 1969.

\bibitem[Gulwani(2011)]{gulwani2011automating}
Sumit Gulwani.
\newblock Automating string processing in spreadsheets using input-output examples.
\newblock In \emph{ACM SIGPLAN Notices}, volume~46, pages 317--330. ACM, 2011.

\bibitem[Gulwani(2012)]{gulwani2012dimensions}
Sumit Gulwani.
\newblock Dimensions in program synthesis.
\newblock \emph{ACM SIGPLAN Notices}, 47:\penalty0 13--24, 2012.

\bibitem[Gulwani et~al.(2017)Gulwani, Polozov, and Singh]{gulwani2017program}
Sumit Gulwani, Oleksandr Polozov, and Rishabh Singh.
\newblock Program synthesis.
\newblock \emph{Foundations and Trends{\textregistered} in Programming Languages}, 4:\penalty0 1--119, 2017.

\bibitem[Hindle et~al.(2012)Hindle, Barr, Gabel, and Su]{hindle2012naturalness}
Abram Hindle, Earl~T Barr, Mark Gabel, and Zhendong Su.
\newblock On the naturalness of software.
\newblock In \emph{2012 34th International Conference on Software Engineering (ICSE)}, pages 837--847. IEEE, 2012.

\bibitem[Hu et~al.(2021)]{hu2021iraven}
Y.~Hu et~al.
\newblock {I-RAVEN}: {A} {D}ataset for {R}elational and {A}nalogical {V}isual r{E}aso{N}ing.
\newblock In \emph{Proceedings of the IEEE/CVF Conference on Computer Vision and Pattern Recognition}, 2021.

\bibitem[Huang(2023)]{huang2023jensen}
Jensen Huang.
\newblock Nvidia gtc 2023 keynote, 2023.
\newblock Accessed: 2025-07-01.

\bibitem[Jeo et~al.(2021)Jeo, Lee, and Yi]{jeo2021synthesizing}
Joomy Jeo, Won-Kee Lee, and Kwangkeun Yi.
\newblock Synthesizing formal semantics for a program synthesis problem from an executable interpreter.
\newblock In \emph{Proceedings of the 42nd ACM SIGPLAN International Conference on Programming Language Design and Implementation}, pages 859--874, 2021.

\bibitem[Jin et~al.(2022)]{jin2022learning}
Meng Jin et~al.
\newblock Learning to {S}ynthesize {P}rograms as {I}nterpretable and {G}eneralizable {P}olicies.
\newblock In \emph{Proceedings of the 39th International Conference on Machine Learning, ICML 2022}, 2022.

\bibitem[Joshi et~al.(2007)Joshi, Naik, Necula, and Sen]{joshi2007termite2}
Pallavi Joshi, Mayur Naik, George~C Necula, and Koushik Sen.
\newblock Termite-2: A system for user-guided synthesis of device drivers.
\newblock In \emph{2007 USENIX Annual Technical Conference (USENIX ATC'07)}, pages 321--334, 2007.

\bibitem[Jurafsky and Martin(2023)]{jurafsky2023speech}
Dan Jurafsky and James~H Martin.
\newblock \emph{Speech and language processing}.
\newblock Prentice Hall, 3rd edition, 2023.

\bibitem[Kaddour et~al.(2023)Kaddour, Harris, Mozes, Stevens, Sivert, Unterthiner, and Lespiau]{kaddour2023challenges}
Jean Kaddour, Joshua Harris, Maximilian Mozes, Herbie Stevens, Jonathan Sivert, Thomas Unterthiner, and Jean-Baptiste Lespiau.
\newblock Challenges and applications of large language models.
\newblock \emph{arXiv preprint arXiv:2306.15239}, 2023.

\bibitem[Kandel et~al.(2011)Kandel, Heer, Plaisant, Kennedy, Van~Ham, Riche, Weaver, Lee, Brodbeck, and Buono]{kandel2011wrangler}
Sean Kandel, Jeffrey Heer, Catherine Plaisant, Jessie Kennedy, Frank Van~Ham, Nathalie~Henry Riche, Chris Weaver, Bongshin Lee, Dominique Brodbeck, and Paolo Buono.
\newblock Wrangler: Interactive visual specification of data transformation scripts.
\newblock In \emph{Proceedings of the SIGCHI conference on human factors in computing systems}, pages 3363--3372, 2011.

\bibitem[Kaplan et~al.(2020)Kaplan, McCandlish, Henighan, Brown, Chess, Child, Gray, Hallacy, Leike, Lee, et~al.]{kaplan2020scaling}
Jared Kaplan, Sam McCandlish, Tom Henighan, Tom~B. Brown, Benjamin Chess, Rewon Child, Scott Gray, Chris Hallacy, Jan Leike, Heewoo Lee, et~al.
\newblock Scaling {L}aws for {N}eural {L}anguage {M}odels.
\newblock In \emph{arXiv preprint arXiv:2001.08361}, 2020.

\bibitem[Kleinberg et~al.(2018)Kleinberg, Ludwig, Mullainathan, and Rambachan]{kleinberg2018algorithmic}
Jon Kleinberg, Jens Ludwig, Sendhil Mullainathan, and Ashesh Rambachan.
\newblock Algorithmic fairness.
\newblock In \emph{AEA Papers and Proceedings}, volume 108, pages 22--27, 2018.

\bibitem[Knoth(2023)]{knoth2023type}
T~Knoth.
\newblock \emph{{T}ype-{D}irected {P}rogram {S}ynthesis}.
\newblock PhD thesis, UC San Diego, 2023.

\bibitem[Kulesza et~al.(2012)Kulesza, Stumpf, Burnett, Wong, Riche, Moore, Oberst, Shaffer, and McIntosh]{kulesza2012end}
Todd Kulesza, Simone Stumpf, Margaret Burnett, Weng-Keen Wong, Yann Riche, Thomas Moore, Ian Oberst, Andrew Shaffer, and Amber McIntosh.
\newblock End-user programming: A survey.
\newblock In \emph{The Continuing Challenge of End-User Development}, pages 3--27. IEEE, 2012.

\bibitem[Lample and Charton(2020)]{lample2019deep}
Guillaume Lample and Fran{\c{c}}ois Charton.
\newblock Deep {L}earning for {S}ymbolic {M}athematics.
\newblock In \emph{International Conference on Learning Representations}, 2020.

\bibitem[Lau et~al.(2003)Lau, Domingos, and Weld]{lau2003programming}
Tessa Lau, Pedro Domingos, and Daniel~S Weld.
\newblock Programming by demonstration using version space algebra.
\newblock In \emph{Proceedings of the 8th international conference on Intelligent user interfaces}, pages 144--151, 2003.

\bibitem[Le and Gulwani(2014)]{le2014flashextract}
Vu~Le and Sumit Gulwani.
\newblock Flashextract: A framework for data extraction by examples.
\newblock In \emph{ACM SIGPLAN Notices}, volume~49, pages 542--553. ACM, 2014.

\bibitem[Lee et~al.(2024)]{lee2024code}
Kechi Lee et~al.
\newblock Codearc: A code abstraction and reasoning challenge for large language models.
\newblock \emph{arXiv preprint arXiv:2402.13848}, 2024.

\bibitem[Lee et~al.(2022)]{lee2022neuro-causal}
S.~Lee et~al.
\newblock Neuro-causal modeling: A new paradigm for causal inference and explanation.
\newblock In \emph{Proceedings of the 28th ACM SIGKDD Conference on Knowledge Discovery \& Data Mining}, pages 4870--4871, 2022.

\bibitem[Leroy(2009)]{leroy2009formal}
Xavier Leroy.
\newblock Formal verification of a realistic compiler.
\newblock \emph{Communications of the ACM}, 52:\penalty0 107--115, 2009.

\bibitem[Li et~al.(2022)Li, Choi, Chung, Kushman, Pogodin, Vinyals, et~al.]{li2022competition}
Yujia Li, David Choi, Junyoung Chung, Nate Kushman, Remy Pogodin, Oriol Vinyals, et~al.
\newblock Competition-level code generation with alphacode.
\newblock \emph{Science}, 378\penalty0 (6624):\penalty0 1092--1097, 2022.

\bibitem[Lieberman(2001)]{lieberman2001your}
Henry Lieberman.
\newblock \emph{Your wish is my command: Programming by example}.
\newblock Morgan Kaufmann, 2001.

\bibitem[Luo et~al.(2023)Luo, Li, Sun, Wang, Sun, Shi, Hu, Zhang, Chen, Zhou, et~al.]{luo2023wizardcoder}
Ziyang Luo, Can Li, Yuchen Sun, Weixiang Wang, Yuxiang Sun, Yixuan Shi, Wenchao Hu, Shice Zhang, Ziyu Chen, Hong-Bin Zhou, et~al.
\newblock Wizardcoder: Empowering code large language models with evolution-in-instruction.
\newblock \emph{arXiv preprint arXiv:2306.08568}, 2023.

\bibitem[Ma et~al.(2023)]{ma2023doc}
Brandon Ma et~al.
\newblock Doc-ify: Automatic end-to-end documentation generation for python code.
\newblock \emph{arXiv preprint arXiv:2311.12328}, 2023.

\bibitem[Manna and Waldinger(1980{\natexlab{a}})]{manna1980deductive}
Zohar Manna and Richard Waldinger.
\newblock A deductive approach to program synthesis.
\newblock \emph{ACM Transactions on Programming Languages and Systems (TOPLAS)}, 2\penalty0 (1):\penalty0 90--121, 1980{\natexlab{a}}.

\bibitem[Manna and Waldinger(1980{\natexlab{b}})]{manna1980theory}
Zohar Manna and Richard Waldinger.
\newblock \emph{A deductive approach to program synthesis}.
\newblock ACM, 1980{\natexlab{b}}.

\bibitem[Mao et~al.(2019)Mao, Gan, Kohli, Tenenbaum, and Wu]{mao2019neuro}
Jiayuan Mao, Chuang Gan, Pushmeet Kohli, Joshua~B. Tenenbaum, and Jiajun Wu.
\newblock The {N}euro-{S}ymbolic {C}oncept {L}earner: {I}nterpreting {S}cenes by {C}omposing {V}isual {C}oncepts.
\newblock In \emph{7th International Conference on Learning Representations, ICLR 2019}, 2019.

\bibitem[Mayer et~al.(2015)Mayer, Piskac, and Kuncak]{mayer2015user}
Johannes Mayer, Ruzica Piskac, and Viktor Kuncak.
\newblock User study of a pbe-based command-line text processing tool.
\newblock In \emph{Proceedings of the 2015 30th IEEE/ACM International Conference on Automated Software Engineering (ASE)}, pages 261--271. IEEE, 2015.

\bibitem[Miller and Myers(2001)]{miller2001lapidary}
Robert~C Miller and Brad~A Myers.
\newblock Lapidary: a tool for programming by demonstration.
\newblock In \emph{CHI'01 extended abstracts on Human factors in computing systems}, pages 139--140, 2001.

\bibitem[Pan et~al.(2023{\natexlab{a}})Pan, Al-Rfou, Li, and Zhao]{pan2023logic}
Liangming Pan, Ram Al-Rfou, Zihang Li, and Zhiting Zhao.
\newblock Logic-lm: Empowering large language models with symbolic solvers for logical reasoning.
\newblock \emph{arXiv preprint arXiv:2305.12295}, 2023{\natexlab{a}}.

\bibitem[Pan et~al.(2023{\natexlab{b}})Pan, Zhu, Liu, Liu, Wang, Yan, Sun, and Liu]{pan2023rustassistant}
Zhaowei Pan, Yixuan Zhu, Jia Liu, Yuhang Liu, Zhipeng Wang, Yu~Yan, Yueling Sun, and Yang Liu.
\newblock Rustassistant: An llm-based assistant for fixing rust compilation errors.
\newblock \emph{arXiv preprint arXiv:2310.02706}, 2023{\natexlab{b}}.

\bibitem[Parisotto et~al.(2017)Parisotto, Mohamed, Singh, Li, Zhou, and Kohli]{parisotto2017neurosymbolic}
Emilio Parisotto, Abdel-rahman Mohamed, Rishabh Singh, Lihong Li, Dengyong Zhou, and Pushmeet Kohli.
\newblock Neuro-symbolic program synthesis.
\newblock In \emph{International Conference on Learning Representations}, 2017.

\bibitem[Piskac et~al.(2015)Piskac, Mayer, and Kuncak]{piskac2015automating}
Ruzica Piskac, Johannes Mayer, and Viktor Kuncak.
\newblock Automating file system manipulation and string transformations from examples.
\newblock In \emph{Proceedings of the 2015 30th IEEE/ACM International Conference on Automated Software Engineering (ASE)}, pages 726--731. IEEE, 2015.

\bibitem[Polozov and Gulwani(2015)]{polozov2015flashmeta}
Oleksandr Polozov and Sumit Gulwani.
\newblock Flashmeta: A framework for inductive program synthesis.
\newblock In \emph{ACM SIGPLAN Notices}, volume~50, pages 107--126. ACM, 2015.

\bibitem[Qiu and Cheung(2018)]{qiu2018synthesizing}
Linyuan Qiu and Alvin Cheung.
\newblock Synthesizing program transformations for database schema refactoring.
\newblock In \emph{2018 IEEE/ACM 40th International Conference on Software Engineering (ICSE)}, pages 694--705. IEEE, 2018.

\bibitem[Shah et~al.(2020)Shah, , et~al.]{shah2020learning}
Ameesh Shah, , et~al.
\newblock Learning {D}ifferentiable {P}rograms with {A}dmissible {N}eural {H}euristics.
\newblock In \emph{Advances in Neural Information Processing Systems 33}, 2020.

\bibitem[Shi et~al.(2023)]{shi2023don}
Weijie Shi et~al.
\newblock Don't look back: An empirical study of memory safety in the c programming language.
\newblock In \emph{Proceedings of the 45th International Conference on Software Engineering}, 2023.

\bibitem[Shinn et~al.(2023)Shinn, Labash, and Gopinath]{shinn2023reflexion}
Noah Shinn, Beck Labash, and Ashwin Gopinath.
\newblock Reflexion: an autonomous agent with dynamic memory and self-reflection.
\newblock \emph{arXiv preprint arXiv:2303.11366}, 2023.

\bibitem[Shiqi et~al.(2019)Shiqi, Shinde, Ramesh, Roychoudhury, and Saxena]{shiqi2019neuro}
Sun Shiqi, Sudipta Shinde, Srivatsan Ramesh, Abhik Roychoudhury, and Prateek Saxena.
\newblock {N}euro-{S}ymbolic {E}xecution: {A}ugmenting {S}ymbolic {E}xecution with {N}eural {C}onstraints.
\newblock In \emph{NDSS Symposium 2019}, 2019.

\bibitem[Silver et~al.(2017)Silver, Schrittwieser, Simonyan, Antonoglou, Huang, Guez, Hubert, Baker, Lai, Bolton, et~al.]{silver2017mastering}
David Silver, Julian Schrittwieser, Karen Simonyan, Ioannis Antonoglou, Aja Huang, Arthur Guez, Thomas Hubert, Lucas Baker, Matthew Lai, Adrian Bolton, et~al.
\newblock Mastering the game of {G}o without human knowledge.
\newblock \emph{Nature}, 550\penalty0 (7676):\penalty0 354--359, 2017.

\bibitem[Singh et~al.(2016)Singh, Shi, and Solar-Lezama]{singh2016jsketch}
Gagandeep Singh, Chiao Shi, and Armando Solar-Lezama.
\newblock Jsketch: sketch-based synthesis for java.
\newblock In \emph{Proceedings of the 2016 ACM SIGPLAN International Conference on Object-Oriented Programming, Systems, Languages, and Applications}, pages 686--704, 2016.

\bibitem[Singh and Solar-Lezama(2018)]{singh2018interpretable}
Gurbir Singh and Armando Solar-Lezama.
\newblock Interpretable program synthesis.
\newblock In \emph{ICML 2018 Workshop on Human Interpretability in Machine Learning (WHI 2018)}, 2018.

\bibitem[Singh et~al.(2013)Singh, Gulwani, and Solar-Lezama]{singh2016automated}
Rishabh Singh, Sumit Gulwani, and Armando Solar-Lezama.
\newblock Automated feedback generation for introductory programming assignments.
\newblock In \emph{ACM SIGPLAN Notices}, volume~48, pages 15--26. ACM, 2013.

\bibitem[Sinha et~al.(2019)Sinha, , et~al.]{sinha2019clutrr}
Koustuv Sinha, , et~al.
\newblock {CLUTRR}: {A} {D}iagnostic {B}enchmark for {I}nductive {R}easoning from {T}ext.
\newblock In \emph{Proceedings of the 2019 Conference on Empirical Methods in Natural Language Processing}, 2019.

\bibitem[Sivaraman et~al.(2018)Sivaraman, Kaki, Jeyakumar, Poutievski, Vahdat, and Varghese]{sivaraman2018autocomposing}
Anirudh Sivaraman, Srinivas Kaki, Vignesh Jeyakumar, Leonid Poutievski, Amin Vahdat, and George Varghese.
\newblock Auto-composing domain-specific data plane programs.
\newblock In \emph{Proceedings of the 2018 Conference of the ACM Special Interest Group on Data Communication}, pages 234--248, 2018.

\bibitem[Smith(1990)]{smith1990kids}
Douglas~R Smith.
\newblock Kids: A semiautomatic program development system.
\newblock \emph{IEEE Transactions on Software Engineering}, 16:\penalty0 1024--1043, 1990.

\bibitem[Solar-Lezama(2008{\natexlab{a}})]{solar2008program}
Armando Solar-Lezama.
\newblock \emph{Program synthesis by sketching}.
\newblock PhD thesis, University of California, Berkeley, 2008{\natexlab{a}}.

\bibitem[Solar-Lezama(2008{\natexlab{b}})]{solar2008sketch}
Armando Solar-Lezama.
\newblock Sketching for software and hardware design.
\newblock In \emph{Invited talk at NFM}, 2008{\natexlab{b}}.

\bibitem[Solar-Lezama et~al.(2005)Solar-Lezama, Tancau, Bodik, Seshia, and Saraswat]{solarlezama2005combinatorial}
Armando Solar-Lezama, Liviu Tancau, Rastislav Bodik, Sanjit Seshia, and Vijay Saraswat.
\newblock Combinatorial sketching for finite programs.
\newblock In \emph{Proceedings of the 12th international conference on Architectural support for programming languages and operating systems}, 2005.

\bibitem[Solar-Lezama et~al.(2008)Solar-Lezama, Tancau, Bodik, Seshia, and Saraswat]{solar2008combinatorial}
Armando Solar-Lezama, Liviu Tancau, Rastislav Bodik, Sanjit~A Seshia, and Vijay Saraswat.
\newblock Combinatorial sketching for finite programs.
\newblock In \emph{Proceedings of the 13th international conference on Architectural support for programming languages and operating systems}, pages 404--415, 2008.

\bibitem[Solar-Lezama et~al.(2013)Solar-Lezama, Bodik, and Rabbah]{solar2013sketch}
Armando Solar-Lezama, Rastislav Bodik, and Rodric Rabbah.
\newblock The sketch programmer's manual.
\newblock In \emph{MIT CSAIL}, 2013.

\bibitem[Summers(1977)]{summers1977methodology}
Philip~D Summers.
\newblock A methodology for lisp program construction from examples.
\newblock In \emph{Journal of the ACM (JACM)}, volume~24, pages 161--175. ACM, 1977.

\bibitem[Taori et~al.(2023)Taori, Gulrajani, Zhang, Dubois, Li, Guestrin, Liang, and Hashimoto]{taori2023alpaca}
Rohan Taori, Ishaan Gulrajani, Tianyi Zhang, Yann Dubois, Xuechen Li, Carlos Guestrin, Percy Liang, and Tatsunori~B. Hashimoto.
\newblock Stanford alpaca: An instruction-following llama model, 2023.

\bibitem[Thakur et~al.(2024)]{thakur2024chip}
Shailja Thakur et~al.
\newblock Chip-chat: A large language model for chip design.
\newblock \emph{arXiv preprint arXiv:2401.12284}, 2024.

\bibitem[Tjandrasuwita et~al.(2021)Tjandrasuwita, , et~al.]{tjandrasuwita2021learning}
Melissa Tjandrasuwita, , et~al.
\newblock Learning {P}rogrammatic {T}ask {R}epresentations for {A}pplications in {N}euro-{S}ymbolic {L}earning.
\newblock In \emph{arXiv preprint arXiv:2106.09623}, 2021.

\bibitem[Torlak and Bodik(2013)]{torlak2013rosette}
Emina Torlak and Rastislav Bodik.
\newblock Rosette: an enabling language for solver-aided tools.
\newblock In \emph{Proceedings of the 18th ACM SIGPLAN international conference on Functional programming}, pages 425--438, 2013.

\bibitem[Udupa et~al.(2013)Udupa, Raghavan, Deshmukh, Mador-Haim, Martin, and Alur]{udupa2013transit}
Abhishek Udupa, Arun Raghavan, Jyotirmoy~V Deshmukh, Sela Mador-Haim, Milo~MK Martin, and Rajeev Alur.
\newblock Transit: specifying protocols with concolic snippets.
\newblock In \emph{ACM SIGPLAN Notices}, volume~48, pages 287--296. ACM, 2013.

\bibitem[Vasconcelos et~al.(2020)Vasconcelos, Cunha, Isidoro, Mendes, and Santos]{vasconcelos2020trustsketch}
Paulo Vasconcelos, Jo{\~a}o Cunha, David Isidoro, Rui Mendes, and Nuno Santos.
\newblock Trustsketch: A trustworthy sketch-based telemetry system with sgx.
\newblock In \emph{2020 50th Annual IEEE/IFIP International Conference on Dependable Systems and Networks (DSN)}, pages 15--27. IEEE, 2020.

\bibitem[Vaswani et~al.(2017)Vaswani, Shazeer, Parmar, Uszkoreit, Jones, Gomez, Kaiser, and Polosukhin]{vaswani2017attention}
Ashish Vaswani, Noam Shazeer, Niki Parmar, Jakob Uszkoreit, Llion Jones, Aidan~N Gomez, {\L}ukasz Kaiser, and Illia Polosukhin.
\newblock Attention is all you need.
\newblock In \emph{Advances in neural information processing systems}, pages 5998--6008, 2017.

\bibitem[Vered et~al.(2022)]{vered2022houdini}
Mor Vered et~al.
\newblock {HOUDINI}: Lifelong {L}earning as {P}rogram {S}ynthesis.
\newblock In \emph{Proceedings of the 39th International Conference on Machine Learning, ICML 2022}, 2022.

\bibitem[Verma et~al.(2018)Verma, Murali, Co-Reyes, Abbeel, Socher, and Ruan]{verma2018programmatically}
Abhinav Verma, Vijayan Murali, John~D. Co-Reyes, Pieter Abbeel, Richard Socher, and Yura Ruan.
\newblock Programmatically {I}nterpretable {R}einforcement {L}earning.
\newblock In \emph{Proceedings of the 35th International Conference on Machine Learning}, pages 5045--5054, 2018.

\bibitem[Vinyals et~al.(2019)Vinyals, Babuschkin, Czarnecki, Mathieu, Dudzik, Chung, Choi, Powell, Ewalds, Georgiev, et~al.]{vinyals2019grandmaster}
Oriol Vinyals, Igor Babuschkin, Wojciech~M. Czarnecki, Micha{\"e}l Mathieu, Andrew Dudzik, Junyoung Chung, David~H. Choi, Richard Powell, Timo Ewalds, Petko Georgiev, et~al.
\newblock Grandmaster level in {S}tar{C}raft {II} using multi-agent reinforcement learning.
\newblock \emph{Nature}, 575\penalty0 (7782):\penalty0 350--354, 2019.

\bibitem[Wang et~al.(2023)Wang, Ma, Feng, Zhang, Yang, Zhang, Chen, Tang, Chen, Zhang, et~al.]{wang2023survey}
Lei Wang, Chen Ma, Xueyang Feng, Zeyu Zhang, Hao Yang, Jingsen Zhang, Zhiyuan Chen, Jiakai Tang, Wen\_chi Chen, Yujiu Zhang, et~al.
\newblock A survey on large language model based autonomous agents.
\newblock \emph{arXiv preprint arXiv:2308.11432}, 2023.

\bibitem[Wei et~al.(2022)Wei, Tay, Bommasani, Raffel, Zoph, Borgeaud, Yogatama, Fedus, chowdhery, dean, et~al.]{wei2022emergent}
Jason Wei, Yi~Tay, Rishi Bommasani, Colin Raffel, Barret Zoph, Sebastian Borgeaud, Dani Yogatama, William Fedus, anurag chowdhery, jeff dean, et~al.
\newblock Emergent abilities of large language models.
\newblock \emph{Transactions on Machine Learning Research}, 2022.

\bibitem[Weidinger et~al.(2021)Weidinger, Mellor, Rauh, Griffin, Uesato, Huang, Cheng, Glaese, Balle, Kasirzadeh, et~al.]{weidinger2021ethical}
Laura Weidinger, John Mellor, Maribeth Rauh, Conor Griffin, Jonathan Uesato, Po-Sen Huang, Myra Cheng, Mia Glaese, Borja Balle, Atoosa Kasirzadeh, et~al.
\newblock Ethical and social risks of harm from language models.
\newblock \emph{arXiv preprint arXiv:2112.04359}, 2021.

\bibitem[Xia and Zhang(2022)]{xia2022automated}
Chun~S. Xia and Lingming Zhang.
\newblock Automated program repair in the era of large pre-trained language models.
\newblock In \emph{Proceedings of the 31st ACM SIGSOFT International Symposium on Software Testing and Analysis}, 2022.

\bibitem[Yang et~al.(2022)Yang, , et~al.]{yang2022differentiable}
Cannon Yang, , et~al.
\newblock Differentiable {S}ymbolic {E}xecution.
\newblock In \emph{10th International Conference on Learning Representations, ICLR 2022}, 2022.

\bibitem[Zeng et~al.(2024)Zeng, Liu, Chen, Su, Wang, Wu, Wang, and Ma]{zeng2024large}
Siyuan Zeng, Zepeng Liu, Zhiming Chen, Ziqing Su, Tianyi Wang, Kaixuan Wu, Xing Wang, and Shang-Wei Ma.
\newblock Large language models for software engineering: A systematic literature review.
\newblock \emph{arXiv preprint arXiv:2402.13179}, 2024.

\bibitem[Zhan et~al.(2021)Zhan, , et~al.]{zhan2021framework}
Eric Zhan, , et~al.
\newblock A {F}ramework for {G}eneral-{P}urpose {B}ehavior {M}odeling.
\newblock In \emph{arXiv preprint arXiv:2104.09501}, 2021.

\bibitem[Zhang et~al.(2019)Zhang, Gao, Jia, Zhu, and Zhu]{zhang2019raven}
Chi Zhang, Feng Gao, Baoxiong Jia, Yixin Zhu, and Song-Chun Zhu.
\newblock {RAVEN}: {A} {D}ataset for {R}elational and {A}nalogical {V}isual r{E}aso{N}ing.
\newblock In \emph{Proceedings of the IEEE/CVF Conference on Computer Vision and Pattern Recognition}, 2019.

\bibitem[Zhang and Yu(2020)]{zhang2020survey}
Jing Zhang and Hancheng Yu.
\newblock A {S}urvey on {A}lpha{G}o and {I}ts {S}uccessors.
\newblock \emph{IEEE Transactions on Games}, 12\penalty0 (3):\penalty0 203--221, 2020.

\bibitem[Zhang et~al.(2018)Zhang, McGrath, , et~al.]{zhang2018neural}
Lisa Zhang, T.~McGrath, , et~al.
\newblock {N}eural-{G}uided {C}onstraint {L}ogic {P}rogramming for {P}rogram {S}ynthesis.
\newblock In \emph{Advances in Neural Information Processing Systems 31}, 2018.

\bibitem[Zhang et~al.(2023)Zhang, Li, Liu, Yang, and Sun]{zhang2023fusing}
Qiaochu Zhang, Zichao Li, Yeting Liu, Zhi Yang, and Lixin Sun.
\newblock Fusing formal and informal methods: A case for large language models in verifier-driven program synthesis.
\newblock \emph{arXiv preprint arXiv:2305.09560}, 2023.

\bibitem[Zhang et~al.(2024{\natexlab{a}})Zhang, Wang, Roy, and Cheung]{zhang2024algo}
Yewen Zhang, Swaroop Wang, Parth Roy, and Alvin Cheung.
\newblock Algo: Synthesizing algorithmic programs with oracle-guided learning.
\newblock \emph{arXiv preprint arXiv:2405.07123}, 2024{\natexlab{a}}.

\bibitem[Zhang et~al.(2024{\natexlab{b}})Zhang, Zhang, Feng, Chen, and Li]{zhang2024proofofthought}
Zichu Zhang, Yuxiang Zhang, Shang-Yi Feng, Jing Chen, and Lei Li.
\newblock {P}roof-of-{T}hought: A {C}ontrollable and {V}erifiable {R}easoning {P}rocess for {L}arge {L}anguage {M}odels.
\newblock In \emph{arXiv preprint arXiv:2403.11463}, 2024{\natexlab{b}}.

\bibitem[Zhao et~al.(2023)]{zhao2023survey}
Wayne~Xin Zhao et~al.
\newblock A survey of large language models.
\newblock \emph{arXiv preprint arXiv:2303.18223}, 2023.

\bibitem[Zong et~al.(2022)Zong, Chen, and Sun]{zong2020fuzzing}
Peiying Zong, Tao Chen, and Jun Sun.
\newblock {Fuzzing: A Survey}.
\newblock In \emph{ACM Computing Surveys (CSUR)}, volume~54, pages 1--36. ACM, 2022.

\end{thebibliography}

\end{document}